\documentclass[useAMS,usenatbib]{mn2e}

\usepackage{graphicx,amssymb,url}
\usepackage{placeins}
\usepackage{amsmath}

\title[Searching for Transient Obscuration Events]{Variability in the 2MASS Calibration Fields: A Search for Transient Obscuration Events}

\author[Quillen et al.]{Alice C. Quillen$^1$,  Marco Ciocca$^2$, Jeffrey L. Carlin$^3$,
Zeyang Meng$^{1,4}$,  \&
\newauthor
Cameron P. M. Bell$^{1,5}$
\\ 
$^1$ Department of Physics and Astronomy, University of Rochester, Rochester, NY 14627, USA\\
$^2$ Department of Physics and Astronomy, Eastern Kentucky University, Richmond, KY 40475, USA\\
$^3$ Department of Physics, Applied Physics and Astronomy, Rensselaer Polytechnic Institute, Troy, NY 12180, USA\\
$^4$ Department of Astronomy \& Key Laboratory of Modern Astronomy and Astrophysics in Ministry of Education, \\
Nanjing University, Nanjing, 210093, China \\
$^5$ School of Physics, University of Exeter, Exeter EX4 4QL, UK\\
}

\begin{document}
\maketitle

\begin{abstract}

We searched the light curves of over 40000 stars in the 2 Micron All Sky Survey (2MASS) calibration database, spanning
approximately 4 years, for objects that have significant day long dimming events.  
We also searched the multi-color light curves
for red dimming events that could be due to transient extinction.
In the color independent sigma-limited search, 
we found 46 previously unknown eclipsing binaries, 6 periodic variable stars 
likely to be intrinsic pulsators and 21 young stellar objects in the $\rho$ Ophiucus 
star formation region previously studied by Parks et al. 2013.
An additional 11 objects exhibited dimming events and most of these are unclassified. 

The search for red dimming events primarily reveals a population of low luminosity active galaxies that become bluer when
they are brighter, and variable young stellar objects exhibiting high cross-correlation coefficients
between color and brightness.   
The young stellar objects exhibit brightness and color variations
in the direction of interstellar extinction 
whereas the active galaxies can have a bowed distribution in color and magnitude with reduced variation in color 
when the object is brightest.
By plotting cross-correlation coefficients (between color and brightness) against the slope fit to points
in the color/magnitude distribution,
we find that active galaxies  tend to have lower slopes and correlation coefficients.

Among the objects that are usually quiescent (not strongly variable), we failed to find any 
dimming events deeper than 0.2 magnitude and lasting longer than a day. Two of the least embedded
young stellar objects, however, dimmed by 0.2 mag for longer than a day without strong color variation.
Having failed to find new exotic objects, we explored ways to eliminate commonly found objects so
that a larger number of objects may be searched.
We find that  all but one eclipsing binary are excluded 
by requiring moderate color variation during a dimming event and most of the
active galaxies and young stellar objects are excluded by placing a limit on the standard deviation 
of the magnitude distribution in the light curve.

\end{abstract}

\section{Introduction}

One star in a young stellar binary system can host a circumstellar disc and the disc
can periodically occult the other star as seen from a distance
(e.g., $\epsilon$ Aurigae,
 \citealt{guinan02,kloppenborg10,chadima11}; EE-Cep, \citealt{mikolajewski99,graczyk03,mikolajewski05,galan12};
  OGLE-LMC-ECL-17782, \citealt{graczyk11};  OGLE-LMC-ECL-11893, \citealt{dong14}). 
Because discs can be large,
the probability that a randomly oriented but young system exhibits eclipses may be fairly high 
 \citep{mamajek12,meng14}. 
 Circumbinary disks can also periodically occult a star 
 (e.g., KH 15D, \citealt{kearns98,winn04}; V718 Per, \citealt{grinin08}).
 Periodic and aperiodic occultations have been detected in young stellar objects \citep{bouvier07,grinin08,moralescalderon11,parks13,wolk13,cody14}.
 Transient single, long, and deep occultation events of pre-main sequence stars have also been reported 
\citep{rodriguez13,mamajek12}.   
Occulting disks are valuable as they would allow the structure of a distant disk to be studied
on smaller spatial scales than possible with direct imaging and it would be possible to study their composition
through absorption of star light. 

The Two Micron All-Sky Survey (\citealt{skrutskie06}; 2MASS) 
imaged the entire sky in three near infrared wavelength bands, $J, H, K$, between 1997 and 2001. 
Photometric calibration for 2MASS relied upon repeated observations of 35 calibration fields that are distributed at approximately 2 hour intervals in right ascension near declinations of 0 and $\pm 30$ deg. 
%
All source extractions from the calibration scans are available int the 2MASS Calibration Point Source Working Database 
(Cal-PSWDB). The database contains over 191 million source extractions. 
Further descriptions of the Cal-PSWDB are available by \citet{cutri06}, and \citet{plavchan08} who found
247 red variable objects, including extragalactic variables and 23 periodic variable stars. 
Using one of the calibration fields,
\citet{parks13} recently presented a variability study of the $\rho$ Ophiucus molecular cloud and star formation region,
discovering 22 new cluster members via their near-infrared variability and periodicity in 35 young stellar objects.

While numerous photometric monitoring projects search for objects that brighten, such 
as novae or micro lenses,  there have been few systematic searches for rare, individual dimming events.
Single, isolated, dimming events are often ignored because they can be caused by poor photometric
measurements that occur because of seeing variations, scintillation, clouds, imaging artifacts, and source confusion. 
Variability studies  search for periodic phenomena, including planetary transits, 
(e.g., \citealt{colliercameron06,palaversa13}),
use the moments of the magnitude distribution to find and classify variables (e.g., \citealt{ciocca13,graczyk10})
or combine periods found in one band with color or spectral information (not necessarily as a function of time) 
to classify variable objects (e.g., \citealt{palaversa13}).
Because photometric pipelines can report false dimming events,  rare and deep dimming events
have only been confirmed when the obscuration is periodic (e.g., the eclipsing disk OGLE-LMC-ECL-11893
with a 468 day period; \citealt{dong14})
or multiple observation platforms observed the same drop in magnitude (e.g., RW Aur A;  \citealt{rodriguez13}).
 
Here we focus on searching for potentially rare events and so we do not restrict our study to periodic
stars or stars that are sufficiently variable that they can be found via statistical measurements of
the magnitude distribution.  
We carry out a multi-band search for obscuration events in a fairly large sample, over 40000 stars, 
the brighter stars in the 2MASS Cal-PSWDB. 
The Cal-PSWDB is uniformly observed and processed, and has  well
characterized astrometric and photometric measurements.  It is also a multi-band photometric survey
with all bands measured at each observation time.
Variability is often studied in a single photometric wavelength band, however, if obscuration is
due to extinction from dust, then the depth of a dimming event or eclipse can depend upon wavelength.
For example it is possible to differentiate between star spots and obscuration using near-infrared  color 
\citep{moralescalderon11,parks13,wolk13,cody14}.
The eclipses of the OGLE-LMC-ECL-11893 and EE-Cep disk systems are deeper in $B$ band than $I$ band \citep{galan12,dong14}.
In a three band photometric study, it is also possible to check the uniformity or noise level of measurements at a particular
time by comparing the measurements in the three bands. 
Here we carry out a search for dimming events in single bands and also search for dimming events that are red and so could be 
due to transient levels of extinction.

\section{Searching for transient levels of obscuration}

Source lists in each 2MASS calibration tile are available at 
\url{http://www.ipac.caltech.edu/2mass/releases/allsky/doc/seca7_4.html}.
Using those source lists we selected all objects with
$J < 16.0$ or  $K < 15.0$ where $J$ and $K$ are the $J$- and $K$-band magnitudes, respectively.
This selection ensures bright enough sources that individual photometric measurements
are accurate with standard deviation approximately 0.1 mag or better; (this procedure was also
followed by \citealt{plavchan08}; see their Figure 7). 
Light curves for 
48,976 objects, satisfying these criteria and sorted in order of time of observation,
were downloaded from the 2MASS Cal-PSWDB 
by querying all observations within 3'' of the positions on the sky
 listed in the source lists.
Not all of the 48,976 objects gave light curves with many data points, 
consequently the actual number of stars that can be studied for variability is 
somewhat smaller (see the discussion by \citealt{plavchan08}).  
We counted 40,486 objects with greater than 300 observation times (all 3 bands are usually observed at each time) 
and of these 22,983 objects with greater than 1000 observation times.

From each light curve we measured the mean, $\mu_J, \mu_H, \mu_K$, and standard deviation, $\sigma_J, \sigma_H, \sigma_K$,
 of the magnitude distribution in each band.
To compute the means and standard deviations, observation points were weighted by the 
variance of each photometric measurement.  
We only used light curve points with non {\it null} listed errors, i.e.,  those with good photometric measurements.
We also box smoothed the light curves over a 1 day long observation interval by computing the weighted mean
and standard deviation of the magnitudes observed within a day 
of each observation time.   Uncertainties  for each data point in the box-smoothed curve 
were calculated from the data points used to
make the averages. 

We carried out two searches.  First in each band, separately, we searched for day long dimming events that are significant
compared to the dispersion of points in the light curves.
Second, we searched for dimming events that are comparatively red.

For objects found in the searches we investigate a possible
correlation between color and brightness. 
We compute the cross correlation coefficient\footnote{\url{http://mathworld.wolfram.com/CorrelationCoefficient.html} } using only data points with  non-null $J$ and $K$ magnitudes
\begin{equation}
R  =\dfrac{ \sum_i (J_i  - \mu_J)(J_i - K_i - (\mu_J - \mu_K))}{ \sqrt{ \sum_i (J_i - \mu_J)^2 \sum_i (J_i - K_i -(\mu_J - \mu_K ))^2}}
\label{eqn:R}
\end{equation}
Here $R=1$ for a light curve that reddens as it dims (with not scatter or noise) and $R=0$ if there is no correlation between
brightness and color.
We also fit (using least mean squares) a line to points on a color/magnitude plot where observations at different times 
in $J$ band are compared to their $J-K$ color (as done by \citealt{wolk13}).
We recorded the slope
\begin{equation}
b \equiv \frac{\Delta J}{\Delta( J-K)}  \label{eqn:b}
\end{equation}
of the best fitting line on the color/magnitude plot.
If the color variations and dimming are due to extinction from interstellar dust then we expect
a slope $b=1.6$.
 
\section{Color insensitive search for dimming events \label{S:CB}}
 
We first describe a color insensitive, but sigma-limited, search for dimming events. 
For each night's observations we measured the median magnitude in each band.
We required there to be greater than 4 good (non {\it null} error) observations during the night.
We also measured the median value and standard deviation for the night's observations, $\mu_n,\sigma_n$.
A candidate dimming event required the night's median value to be 
1) fainter than 2 $\sigma$ (dispersion of entire light curve)
from the mean of the entire light curve, 
2) fainter than 1.5$\sigma_n$ (night's dispersion) 
from the mean of the entire light curve and  
3) greater than 0.1 mag from the mean of the entire light curve.  
Here $\sigma$ is the standard deviation computed from
the entire light curve in the band used for the search.  This procedure pulls out rare dimming events compared
to the distribution of the entire light curve.

At each candidate dimming event, we visually inspected the light curves.
Dimming events in noisy light curves were discarded. The noise was usually due to faintness of the star, 
but was sometimes due to confusion with nearby sources.
Care was taken to check the time of individual dimming events. 
 If more than 1 star in the same calibration tile showed a dimming event
on the exact same night, the events were discarded.  In some calibration tiles, many stars
exhibited false dimming events on the same night. 

For each star with a candidate dimming event, we searched the 
VizieR database\footnote{The VizieR database is hosted and developed by 
Centre de Donn\'ees astronomiques de Strasbourg \url{http://vizier.u-strasbg.fr/viz-bin/VizieR}}
for variability.  Excluding the $\rho$ Ophiucus region, we found 15 previously known periodic systems, listed in the 
American Association of Variable Star Observer's (AAVSO)
 International Variable Star Index\footnote{Watson, C., Henden, A.A., \& Price, A. 2006--2014 \url{http://www.aavso.org/vsx}}
  and one previously known variable  
extragalactic source (2MASS 14584479+3720216), previously noted by \citet{plavchan08}.  
The previously known periodic stars are
2MASS 03315574+3703122 and 2MASS 03314856+3723373
(eclipsing binaries found by J. Greaves\footnote{\url{
http://www.astronet.ru/db/varstars/msg/1234323}}),
2MASS J08255405-3908441,
2MASS J18510479-0442005,
2MASS J18512034-0426311,
2MASS J18512261-0409084, 
2MASS J18512929-0412407, and
2MASS J19020989-0439440, that were previously noted as eclipsing binaries by \citet{plavchan08},
and the following that are listed in the AAVSO index:
2MASS J03322127+3701153 (T-Per1-19957; eclipsing binary)
2MASS J08511335+1151401 (HU Cnc; RS Canum Venaticorum-type system),
2MASS J08511799+1145541 (HV Cnc; eclipsing binary)
2MASS J08512079+1153261 (ES Cnc; eclipsing binary), 
2MASS J08512814+1149274 (EV Cnc; eclipsing binary),
2MASS J20313481-4916270 (ASAS J203135-4916.5, Delta-Scuti type variable)
2MASS J22004033+2056425 (GSC 01692-01074; eclipsing binary, found by Patrick Wils using the
 2MASS Cal-PSWDB).

In the $\rho$ Ophiucus molecular cloud region 21 stars were found in our search for dimming events.  
This region has been comprehensively studied previously for variability with the same data \citep{parks13},
and in this region, we only found dimming events in highly variable stars that were previously
identified by \citet{parks13}.  We found no quiescent star in the region, spending most of
its time with little or no variability, that exhibited a dimming event.
We defer the discussion of these objects until after we discuss the red dimming event search below. 

\subsection{Periodic stars}

\begin{figure*}
\includegraphics[width=5.0in, trim= 0 0 0 0 ]{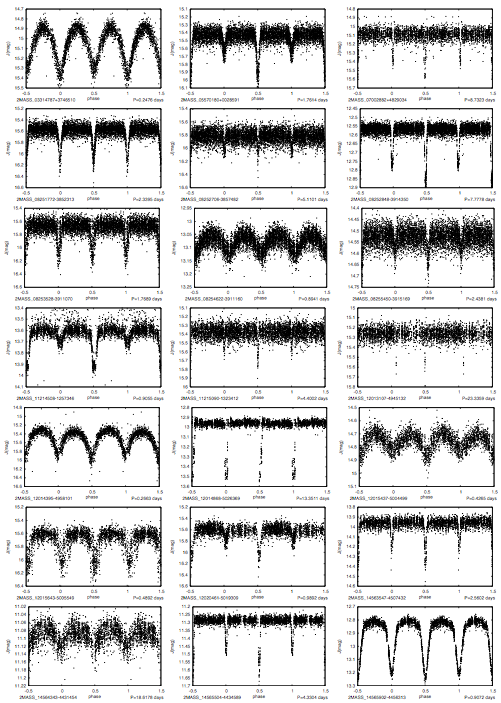}%
\caption{Phased (period-folded) light curves of eclipsing binary candidates in J-band found in a search
for dimming events.
}
\label{fig:EB}
\end{figure*}

For stars with candidate dimming events that were not previously identified as variable stars, (and excluding
the $\rho$-Ophiucus region) 
we searched for periodicity.  We  computed the frequency spectrum of the light curve and then examined 
the phased or period-folded light curves at the locations of peaks or multiples of peaks in the spectrum.
Periods were measured with the software package
Peranso Light Curve and Period Analysis Software\footnote{\url{http://www.peranso.com/} by Tonny Vanmunster of
the CBA Belgium Observatory} or with the 
Period04 Java package\footnote{P. Lenz and M. Breger, 2005, Period04 User Guide, 
Communications in Asteroseismology, 146, 53}.
46 new eclipsing binary candidates and 6 stars with periodic light curves (intrinsic pulsators such as RR-Lyrae type variables) 
were found in this search.  
Phased or period-folded light curves for these stars are shown in Figure \ref{fig:EB} (eclipsing binaries) and 
Figure \ref{fig:RR} (intrinsic pulsators).
Periods for these stars are listed with their 2MASS J identifier in Tables \ref{tab:EB} and \ref{tab:RR}.
In these tables, J-band photometry and $J-K$ color were taken from the 2MASS Point Source Catalog.
Estimated primary eclipse depth, in $J$ magnitudes,  and the Julian dates of a primary eclipse 
are listed  in \ref{tab:EB} for the eclipsing binaries.  
Tentative eclipsing binary types 
EA (Algol type; detached) or EB ($\beta$ Lyrae type; semi-detached) or EW (W Ursae Majoris type or contact) 
were assigned based on the shape of the phased light curves.
For the intrinsic pulsators, peak to peak amplitudes in $J$ band and the Julian date of a 
time of minimum are listed in Table \ref{tab:RR}.  
We also list colors computed from mid infrared photometric measurements from the
Wide-Field Infrared Survey Explorer (WISE) \citep{wise,wright12}.    The WISE observatory mapped
the sky at 3.4, 4.6, 12 and 22~$\mu$m and these are referred to as the W1, W2, W3 and W4 bands, respectively.
Zero points for the bands in Jy are given by \citet{jarrett11}.

From Table \ref{tab:EB} a couple of the eclipsing binary candidates (e.g., 2MASS J03314787+3746510) 
have red enough colors that they are
in the sample searched by  \citet{plavchan08}, however they were not identified as variable stars.
A few of the eclipsing binaries, 2MASS J03314787+3746510, 2MASS J12014395-4958101, 2MASS J18391249+4845394
 are contact eclipsing binaries with short periods $P \sim 0.25$ days.
They are  contact main sequence eclipsing binaries or W UMa systems.   They are of interest
as distance indicators and for constraining stellar models (see for example \citealt{dimitrov10,norton11,vaneyken11,nefs12}).
The periods of these three contact eclipsing binaries  are not so short as to make them exceptional compared 
to those previously discovered. 
 2MASS J16265638+0552039  has a period of 0.339974 days but it is not a contact binary or a W Uma system, 
 it is detached.
 Its galactic longitude is 34 degrees implying negligible extinction in the near-IR bands and its color fairly red, $J-K=0.92$.  
The infrared colors would be consistent with an M3 and not later spectral type (placing it on Figure 11 by \citealt{vaneyken11}).
This system is similar to the interesting short period eclipsing M-dwarf binaries, including detached binary systems, recently 
found by \cite{nefs12}.

We looked in the EB sample to see if any had interesting infrared excesses and so were likely to be young.  
There were two 
with WISE mid-infrared fluxes, 2MASS J19015103-0419487 and 2MASS  J19020201-0430160 listed in the all sky 
catalog \citep{wise}.
However both are near the Galactic plane and inspection of the WISE imaging atlas implied
that there was confusion with mid-infrared extended emission.

Four of the periodic variables that do not resemble eclipsing binaries have short periods (less than 4 days) and
folded-light curves resembling RR-Lyrae type variables.
However, a semi-regular or long period variable star was found, 2MASS J19015057-0431087, 
with a period of $\sim$ 220 days.  
This object has red WISE colors [3.4]-[4.6]=0.41 (Vega magnitudes) and [4.6]-[12]= 0.77 mag, implying that the object
has a dusty envelope.  Its period and mid-infrared colors are similar
to the semi-regular or long period variables discussed by \citet{palaversa13} that 
could be AGB stars.
The scatter in the period-folded light curve is larger than the photometric errors because 
 the light curve is not  reproducible between different cycles.  This behavior was also seen in the long
 period variables identified by \citet{palaversa13} (see their figure 16).




\setcounter{figure}{0}
\begin{figure*}
\includegraphics[width=5.0in, trim= 0 0 0 0 ]{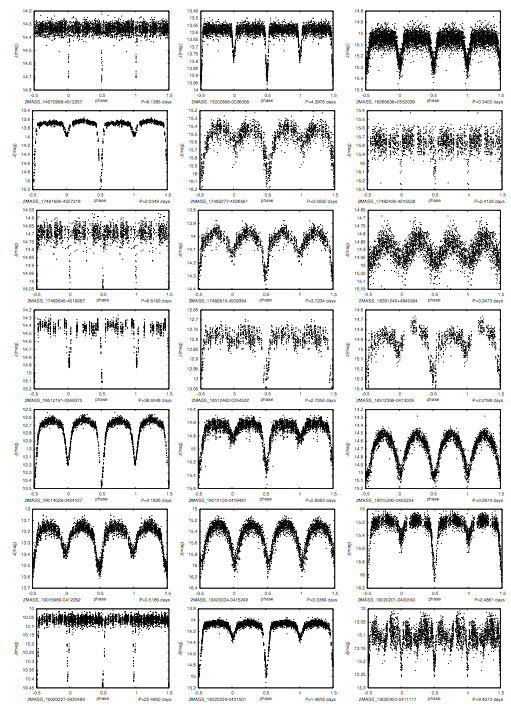}%
\caption{Figure continued.}
\end{figure*}

\setcounter{figure}{0}
\begin{figure*}
\includegraphics[width=5.0in, trim= 0 0 0 0 ]{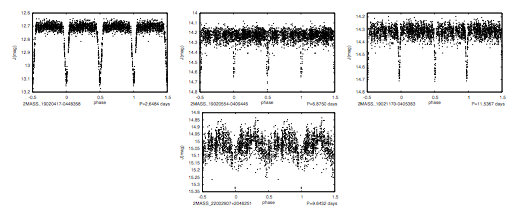} %
\caption{Figure continued.}
\end{figure*}

\begin{table*}
\begin{minipage}{150mm}
\caption[]{Eclipsing binary candidates \label{tab:EB}}
 \begin{tabular}{l l  l l l l l }
 \hline
 Object  & $J$           & $J-K$   & Period     & Depth     & $t_{min}$ & Type\\
              & (mag)       & (mag)   & (days)     & (mag)     & (days)      &         \\
 (1)        & (2)        & (3)            & (4)            &(5)           & (6)             & (7) \\
\hline
 2MASS J03314787+3746510  & 15.248$\pm$0.056 &  1.01 &  0.2475799$\pm$9e-06   & 0.5 &0.204 & EW\\
 2MASS J05570180+0028591 & 15.454$\pm$0.051 & 0.71 & 1.761380$\pm$0.000418  & 0.7 & 2.06  &EA\\
 2MASS J07002882+4829034 & 15.155$\pm$0.044 & 0.51 & 8.732304$\pm$0.014124  & 0.4 & 2.52 & EA \\
 2MASS J08251772$-$3852313  &15.565$\pm$0.067 &  0.56 & 2.33949$\pm$0.00014   &   0.8  & 1.27  & EA\\  
 2MASS J08252706$-$3857482 & 15.824$\pm$0.042 &  0.80 & 5.1101 $\pm$0.0005      &  0.6   &  2.70   & EA \\
 2MASS J08252848$-$3914350  &12.550$\pm$0.026 &  0.36 &  7.777814$\pm$0.0068   & 0.4  & 3.9  &EA \\ 
 2MASS J08253528$-$3911070 & 15.565$\pm$0.066 &  0.28 &  1.76891$\pm$0.00011    & 0.5 & 1.35 &EA \\ 
2MASS J08254622$-$3911160 & 13.149$\pm$0.028 &   0.92 & 0.89405$\pm$ 0.00016    & 0.1 & 0.77 & EW\\
2MASS J08255450$-$3915169  & 14.534$\pm$0.032 & 1.39 &  2.43813$\pm$0.00030   &  0.15  &  2.77   & EA \\
2MASS J11214508$-$1257346 & 13.597$\pm$0.030 &0.46 &  0.90553$\pm$0.0001         & 0.35 & 0.60 & EB \\
 2MASS J11215090$-$1323412  &15.373$\pm$0.042 & 0.82 & 4.40021$\pm$0.0016       & 0.5 & 2.26& EA\\
 2MASS J12013107$-$4945132 & 15.317$\pm$0.052 & 0.32 & 23.3359$\pm$0.0049        & 0.4  &  14.67 &EA\\
 2MASS J12014395$-$4958101 &15.421$\pm$0.056 & 0.8    & 0.266297$\pm$7e-06        & 0.7 & 0.08 &EW\\
 2MASS J12014868$-$5026369 & 12.957$\pm$0.027 & 0.68 &  13.3511$\pm$0.00890    & 0.4 & 3.2& EA \\
 2MASS J12015437$-$5004499  &14.729$\pm$0.051 &  0.52 & 0.426512$\pm$5e-05       & 0.55 & 0.45 & EA\\
 2MASS J12015643$-$5005549  & 15.515$\pm$0.063 & 0.76 & 0.489237$\pm$2e-05       & 0.5 & 0.50 & EB\\
 2MASS J12020461$-$5019309   &15.645$\pm$0.060 & 0.65 &0.989157$\pm$0.000053 & 0.2 &  0.45 & EB\\
 2MASS J14563547$-$4507432 & 13.955$\pm$0.032 & 0.84 & 2.560226$\pm$ 0.00034   &0.45 & 2.18 &EA\\
 2MASS J14564343$-$4431454 & 11.094$\pm$0.021 & 0.88 &18.618$\pm$0.041            & 0.1 & 9.31 & EW \\
 2MASS J14565504$-$4434589 & 11.294$\pm$0.022 & 0.32 & 4.330373$\pm$0.001351  & 0.35 & 4.43 & EA\\
 2MASS J14565902$-$4456313 & 12.824$\pm$0.026 & 0.20 &  0.907237$\pm$0.000006 &0. 4 & 1.14 & EB \\ 
 2MASS J14570998$-$4512257 & 14.508$\pm$0.032 & 0.51 &  6.1384$\pm$0.0120       &0.35  & 2.47 & EA \\
 2MASS J15002898$-$0026058 &13.629$\pm$0.032 & 0.50 & 4.397585$\pm$0.001950  & 0.35 & 4.40 &EA \\
 2MASS J16265638+0552039 & 15.065$\pm$0.050& 0.92  & 0.339974 $\pm$ 0.000004 & 0.3 & 0.148 & EA\\
 2MASS J17481695$-$4507218  & 13.713$\pm$0.029  & 0.50 &  2.03489$\pm$0.0003      & 1.3 & 2.17 & EB\\
 2MASS J17482277$-$4526561 &  15.438$\pm$0.048 &  0.46 & 0.595025$\pm$7e-05     & 0.8 & 0.357 & EB\\
 2MASS J17482426$-$4515538 & 15.724$\pm$0.067 & 0.82 &2.412559$\pm$0.00050  &  0.4 & 0.66 &EA\\
 2MASS J17482645$-$4519087 & 14.732$\pm$0.032  & 0.44 & 6.61616$\pm$0.00069    & 0.35 & 8.24 &EA\\
2MASS J17482815$-$4509394  & 14.064$\pm$0.029 & 0.65 & 3.72342$\pm$0.00097      & 0.4 & 3.06 &EB\\
2MASS J18391249+4845394 & 14.815$\pm$0.037 &  0.51 & 0.2473227$\pm$2e-05     & 0.15 & 0.140 &EW\\
2MASS J18512191$-$0349375  & 14.362$\pm$0.028  & 1.39 &  38.6548$\pm$0.116      &  0.6   &    31.6    & EA \\ 
2MASS J18512369$-$0413009  & 14.812$\pm$0.029  & 1.82 & 2.21975$\pm$0.00049   &   0.5 &    2.08    & EB \\
2MASS J18512462$-$0354532  & 12.828$\pm$0.045 &  0.58 & 2.70563 $\pm$0.00073  &   0.25 &  2.95    & EB \\
2MASS J19014028$-$0424137 & 12.687$\pm$0.032  & 0.95  & 4.182611$\pm$0.0003  &  0.8    & 1.89   &  EB \\
2MASS J19015103$-$0419487  & 14.661$\pm$0.026 & 0.61  & 0.856174$\pm $1e-05   &  0.7  & 0.13  &  EB \\
2MASS J19015390$-$0455254  &14.893$\pm$0.041  & 0.62 & 0.36737$\pm$7e-06        & 0.6 & 0.26 &EW\\
2MASS J19015989$-$0412262  & 13.175$\pm$0.029 &  0.64 & 0.518462$\pm$1e-05       &  0.4   &  0.46 & EB \\
2MASS J19020004$-$0415249  & 15.458$\pm$0.084 & 0.66  & 0.336916$\pm$2e-06        &  0.9   & 0.22 & EW \\
2MASS J19020201$-$0430160 &  15.322$\pm$0.067 & 0.97  & 2.486081$\pm$ 0.00035 & 0.9  &  0.84  &  EB \\
2MASS J19020221$-$0420480 &  10.074$\pm$0.026 &  0.29 &  22.49$\pm$0.36             &  0.35 &  19.85   &  EA \\
2MASS J19020224$-$0431501  & 14.096$\pm$0.036 & 0.64  & 1.481571$\pm$9e-05      & 0.9   &  1.02 &  EB\\
2MASS J19020417$-$0448358  &12.703$\pm$0.026  & 0.48 &  2.64844$\pm$0.00014     & 0.45 & 3.14 &EA\\
2MASS J19020451$-$0411117  & 13.148$\pm$0.024 & 0.87 & 9.4073$\pm$0.0026           & 0.15  &  6.90 &  EA \\
2MASS J19020554$-$0409446  & 14.271$\pm$0.030 &  0.59 & 6.87500$\pm$0.00347      &   0.5  &   3.3   &   EA \\
2MASS J19021170$-$0405383  & 14.276$\pm$0.037 & 0.47 & 11.5367$\pm$0.004           &  0.4    & 6.17  &  EA \\
2MASS J22002907+2046251 & 15.041$\pm$0.052 &  0.69 & 9.643$\pm$0.019              & 0.15 & 9.64 &EW\\

%
%
   \hline
\end{tabular}
\\
Here we list previously unknown eclipsing binaries found in a search for dimming events.
Columns: 
(1) Identifier from the 2MASS point source catalog.
(2,3) $J$ magnitudes and colors are from the 2MASS point source catalog \citep{cutri03}.
(4) The estimated period in days.
(5) The estimated depth of primary eclipse is given here in $J$ magnitudes.
(6) A time of primary eclipse.  We list the Julian Date in days subtracted by 2450000.0.
(7) Binaries are roughly classified as detached (EA), experiencing gravitational deformation (EB), 
or in contact (EW), based on the shapes of the light curves.
\end{minipage}
\end{table*}


\begin{figure*}
\includegraphics[width=5.0in, trim= 0 0 0 0 ]{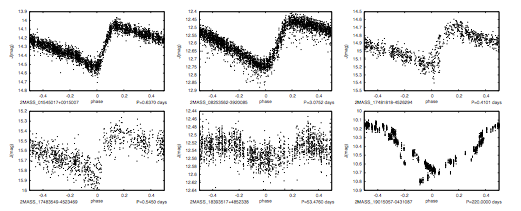}
\caption{Phased (period-folded) light curves (similar to Figure \ref{fig:EB}) 
in $J$-band of previously unknown periodic stars (excluding eclipsing binaries) found in a search
for dimming events.  
2MASS J01545017+0015007, 2MASS J08253562-3920085, 2MASS J17481818-4526294
and  2MASS J17483549-4523469 are probably RR-Lyrae type variables.
2MASS J18393517+4852338 could be a Cepheid type variable and 
2MASS J19015057-0431087 is a long period variable with an infrared excess that is not precisely periodic.
}
\label{fig:RR}
\end{figure*}

\begin{table*}
\begin{minipage}{150mm}
 \caption[]{Periodic stars that are not eclipsing binaries\label{tab:RR}}
 \begin{tabular}{l l  l l l l r l l }
 \hline
 Object                                   & $J$        & $J-K$      & Period           & Amplitude             & $t_{min}$ & [3.4]-[4.6] & [4.6]-[12]\\
 (1)                                         & (2)        & (3)            & (4)                 &(5)                        & (6)          & (7)           & (8)        \\
\hline
  2MASS J01545017+0015007 & 14.626$\pm$0.041  & 0.48 & 0.636980$\pm$2e-05 & 0.3 & 0.318 & 0.00 & $<$1.49\\
  2MASS J08253562$-$3920085  &12.568$\pm$0.030 & 1.36 & 3.07517 $\pm$0.00081& 0.5 &0.938 & 0.04 & 0.58\\ 
  2MASS J17481818$-$4526294  &14.806$\pm$0.033  & 0.12 & 0.4101433$\pm$2e-05& 0.5 &0.285 & -0.90 & $<$3.2\\
  2MASS J17483549$-$4523469  &15.353$\pm$0.060  & 0.50 & 0.545027$\pm$ 3e-05 & 0.5 &0.5225   & -        & - \\
  2MASS J18393517+4852338 & 12.537$\pm$0.021  &  0.85  &  53.476$\pm$0.87 & 0.08&41.74      & -0.06 & 0.043\\
  2MASS J19015057$-$0431087  & 10.382$\pm$0.205 &  1.82 &   220$\pm$30          &  0.55 &  0.0  & 0.41  & 0.77 \\
  \hline
\end{tabular}
 \\
Here we list  previously unknown periodic variable stars (excluding eclipsing binaries) 
found in a color insensitive search for dimming events.
Columns are similar to those of Table \ref{tab:EB} but listing variable stars likely to be intrinsic pulsators.
The period folded light curves are shown in Figure \ref{fig:RR}.
Columns:
(5) The amplitude (peak to peak) of variation in $J$ band magnitude.
(6) A time of a  minimum in the light curve.  
We list the Julian Date in days subtracted by 2450000.0.
(7, 8)  Colors in Vega magnitudes
computed from broad band photometry from the  WISE satellite \citep{wise,wright12}.
For  2MASS J01545017+0015007   and 2MASS J17481818-4526294
 the [4.6]-[12] colors are upper limits as the sources were not detected at 12~$\mu$m.
 2MASS J17483549-4523469 was not in the WISE all sky catalog.
%
\end{minipage}
\end{table*}


\subsection{Additional variable objects}

\begin{figure*}
\includegraphics[width=2.8in, trim= 0 0 0 0 ]{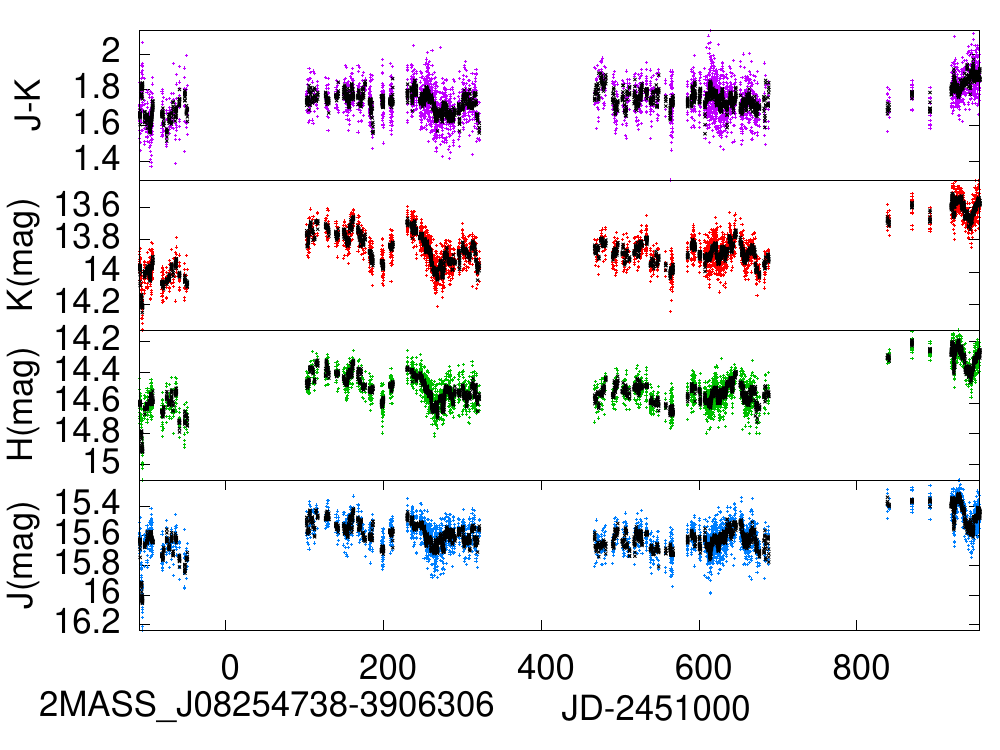} %
\includegraphics[width=2.1in, trim= 0 0 0 0 ]{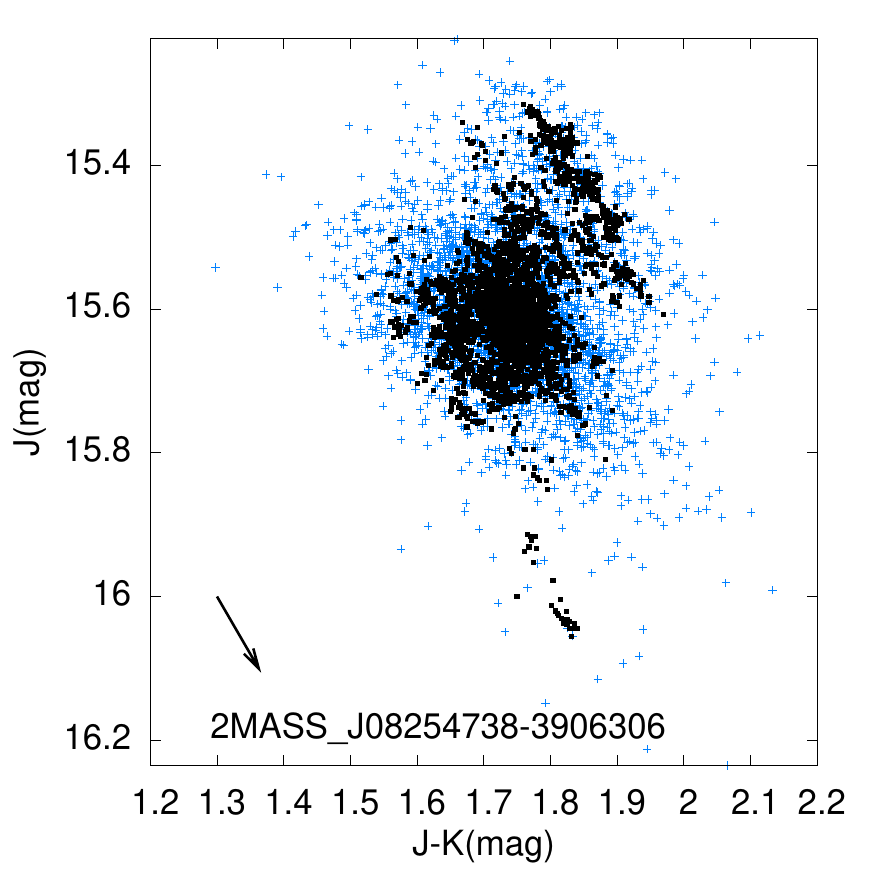} 
\includegraphics[width=2.8in, trim= 0 0 0 0 ]{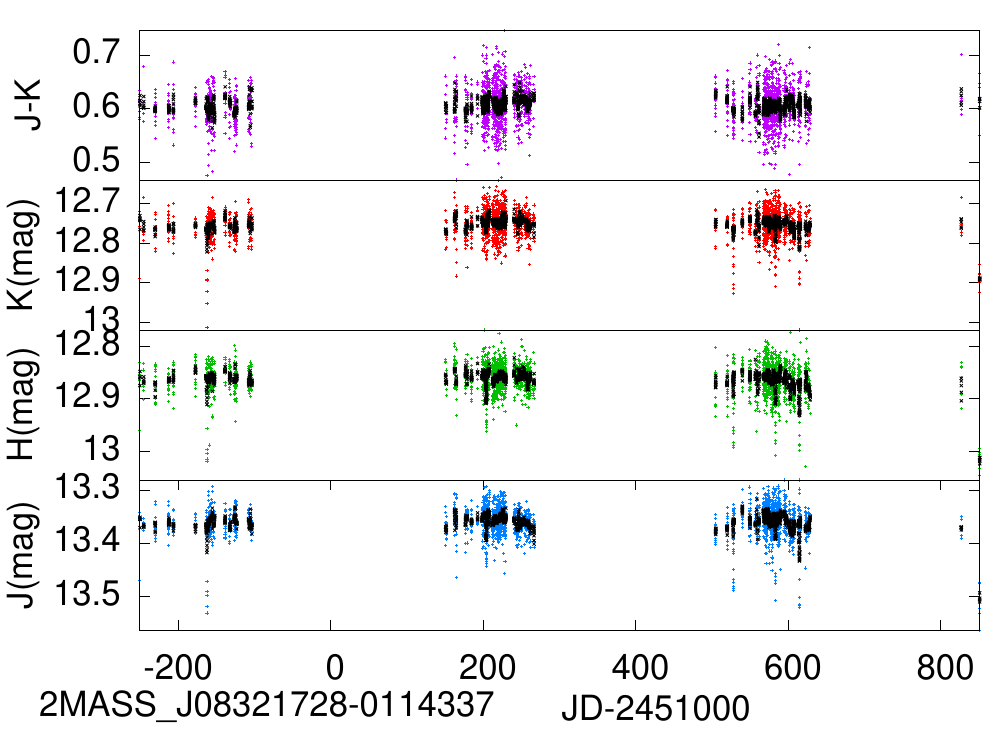}
\includegraphics[width=2.1in, trim= 0 0 0 0 ]{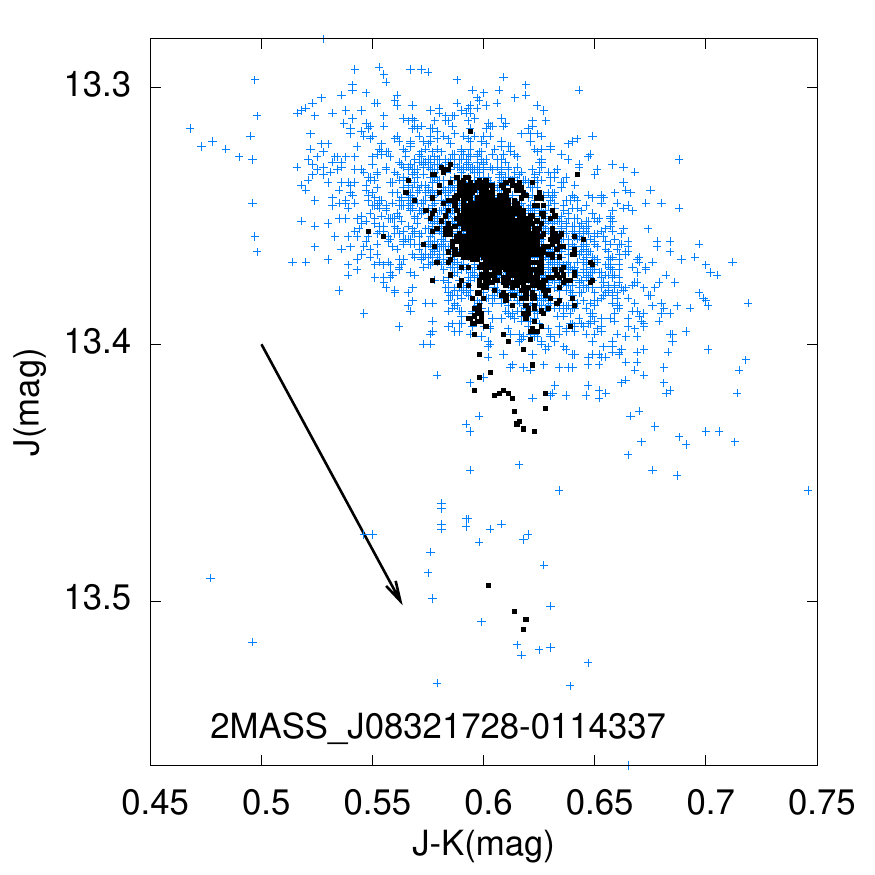} 
\includegraphics[width=2.8in, trim= 0 0 0 0 ]{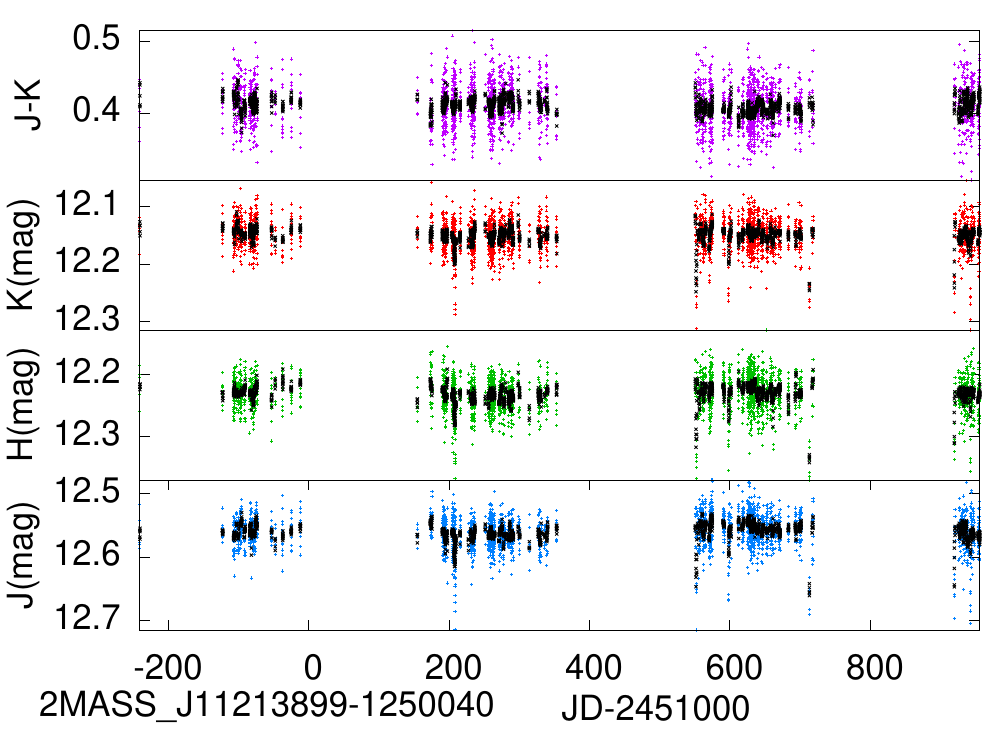} 
\includegraphics[width=2.1in, trim= 0 0 0 0 ]{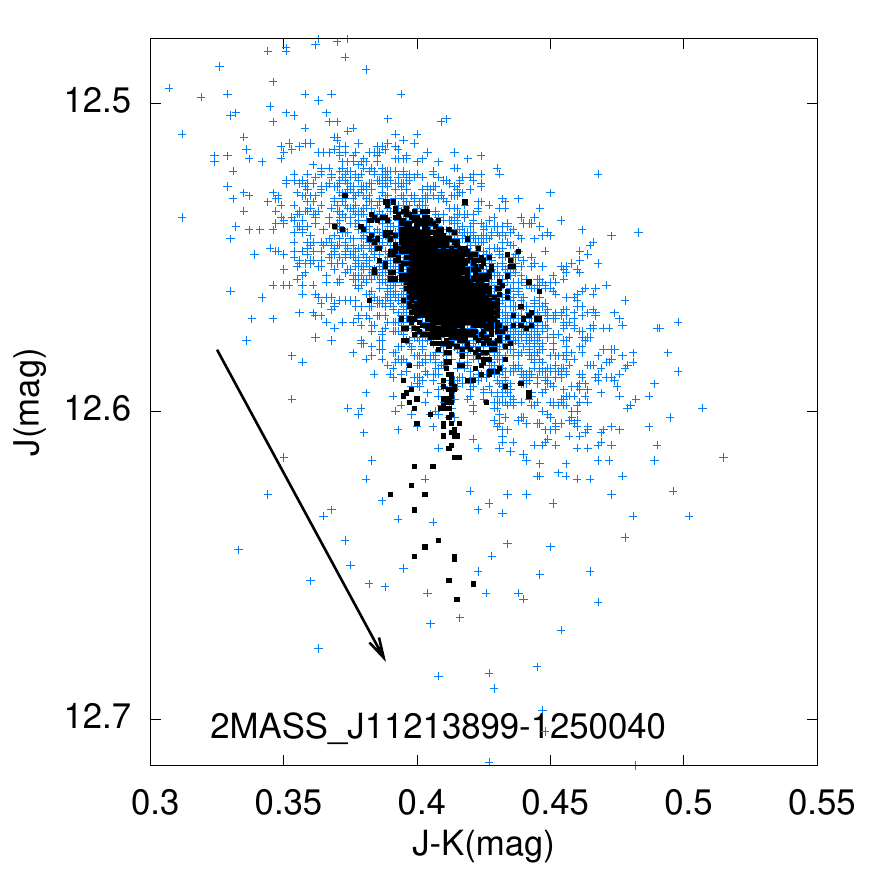} 
\includegraphics[width=2.8in, trim= 0 0 0 0 ]{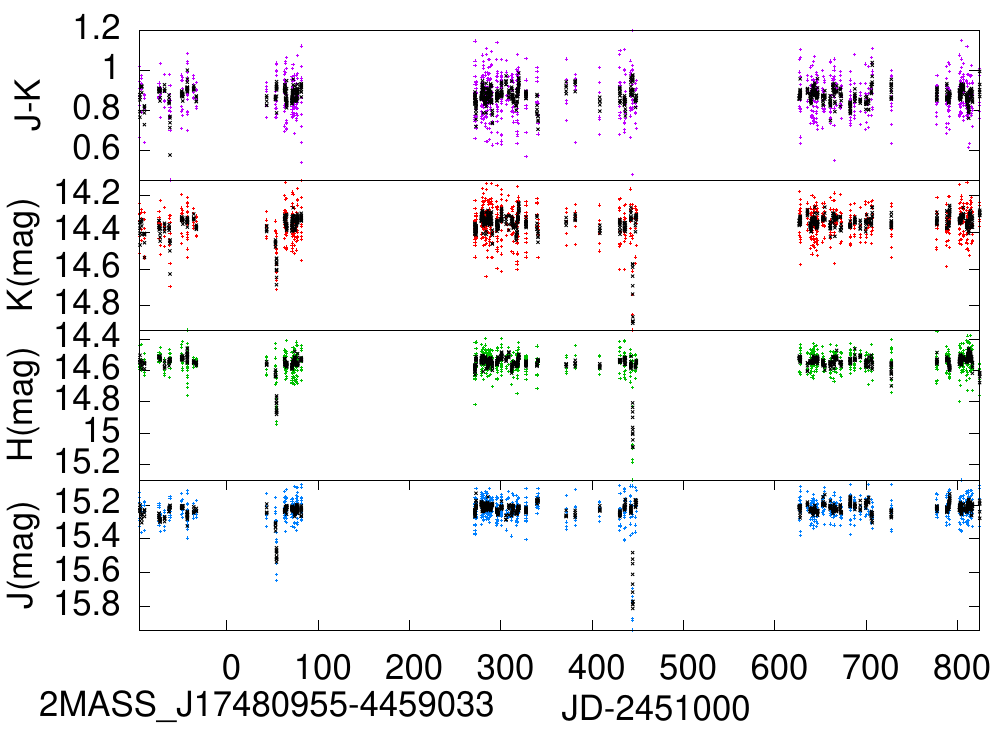}
\includegraphics[width=2.1in, trim= 0 0 0 0 ]{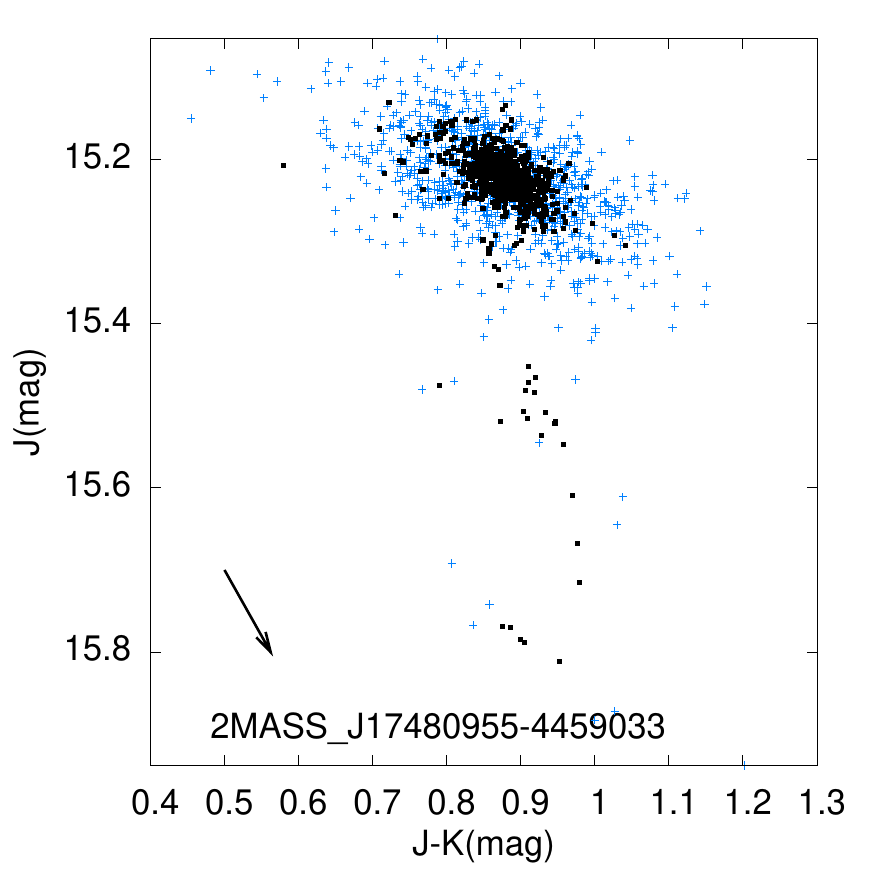}
\caption{Variable objects listed in Table \ref{tab:aperiodic} 
with deep dimming events. We failed to find periods for these objects.
The entire light curves are shown in $J, H, K$ and in $J-K$ magnitudes as a function on time on the left.   
On the right we show a color magnitude plot with $J$ vs $J-K$ magnitudes at different times.
A reddening vector with length $A_J=0.1$ is shown on the lower left in the color-magnitude plots.
Photometric measurements are shown as colored
points.  Box-day averages are shown as black points.
Some of these objects do display significant color variations
during dimming events,     
suggesting that these are eclipsing binary stars
but we have not observed enough eclipses to identify a period.
For objects with  correlated color/magnitude variations, they could be active galaxies or young
stellar objects.
}
\label{fig:aperiodic}
\end{figure*}

Some objects exhibited deep dimming events, 
but we could not identify periods for them.
Excluding the $\rho$ Ophiucus region,
these objects are listed in Table \ref{tab:aperiodic} and their light curves are shown in Figure \ref{fig:aperiodic}
or if they are extragalactic in  Figure \ref{fig:aperiodic_gal}.  
On the left hand side, the day averages are plotted as black points on top of individual measurements (colored points).
From top to bottom are the $J-K$ color, $J, H, $ and $K$ magnitudes as a function of time.
On the right hand side we show color-magnitude plots ($J$ vs $J-K$) but with each point corresponding to a
different observation time.  On the color-magnitude plots, we also show a reddening vector with length
corresponding to an extinction of $A_J = 0.1$.
%

2MASS J12141394+3541567 and 2MASS J16311840+2946322 are extragalactic objects
 based on their membership in the galaxy catalog by \citet{paturel03}. 
Most of the non-extragalactic aperiodic objects exhibiting deep dimming events have WISE colors consistent with stellar sources. 
The object  2MASS J17482764-4531581 has W2-W3 color $[4.6\mu{\rm m}] - [12\mu{\rm m}]= 2.51$ (Vega magnitudes),
and 2MASS J08254738-3906306 has a color [4.6]-[12] of 1.75 mag.  These red colors are more typical
of a Seyfert galaxy than a star (placing it on Figure 14 by \citealt{yan13}) but would also be typical of a class I young
stellar object (see Figure 2 by \citealt{wolk13}).
We checked the WISE image server and found that the 12$\mu$m flux for 2MASS J17482764-4531581 is confused with a
nearby shell, so the red $[4.6] - [12]$ color is probably not that of the object.
2MASS J19020170-0405561 with [4.6]-[12]$\mu$m 0.80 also has a flux at 12$\mu$m and 22$\mu$m
of 8.3 and 10 mJY, respectively.  We checked using the WISE image server that there was a source
at longer wavelengths.   This object likely has an infrared excess at 22$\mu$ even though its
 [4.6]-[12]$\mu$m  is only moderately red.  Based on its mid-infrared excess, 
 this object could be an active galaxy or a young stellar object. 

Even though 2MASS J19020170-0405561 and 2MASS J19020170-0405561 are not identified as extragalactic, their brightness and color variations are strongly correlated, similar to the two galaxies.
2MASS J18394250+4856177 also exhibits correlated brightness and color variations but has no sign of infrared excess. 
It is common for young stellar objects to  exhibit transient dimming events \citep{moralescalderon11,cody14,parks13}.
 20\% of the young disk-bearing stars in NGC2264 surveyed by \citet{cody14} were classified as ``dippers''. 
Dimming in young stellar objects  is often, but not always, associated with reddening and so due to extinction \citep{cody14}.
2MASS J08254738-3906306, and 2MASS J19020170-0405561, with infrared excesses, might be young stellar objects rather than active galaxies.
For 2MASS J18394250+4856177, exhibiting correlated color/magnitude variations but no infrared excess, 
is difficult to classify,  even though it is not faint, at 12.7 mag in $J$ band.

2MASS J19014896-0453423 is a bright object (9.8 mag at $J$ band) 
varying by 0.25 magnitudes, but it does not display strong color variations. 
It clearly has an infrared excess as it was detected with WISE at 22$\mu$m with a flux of 11 mJy.
Unfortunately, it was not observed with WISE at 12$\mu$m, making it difficult to classify it based on its mid-infrared color.

The remaining objects, 2MASS J08321728-0114337,  2MASS J11213899-1250040, 2MASS J17480955-4459033, 2MASS  J17482764-4531581, and 2MASS J22001106+2032472,
display isolated dimming events that do not strongly change the color.
The lack of color variation during dimming events implies that extinction is
not the cause of the dimming. We examined the data points during each dimming even but
did not detect  sub-structure in the shape of the light curve.   They are either dimming, as if at the beginning
of an eclipse, flat, or brightening, as if at the end of an eclipse.   
If these dimming events are real, then most of these objects are 
likely to be stellar eclipsing binaries, and we have not able to identify
their periods because we have not seen enough eclipses.
They could also be more exotic objects.

\setcounter{figure}{2}
\begin{figure*}
\includegraphics[width=2.8in, trim= 0 0 0 0 ]{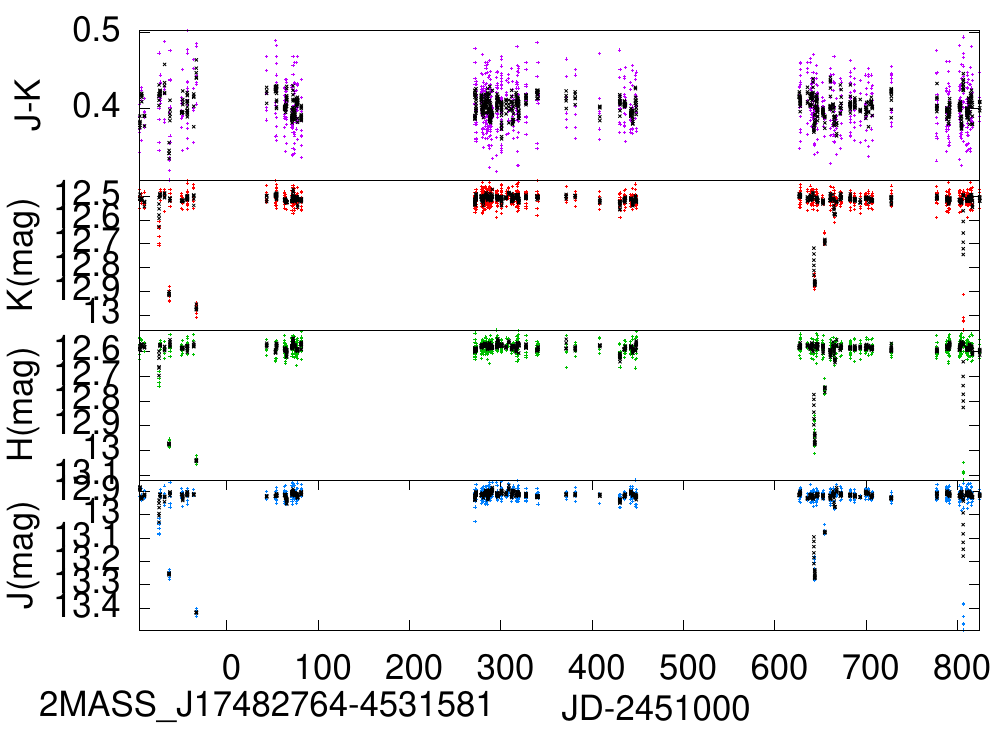}
\includegraphics[width=2.1in, trim= 0 0 0 0 ]{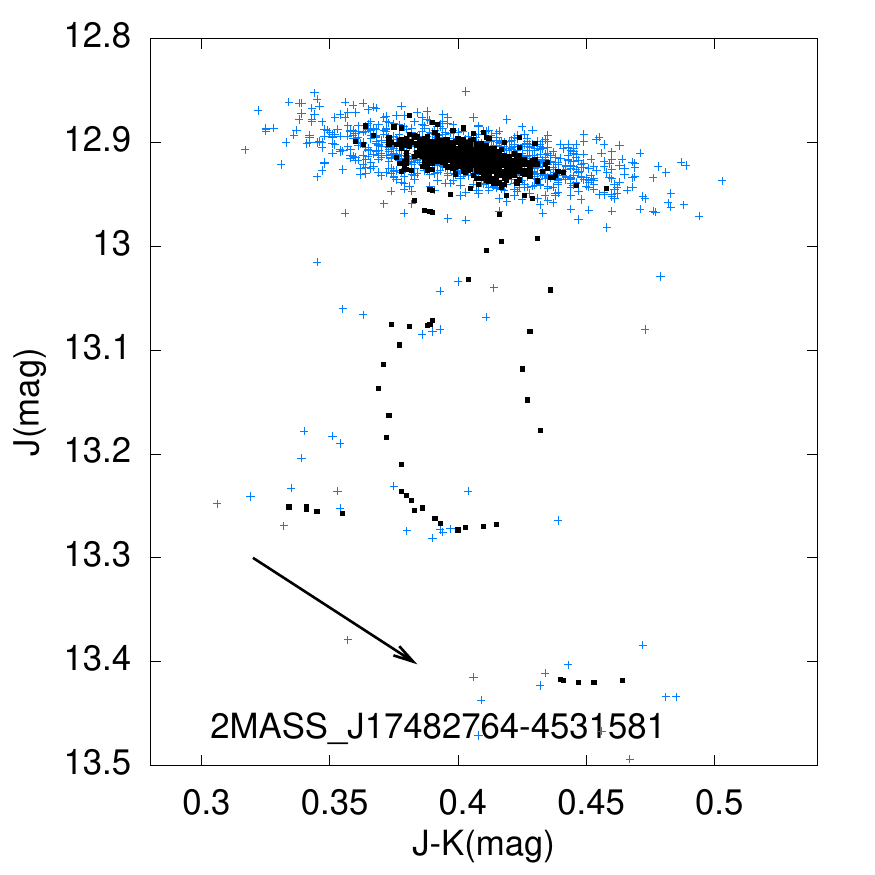}
\includegraphics[width=2.8in, trim= 0 0 0 0 ]{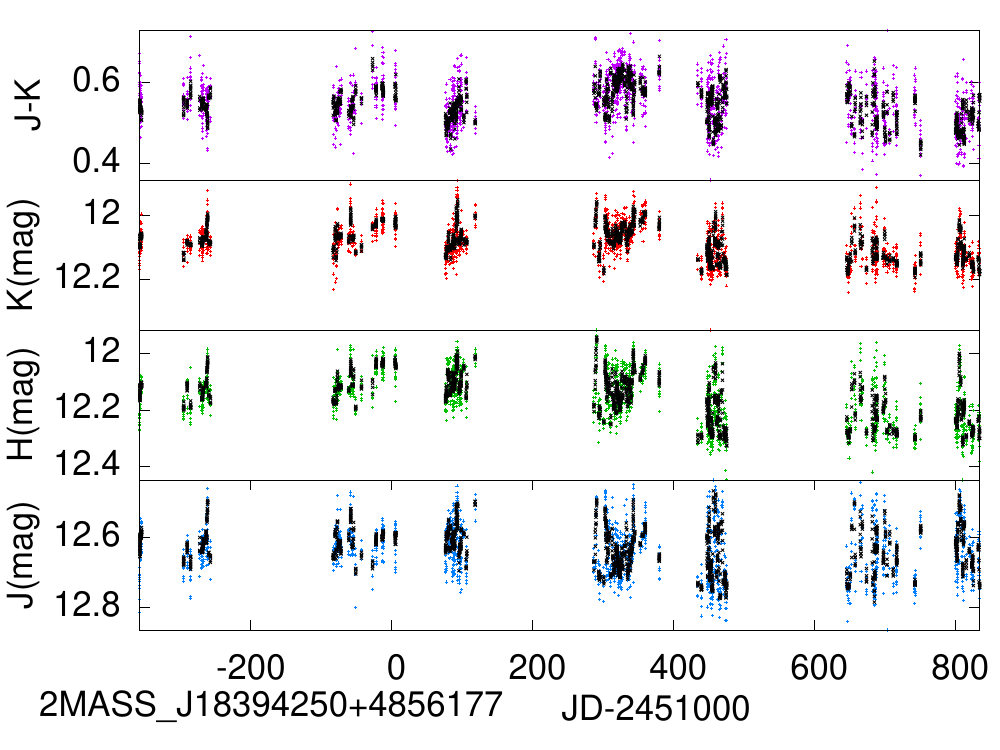}
\includegraphics[width=2.1in, trim= 0 0 0 0 ]{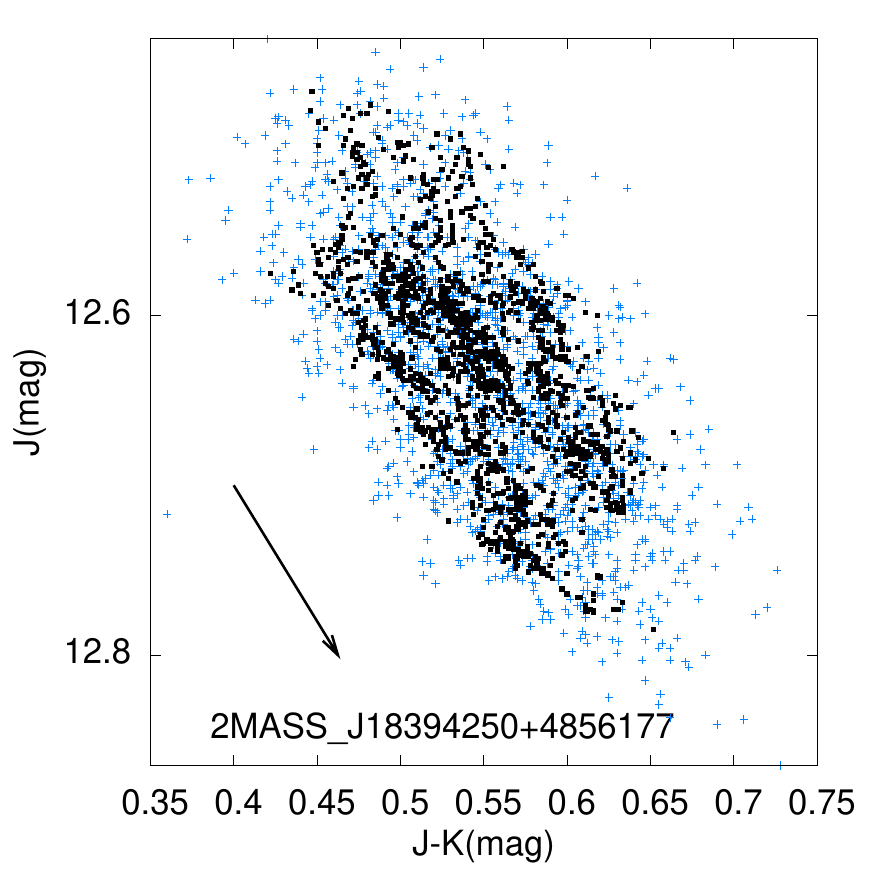}
\includegraphics[width=2.8in, trim= 0 0 0 0 ]{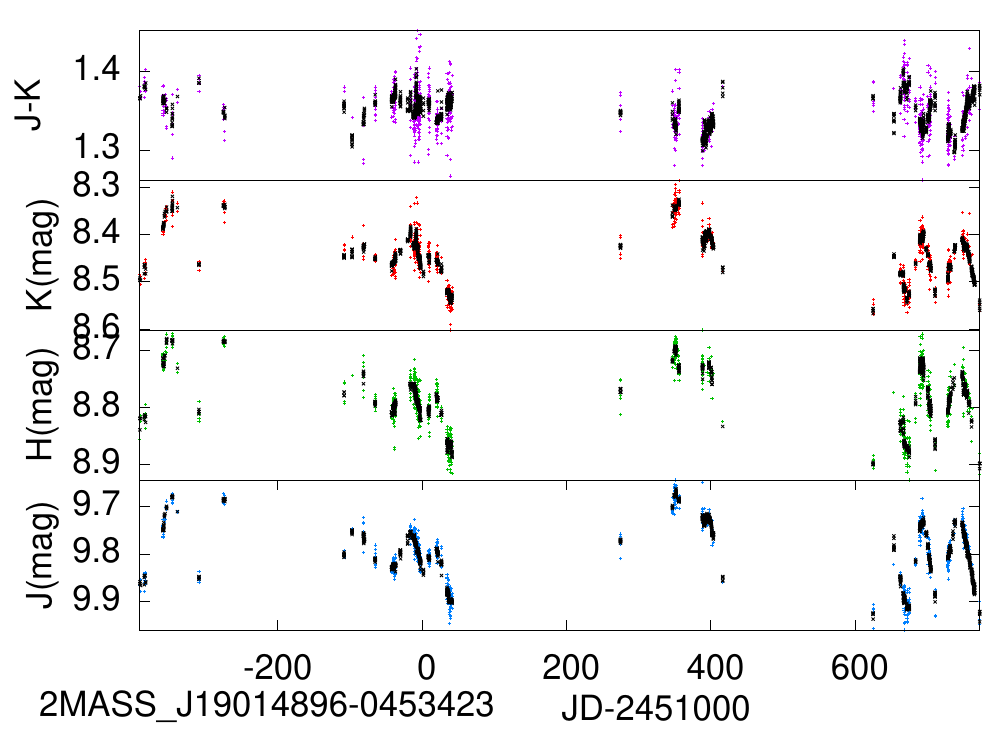}
\includegraphics[width=2.1in, trim= 0 0 0 0 ]{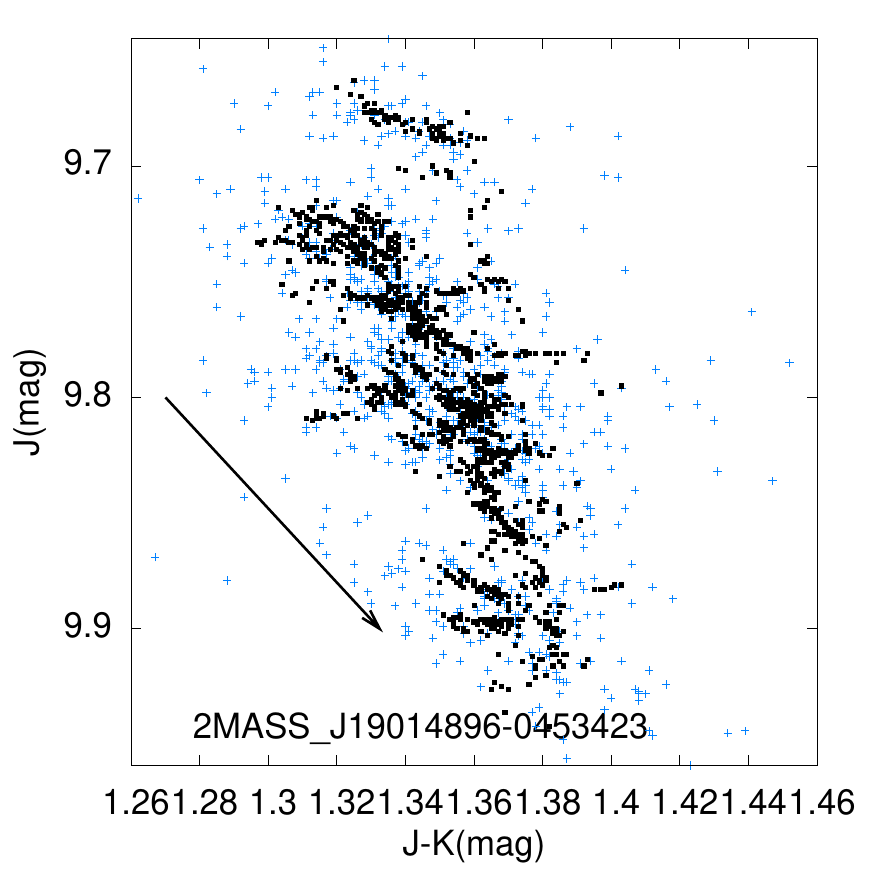}
\includegraphics[width=2.8in, trim= 0 0 0 0 ]{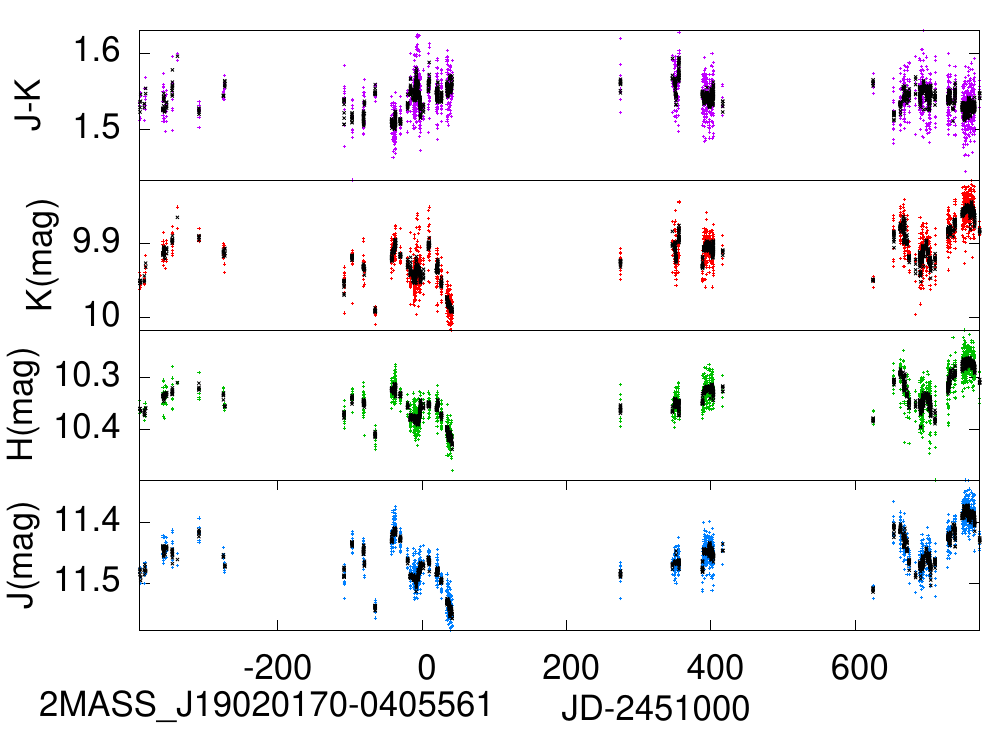}
\includegraphics[width=2.1in, trim= 0 0 0 0 ]{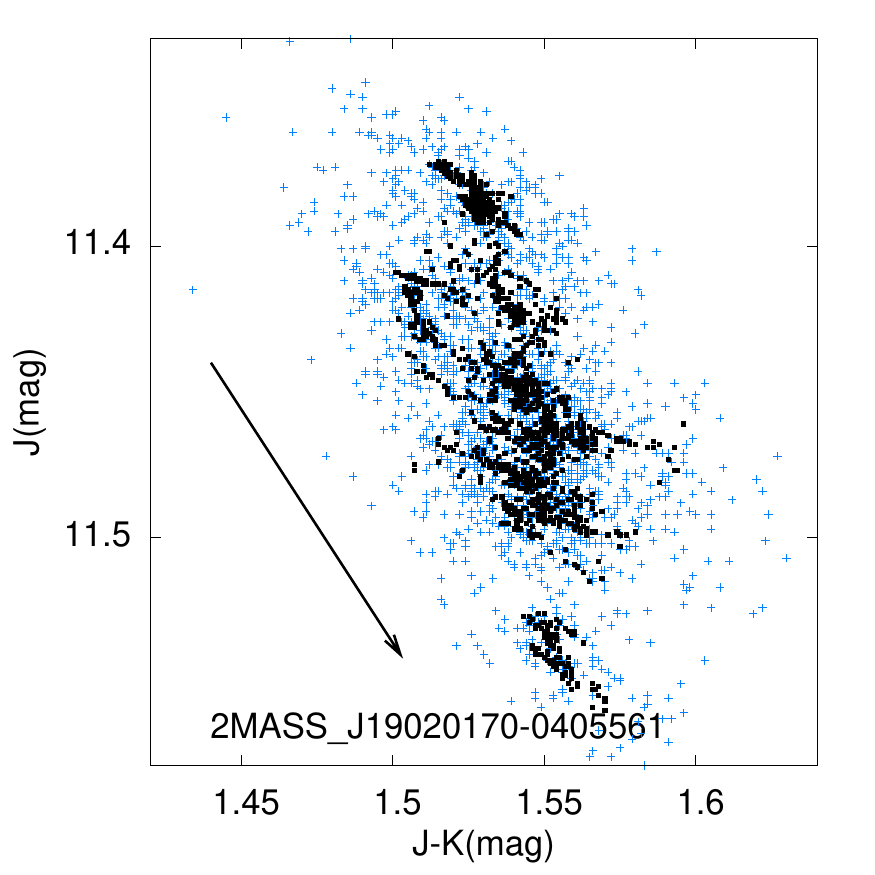}
\caption{Figure continued.}
\end{figure*}

\setcounter{figure}{2}
\begin{figure*}
\includegraphics[width=2.8in, trim= 0 0 0 0 ]{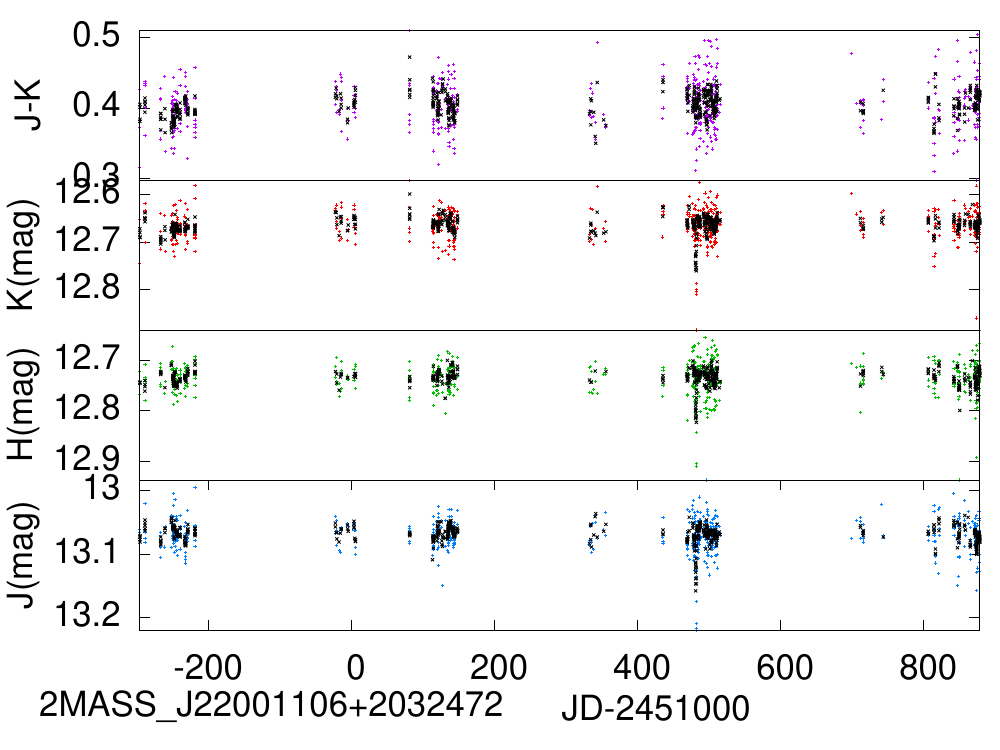}
\includegraphics[width=2.1in, trim= 0 0 0 0 ]{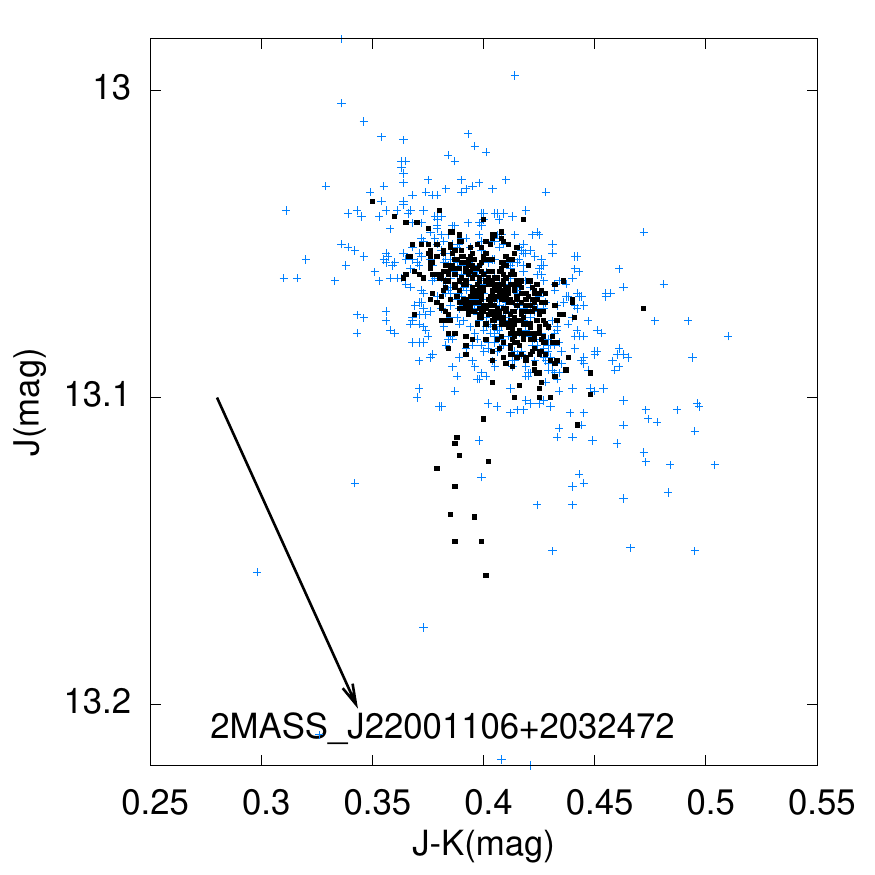}
\caption{Figure continued.}
\end{figure*}

\begin{figure*}
\includegraphics[width=2.8in, trim= 0 0 0 0 ]{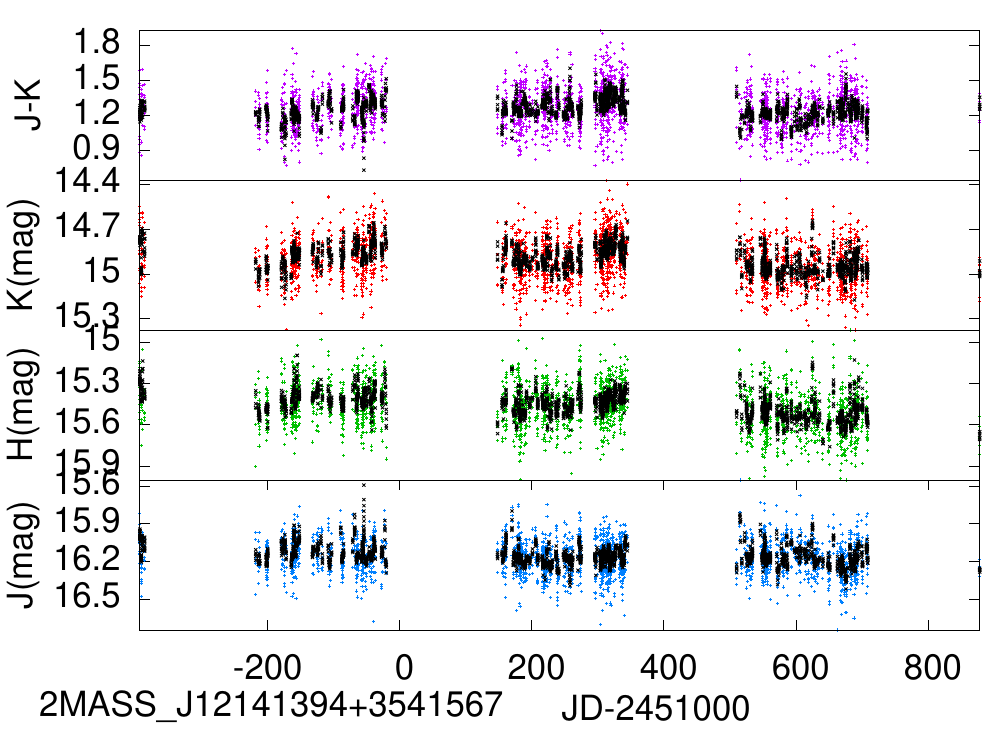} 
\includegraphics[width=2.1in, trim= 0 0 0 0 ]{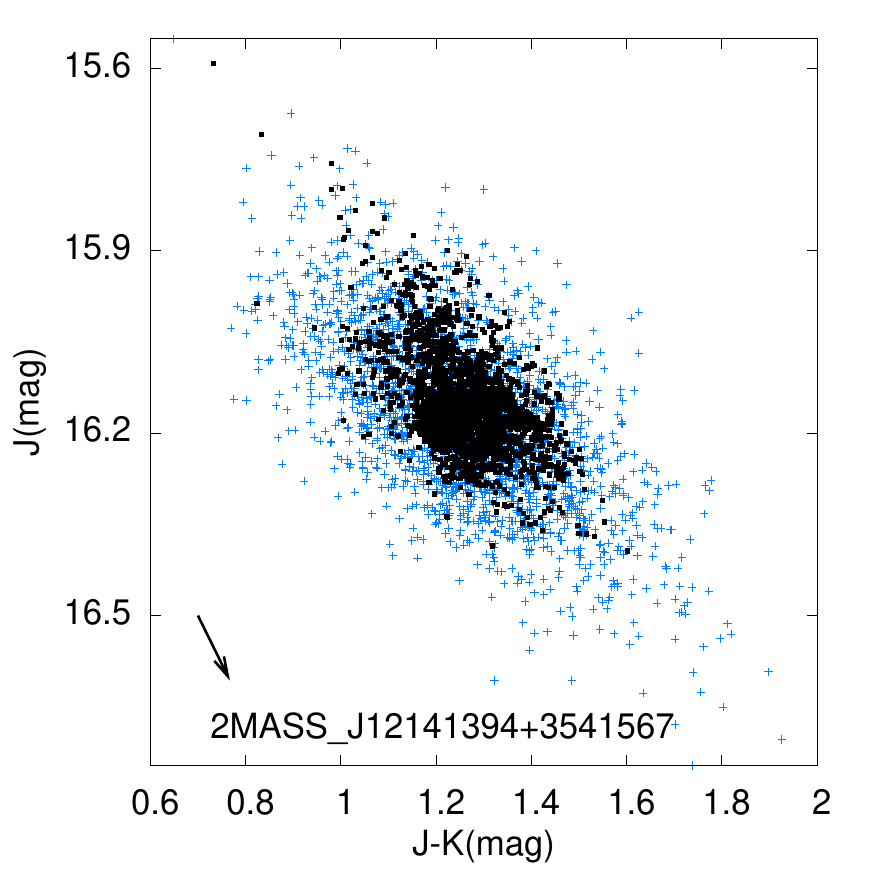} 
\includegraphics[width=2.8in, trim= 0 0 0 0 ]{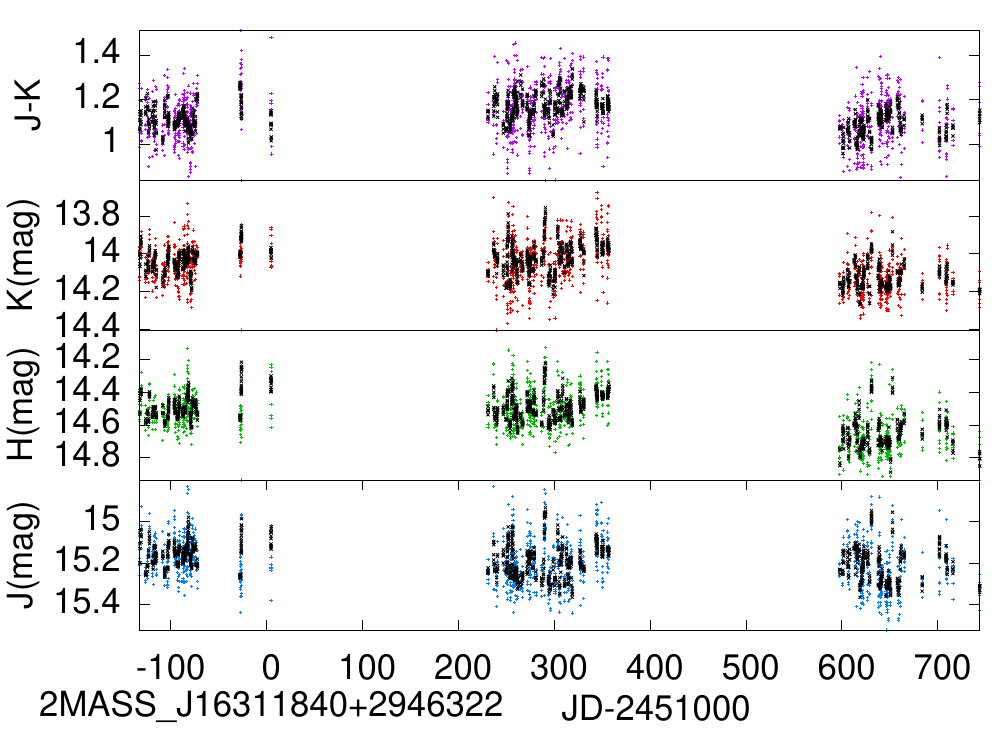} 
\includegraphics[width=2.1in, trim= 0 0 0 0 ]{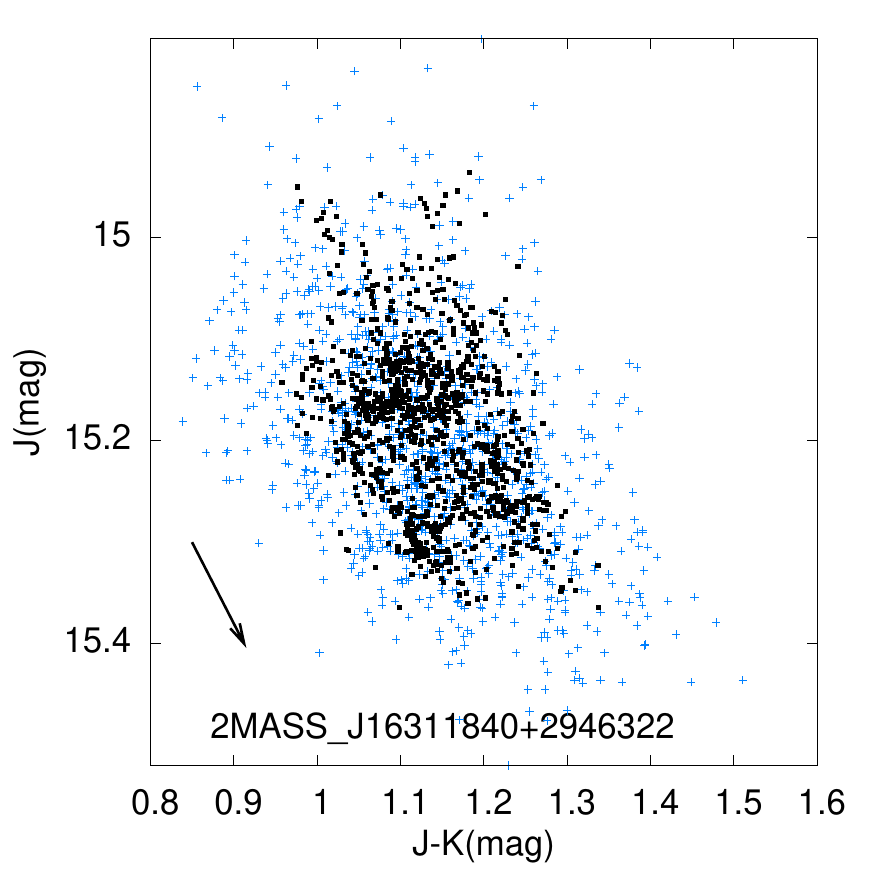}
\caption{Similar to Figure \ref{fig:aperiodic} but for extragalactic variable objects
  listed in Table \ref{tab:aperiodic} with deep dimming events. 
\label{fig:aperiodic_gal}}
\end{figure*}

\begin{table*}
 \begin{minipage}{190mm}
 \caption[]{Variable object candidates without measured periods and with dimming events\label{tab:aperiodic}}
%
 \begin{tabular}{l r  l r l r l r r l l l l}
 \hline
Object                           & $J$ & $\sigma_J$ & $H$ & $\sigma_H$ & $K$ & $\sigma_K$ & $J-K$ & [3.4]-[4.6] & [4.6]-[12]&  $R$ & $b$\\
      (1)                           &  (2) & (3)              & (4)    & (5)               & (6)   &  (7)              & (8)      & (9)          & (10)     & (11)  & (12)  \\
\hline
2MASS J08254738$-$3906306&15.586& 0.116& 14.493& 0.128 & 13.833 & 0.142 &1.75&0.49 & 1.75  & 0.20 & 0.221$\pm$0.019\\
2MASS J08321728$-$0114337 & 13.360& 0.028 & 12.863 & 0.030 &  12.753 & 0.036 &0.61 & -0.05 & 0.13 & 0.38 & 0.304$\pm$0.016\\
2MASS J11213899$-$1250040 & 12.560& 0.025 & 12.231 & 0.027 &  12.150 & 0.027 & 0.41 & -0.03 & 0.54 & 0.52 & 0.459$\pm$0.015\\
2MASS J17480955$-$4459033 & 15.228 &0.074 & 14.548 & 0.075 & 14.350 & 0.091 & 0.88 &-0.18 &$<2.5^a$ & 0.44 & 0.350$\pm$0.023\\
2MASS J17482764$-$4531581 &12.929 &0.076 & 12.599 & 0.077 & 12.524 & 0.077 &0.41 & -0.06 & 2.5$^b$   & 0.12 & 0.285$\pm$0.074\\
2MASS J18394250+4856177 & 12.623 & 0.073 & 12.142 & 0.090 & 12.080 & 0.057 & 0.54 & -0.09 & 0.05  &0.62 & 0.830$\pm$0.025 \\
2MASS J19014896$-$0453423& 9.786 &0.059 & 8.780& 0.053  &8.440 &0.049&1.35&0.05  & 0.5$^c$  & 0.56 & 1.189$\pm$0.046\\
2MASS J19020170$-$0405561&11.453 &0.044&  10.339& 0.044&  9.913 &0.038 &1.54 & 0.02 & 0.8  & 0.53 &0.929$\pm$0.035 \\
2MASS J22001106+2032472&13.069 &0.025 & 12.735 &0.032 &12.665 &0.033  &0.40 &-0.04 &$<0.2^a$&0.42 & 0.321$\pm$0.029\\
 \hline  
  2MASS J12141394+3541567$^d$&16.151 & 0.140 & 15.456 & 0.154 & 14.902 & 0.145 &1.25 & 0.28 & 3.64 & 0.59 & 0.464$\pm$0.012\\
 2MASS J16311840+2946322$^d$&15.185 & 0.110 & 14.526 & 0.135 & 14.057 & 0.115 & 1.13 & 0.29 & 3.97 & 0.43 & 0.436$\pm$0.027\\
 \hline
\end{tabular}
 \\
Here we list objects found in a color insensitive search for dimming events that lack measured periods.
In this table we have excluded young stellar objects in the $\rho$ Ophiucus region.
Objects are sorted into two groups, with the second group identified as extragalactic objects.
Columns: 
(1): 2MASS point source catalog identifier.
(2)-(7):  Mean and standard deviations, in magnitudes, computed from the light curves in $J, H,$ and $K$ bands, respectively.
(8): $J-K$ Color in magnitudes was computed from the mean magnitudes of the light curve.
(9,10):  Colors in Vega magnitudes from the WISE survey \citep{wise}.
The four bands of the WISE survey bands are centered at $3.4,4.6, 12$ and 22~$\mu$m.
(11): The cross correlation coefficient, $R$, (equation \ref{eqn:R}) between color and magnitude.
(12): The slope, $b$, of the best fitting line on the color/magnitude plot (equation \ref{eqn:b}).
 \\
$^a$ For  2MASS J17480955-4459033   and    2MASS J22001106+2032472 
 the [4.6]-[12] color is an upper limit as the sources were not detected at 12~$\mu$m.\\
 $^b$The WISE image server showed that the 12$\mu$n flux for 2MASS J17482764-4531581 is confused with a
nearby shell, so the red $[4.6\mu{\rm m}] - [12\mu{\rm m}]$ is probably not that of the source.\\
 $^c$2MASS J19014896-0453423 was not observed at 12$\mu$m with the WISE satellite, however the source
 was detected at 22$\mu$m. The [4.6]-[12] color is estimated from the fluxes at 4.6 and 22$\mu$m.\\
$^d$2MASS J12141394+3541567 and 2MASS J16311840+2946322 are extragalactic objects
 based on their membership in the galaxy catalog by \citet{paturel03}.
\end{minipage}
\end{table*}


To summarize, our color insensitive but sigma-limited search for dimming events primarily found eclipsing binaries.
Including the previously known eclipsing binaries we found a total of 59 eclipsing binaries from approximately 40000 stars
implying that approximately 1/700 stars searched is an eclipsing binary.  We also found 6 periodic or quasi periodic intrinsic pulsators.  
An 11 additional objects were found with aperiodic dimming events.   Of these two are extragalactic objects and so probably
active galaxies.  A few exhibited dimming events without strong color variation.   These could be eclipsing binaries but we
have failed to see enough eclipses to identify a period.  A few exhibit correlations between dimming and color suggesting
that they are active galaxies or young stellar objects.  The 11 objects found exhibiting dimming events could be interesting 
objects, but further observations are needed to classify them.

\section{A search for red dimming events}

Eclipsing binaries usually do not exhibit strong color variations during eclipses.  However, transient extinction would
change the color of a star during a dimming event.  For example, OGLE-LMC-ECL-11893 disk eclipses are 0.3 magnitude
deeper in $B$ band than $I$ band (\citealt{dong14}; see their Figure 3).  Even though this object was first identified as an
eclipsing binary,  its light curve during eclipse is complex  (asymmetric and W-shaped) 
and so is  likely due to a disk that is hosted by the secondary star.
For a standard Milky Way dust extinction law the change in color $E(J-K)  \sim  0.625 A_J$, with $A_J$  
the extinction in $J$ band and both $A_J$ and $E(J-K)$ in magnitudes.
During a dimming event caused by extinction,  
we expect an increase in $J-K$ color about half of the depth in magnitudes of the event in $J$ band.

We carried out a search independent of the magnitude dispersions (of the entire light curves),  
by searching   for dimming events in $J$ band greater than 0.15 mag that
are accompanied by reddening in color $J-K$ of greater than 0.07 with both values compared to the means, $\mu_J$, 
and $\mu_J- \mu_K$, of the entire light curve. 
Because we are looking for small color changes, we used the day averaged light curves for this search.
We discarded lower equality data points with a sum of standard deviations $\sigma_{Jb} + \sigma_{Kb} > 0.08$,
where  $\sigma_{Jb},  \sigma_{Kb}$ are the uncertainty of the box-averaged light curve in $J$ and $K$ bands, respectively. 
We discarded events with nearby source confusion, and sparse data (a particular problem for the 90547 tile)
and at times where more than 1 star in the same calibration tile exhibited an event.
Our search criteria do not depend on the standard deviation of the entire light curve so we are
sensitive to strongly variable objects.

This search revealed a number of variable objects in which the object systematically is bluer when brighter.
The objects that are extragalactic objects, we have listed in Table \ref{tab:AGN}.
Table \ref{tab:AGN} also lists mean magnitudes and standard deviations of the magnitude distributions we measured from 
the entire light curves.
An object was designated extragalactic if we found it in a galaxy catalog such as \citet{paturel03,paturel05} or
it was classified as a galaxy in the Sloan Digital Sky Survey (SDSS) or found  in the 2MASS extended source catalog (see 
the references listed in column 13 and notes in Table \ref{tab:AGN}).
An additional object we classify as extragalactic based on its mid-infrared color as measured with WISE.
This object, 2MASS  J08255140-3907590, is quite red, with $J-K \sim 2.2$ and nearly in the Galactic plane, 
with a Galactic latitude of only -0.65 degrees, but it is not resolved, found to be as extended nor listed in any extragalactic catalog.
However, its mid-infrared flux is 0.028~Jy at 12~$\mu$m and 0.06~Jy at 22~$\mu$m (based on the WISE
measurements) and is increasing
with increasing wavelength between 12 and 22 microns.
Mira variables and young stellar objects that can be seen in the near-infrared tend to have flux
decreasing with increasing wavelength in the 12--22 $\mu$m region. 
The 3.4 and 4.6 $\mu$ color, based on the WISE W1 and W2 bands, respectively, 
$[3.4]-[4.6] = 1.05$ (in Vega magnitudes), is consistent with $[3.4] - [4.6] > 0.7$ mag
found for quasars \citep{yan13}.
Consequently 2MASS  J08255140-3907590 is most likely an active galaxy 
and we have designated it as extragalactic.

Fainter AGNs vary more strongly than their bright counterparts and they vary more strongly on shorter 
wavelengths (e.g., \citealt{diclemente96}).
Brightness and color are correlated with objects becoming bluer when brighter.  This is usually attributed
to a contribution to the flux in red light from non-variable emission source (narrow line region, galaxy or star light) 
combined with a variable bluer
component associated with the active nucleus \citep{winkler97,webb00,sakata10}.
Since active galaxies become bluer when they are brighter, they also are redder when they are fainter.
This explains why we mostly found faint AGNs when searching for red dimming events.
The object, 2MASS J08255140-3907590 (that is bright in the mid-infrared) is variable on longer 
timescales than the other extragalactic objects, suggesting that
it is a more luminous active galaxy than the others, perhaps a quasar. 

\begin{figure*}
\includegraphics[width=2.8in, trim= 0 0 0 0 ]{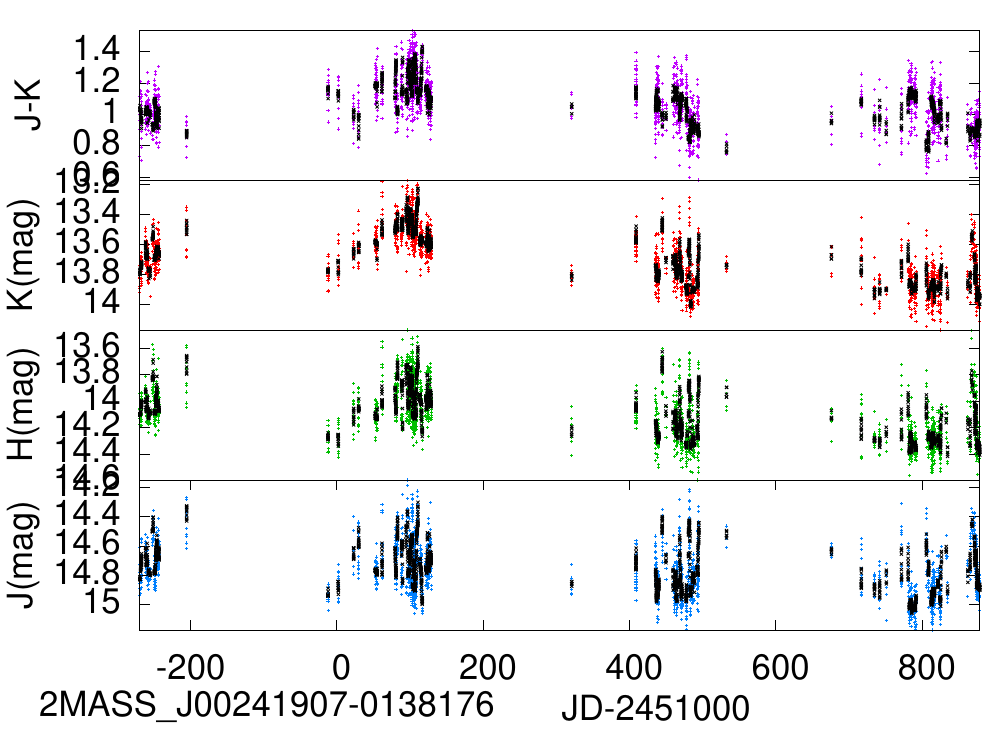}
\includegraphics[width=2.1in, trim= 0 0 0 0 ]{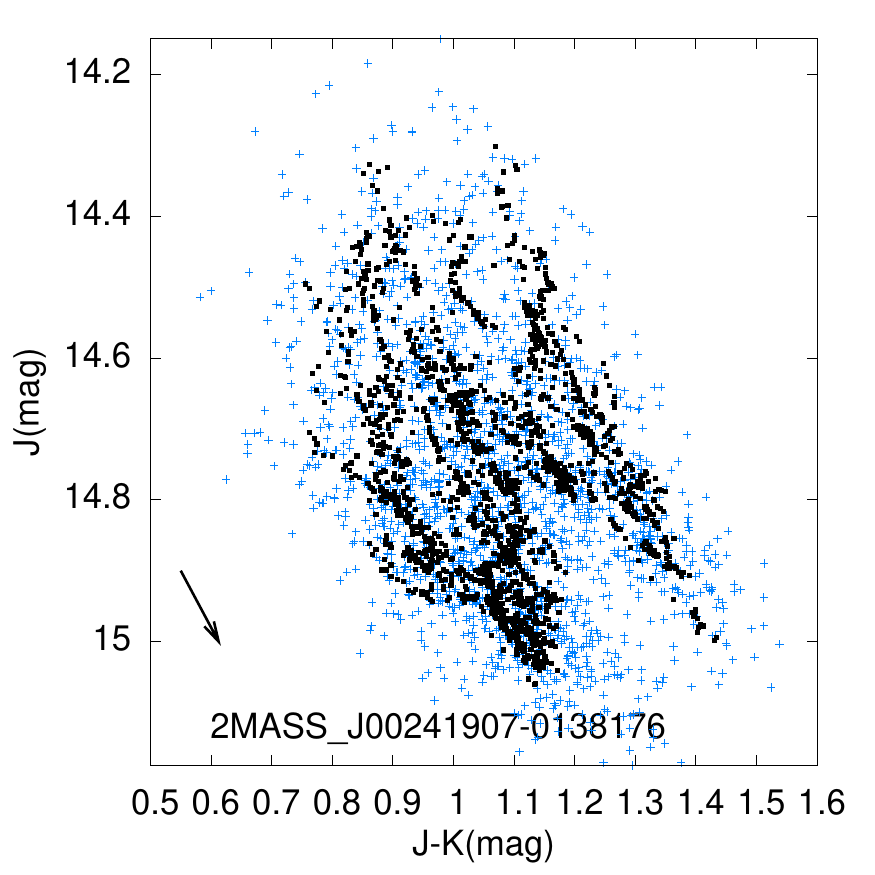}
\includegraphics[width=2.8in, trim= 0 0 0 0 ]{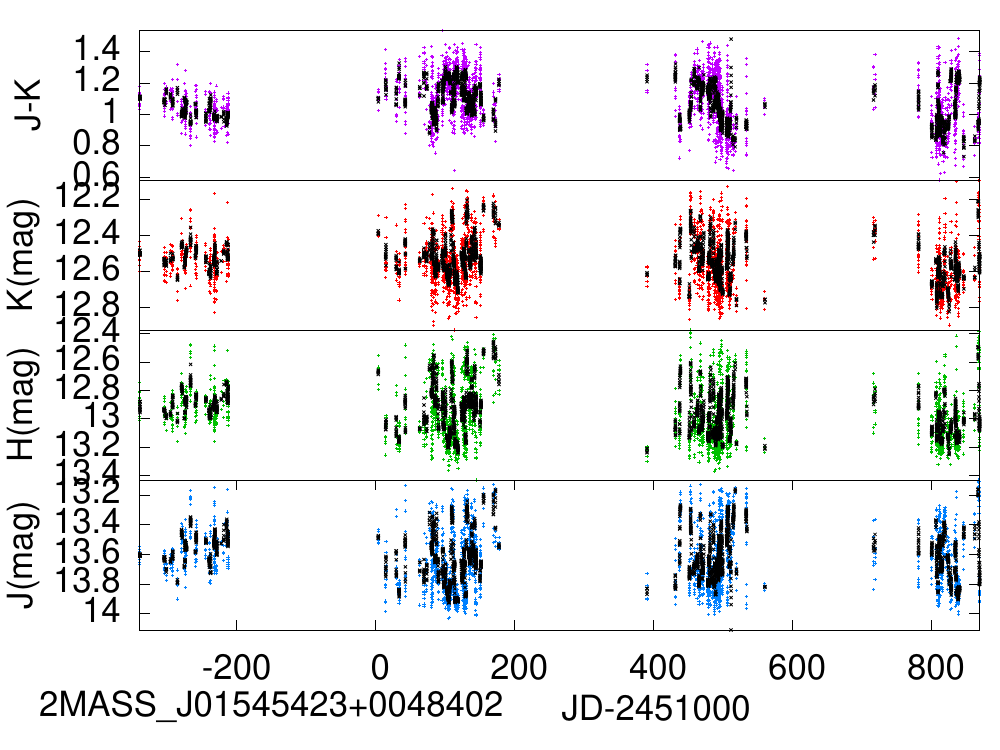} 
\includegraphics[width=2.1in, trim= 0 0 0 0 ]{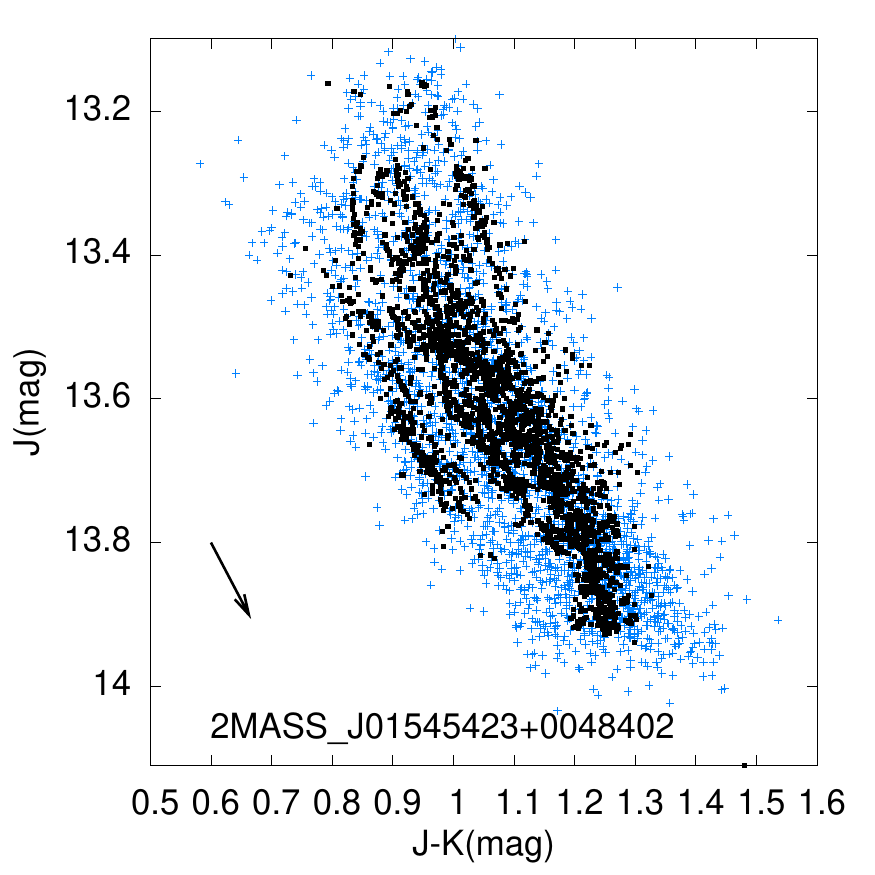} 
\includegraphics[width=2.8in, trim= 0 0 0 0 ]{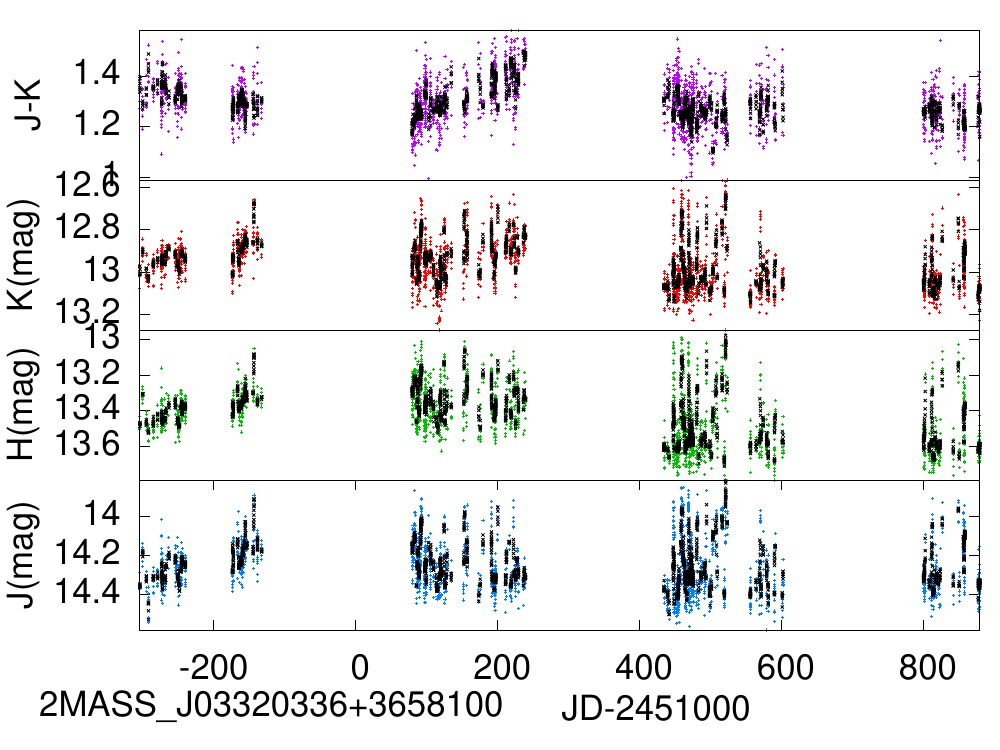} 
\includegraphics[width=2.1in, trim= 0 0 0 0 ]{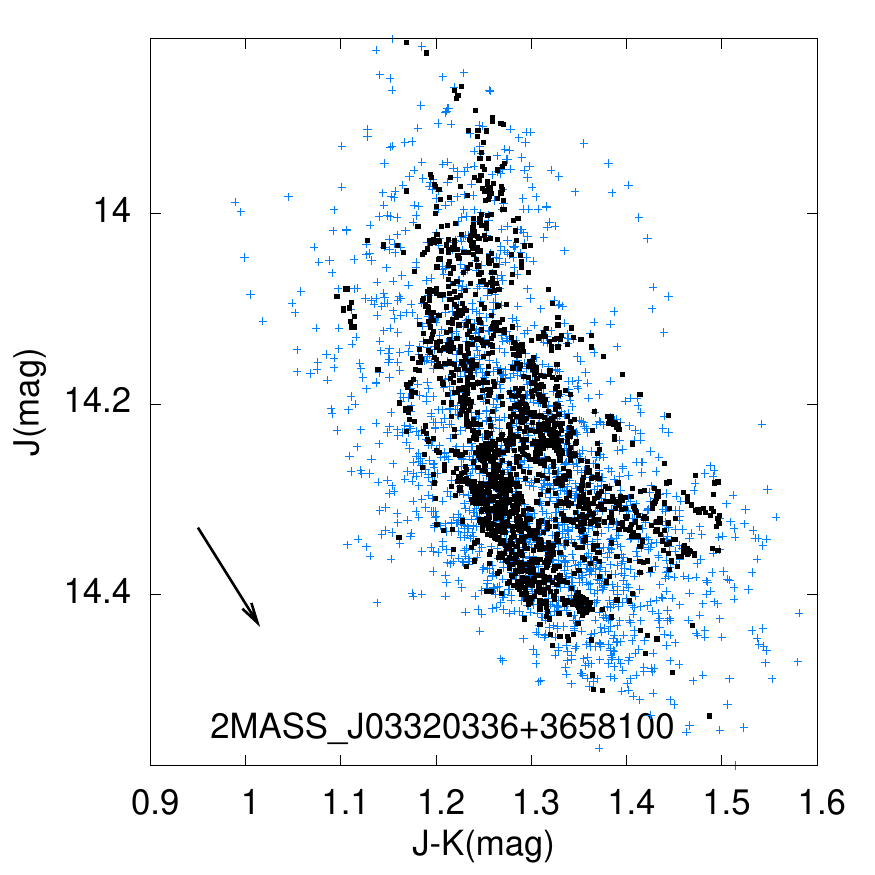} 
\includegraphics[width=2.8in, trim= 0 0 0 0 ]{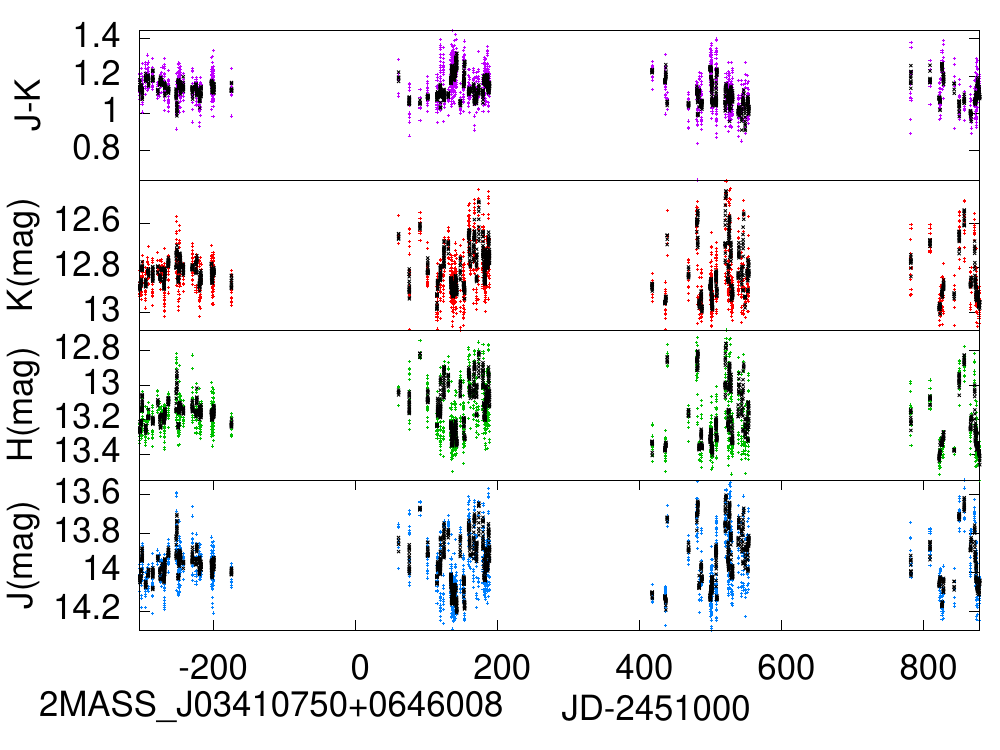}
\includegraphics[width=2.1in, trim= 0 0 0 0 ]{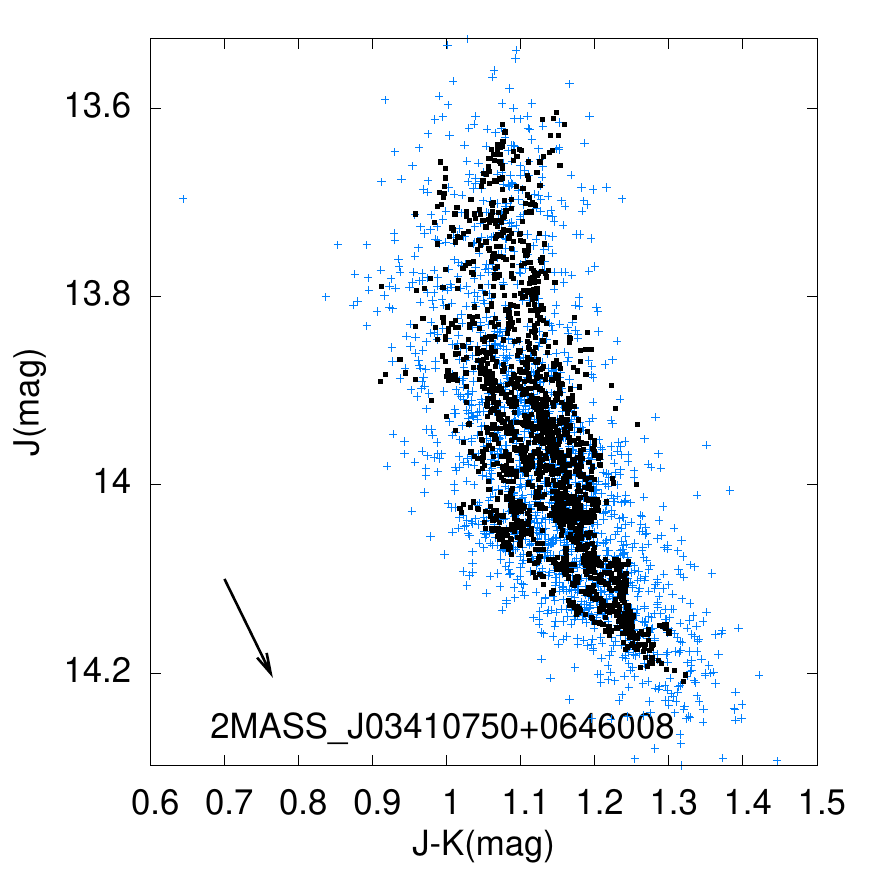}
\caption{Light curves and colors (similar to figure \ref{fig:aperiodic}) 
but of active galaxies found by searching for red dimming events.  These become bluer when they are brighter.
The objects are listed in Table \ref{tab:AGN}
}
\label{fig:AGN}
\end{figure*}

The color magnitude plots for the active galaxies (shown in Figure \ref{fig:AGN}) show that the
the color and magnitude variations are strongly correlated.   However the distribution in these plots
is often bowed, with less color variation taking place when the object is brightest
and more color variation when the object is faintest (e.g., the color magnitude plot of 2MASS J0341750+0646008).
In some cases, color magnitude variations are not restricted to a single one-dimensional line (e.g., 2MASS J00241907-0138176)
implying that there is more than one process contributing to the color and brightness variations.
For example, there might be both flares and extinction.

An additional two objects were found in the red dimming search for which we don't have classification.
 These are shown in Figure \ref{fig:additional} and listed in Table \ref{tab:additional}.
 2MASS J18395818+4903590 was not observed by WISE so we have no information on its mid-infrared colors.
 2MASS J07004330+4828115 has a red [4.6]-[12] color of 3.03 mag suggesting that
 the object is extragalactic. 
 A single eclipsing binary was found in the red-dimming search that was also found in the color insensitive search.
 We discuss this object below when we discuss the objects found in both dimming searches.
 
 To summarize, after excluding the $\rho$ Ophiucus region, the red dimming search found
 a single eclipsing binary (discussed below), 14 active galaxies, and 3 additional objects that exhibit
 correlations between color and brightness similar to those seen in the active galaxies
 (one of these was also found in the color insensitive search and was previously discussed).
 
\setcounter{figure}{3}
\begin{figure*}
\includegraphics[width=2.8in, trim= 0 0 0 0 ]{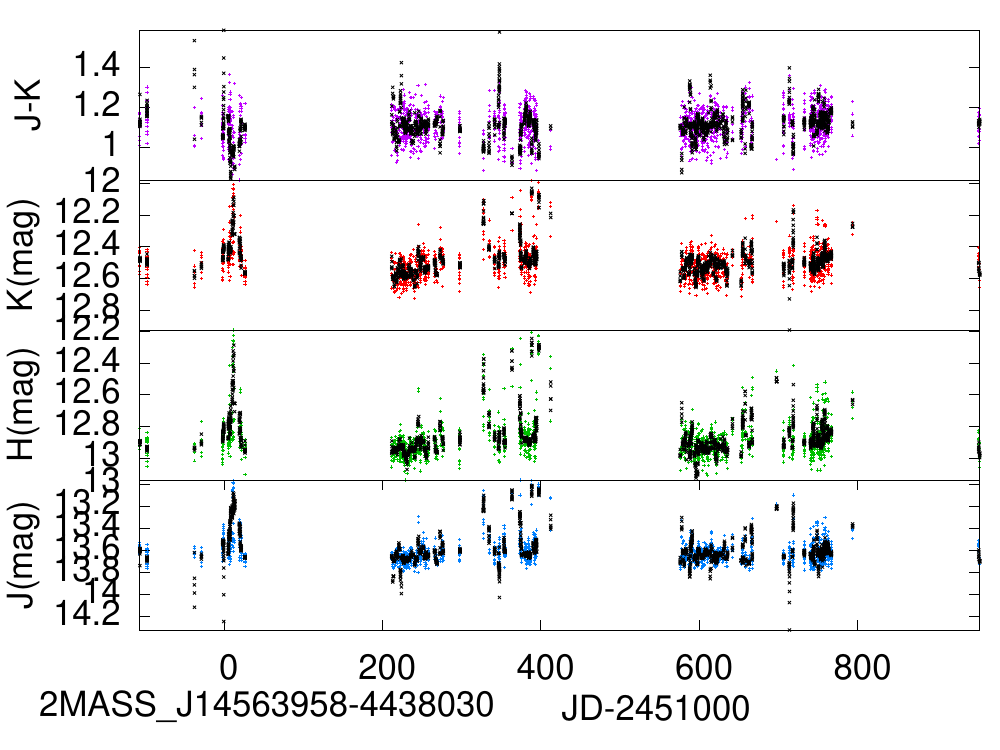}
\includegraphics[width=2.1in, trim= 0 0 0 0 ]{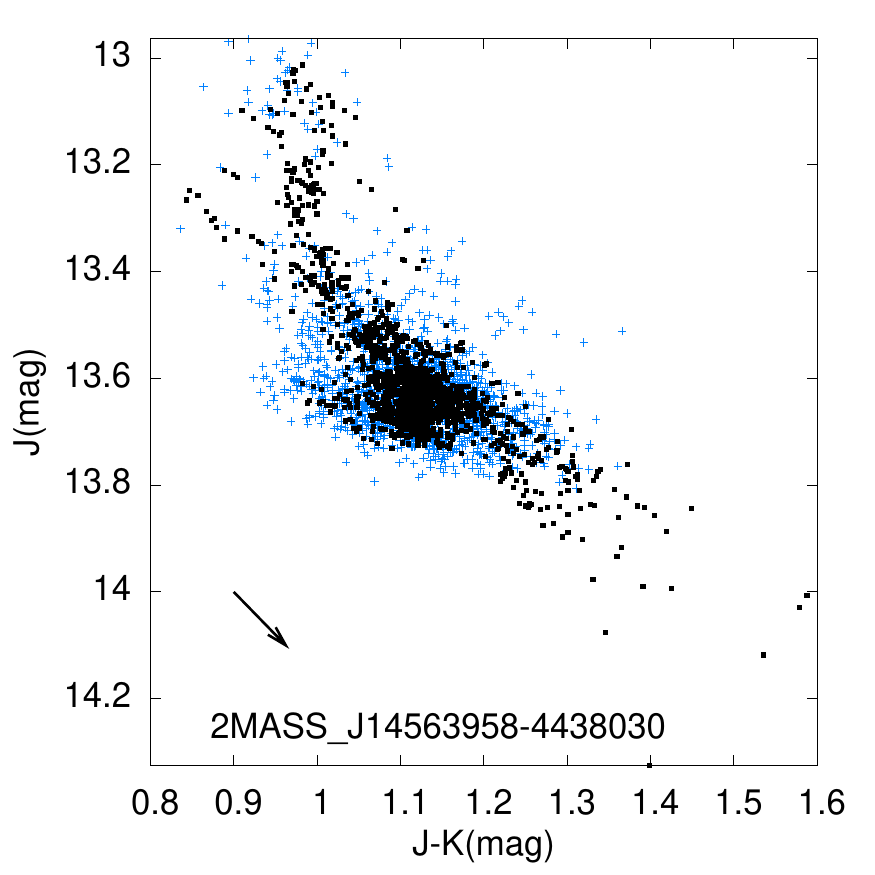}
\includegraphics[width=2.8in, trim= 0 0 0 0 ]{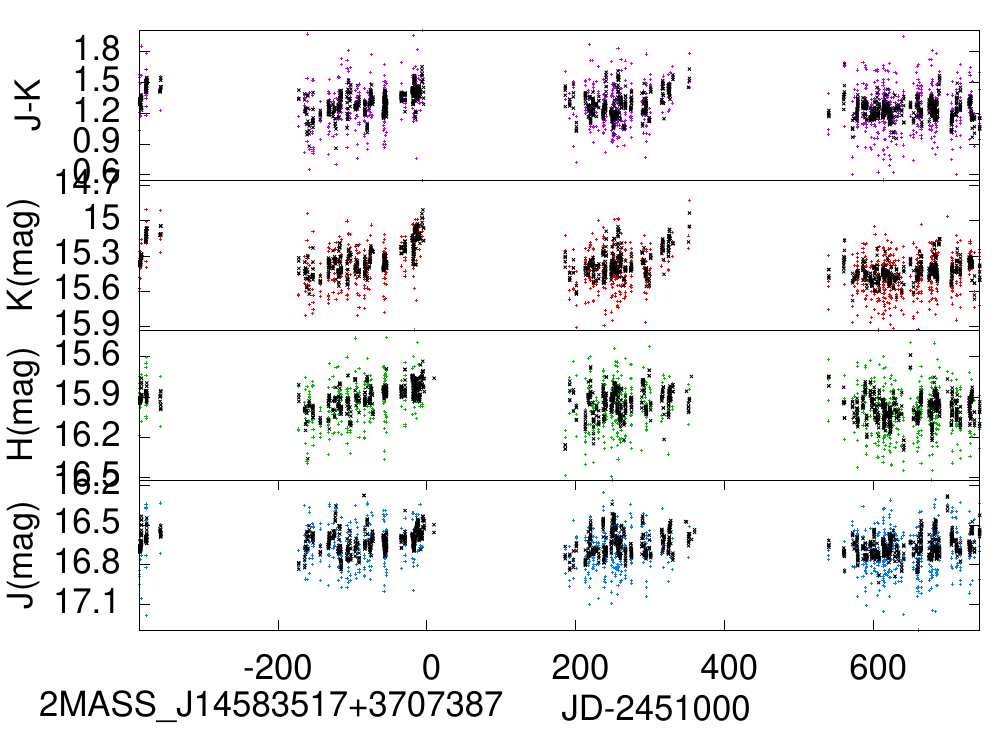} 
\includegraphics[width=2.1in, trim= 0 0 0 0 ]{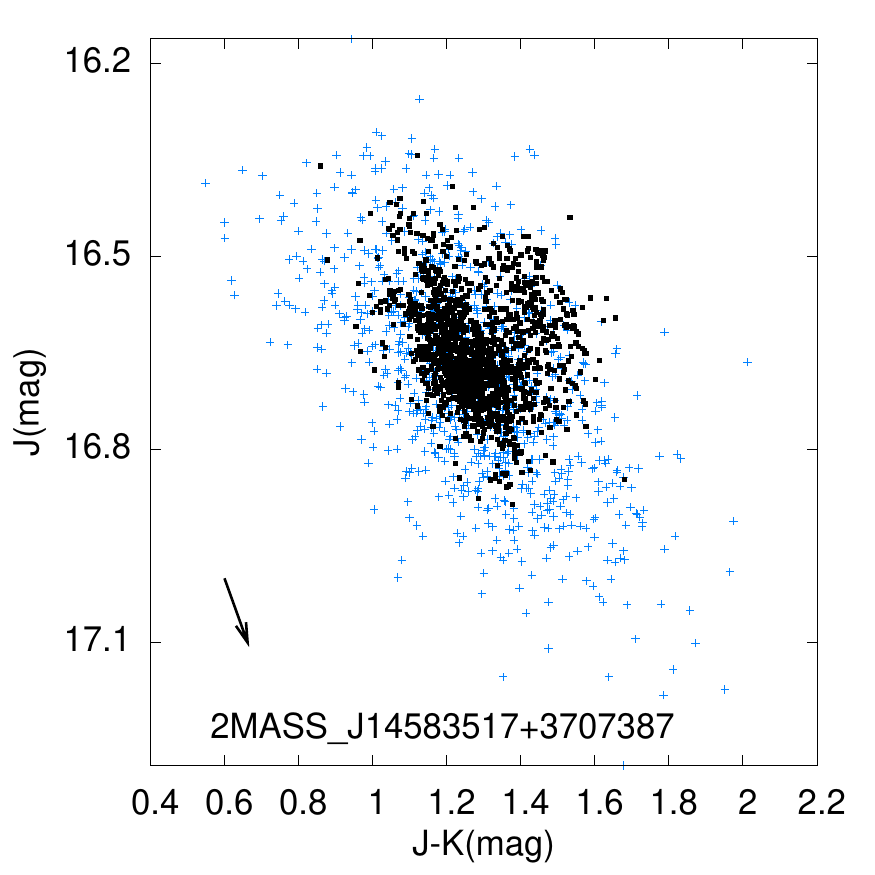} 
\includegraphics[width=2.8in, trim= 0 0 0 0 ]{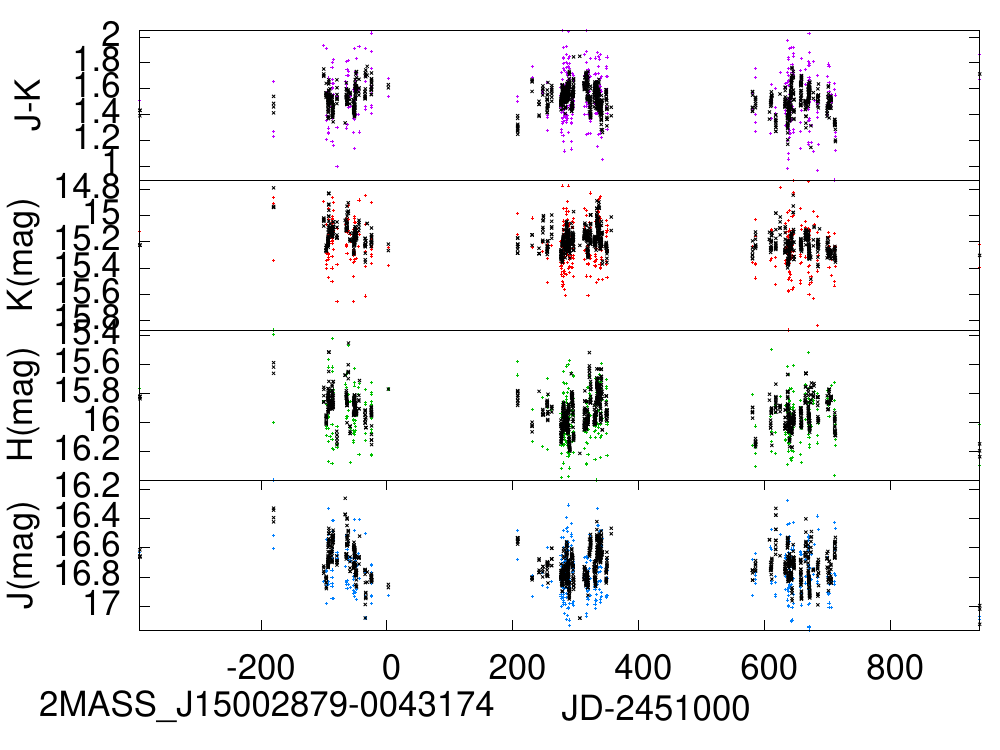} 
\includegraphics[width=2.1in, trim= 0 0 0 0 ]{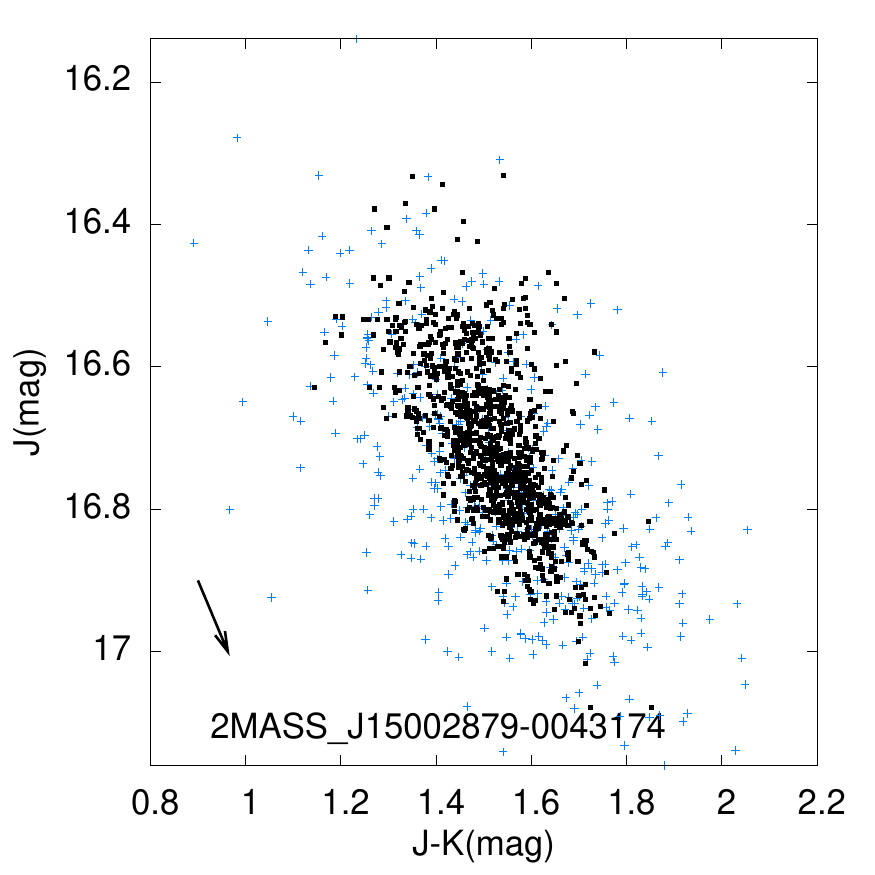} 
\includegraphics[width=2.8in, trim= 0 0 0 0 ]{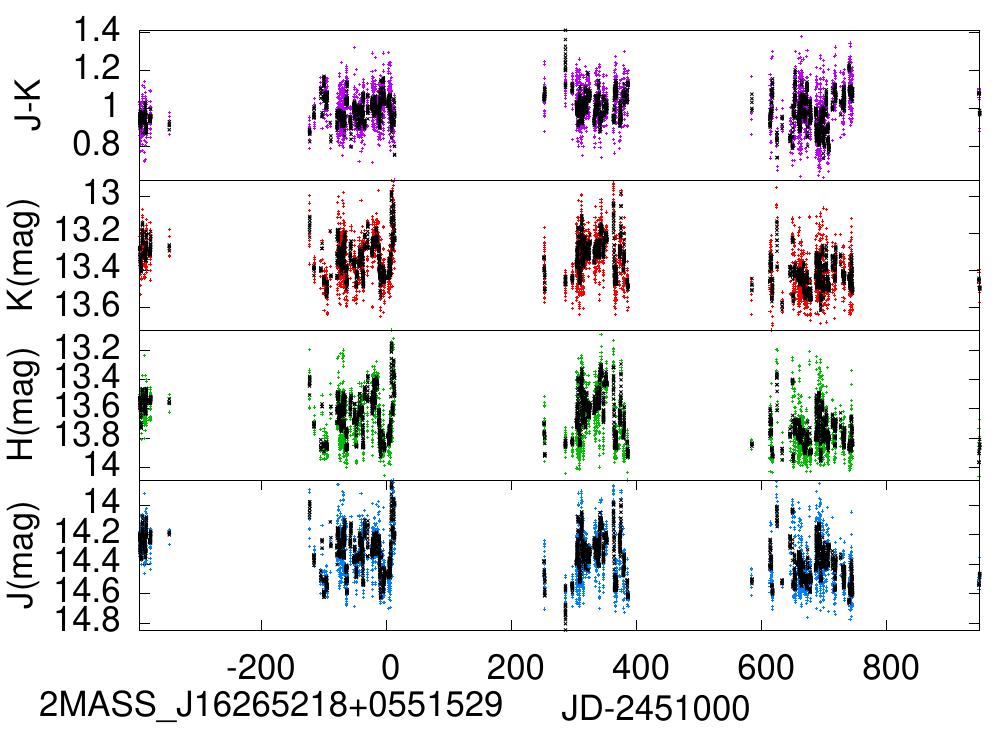}
\includegraphics[width=2.1in, trim= 0 0 0 0 ]{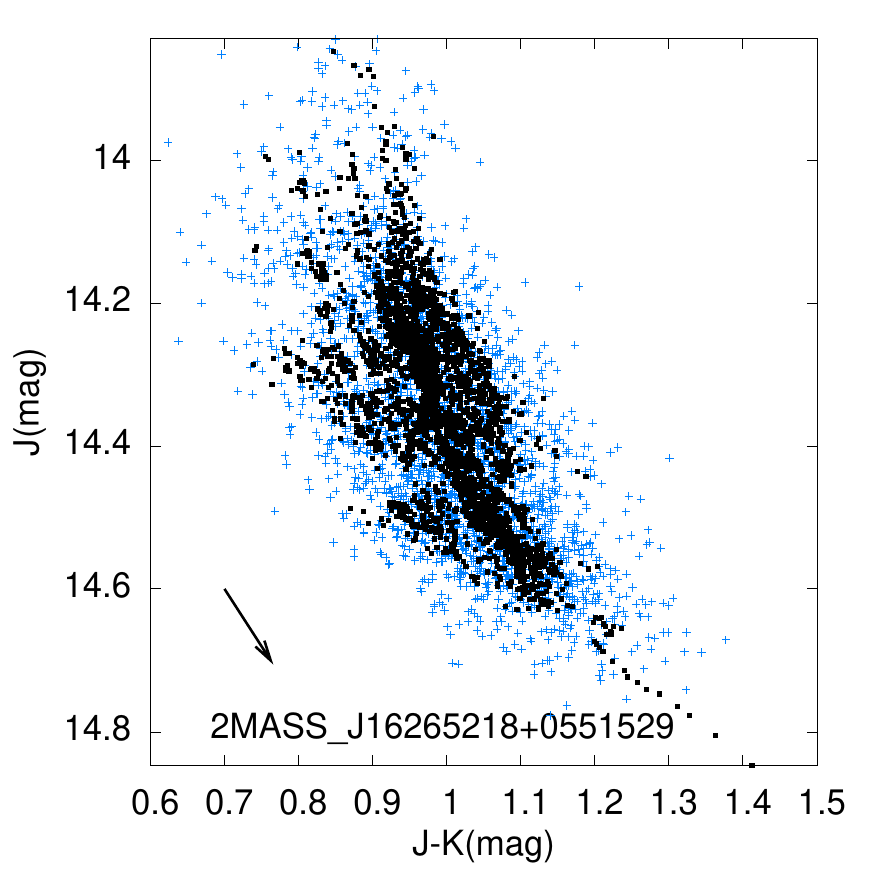}
\caption{Figure continued.
}
\end{figure*}

\setcounter{figure}{3}
\begin{figure*}
\includegraphics[width=2.8in, trim= 0 0 0 0 ]{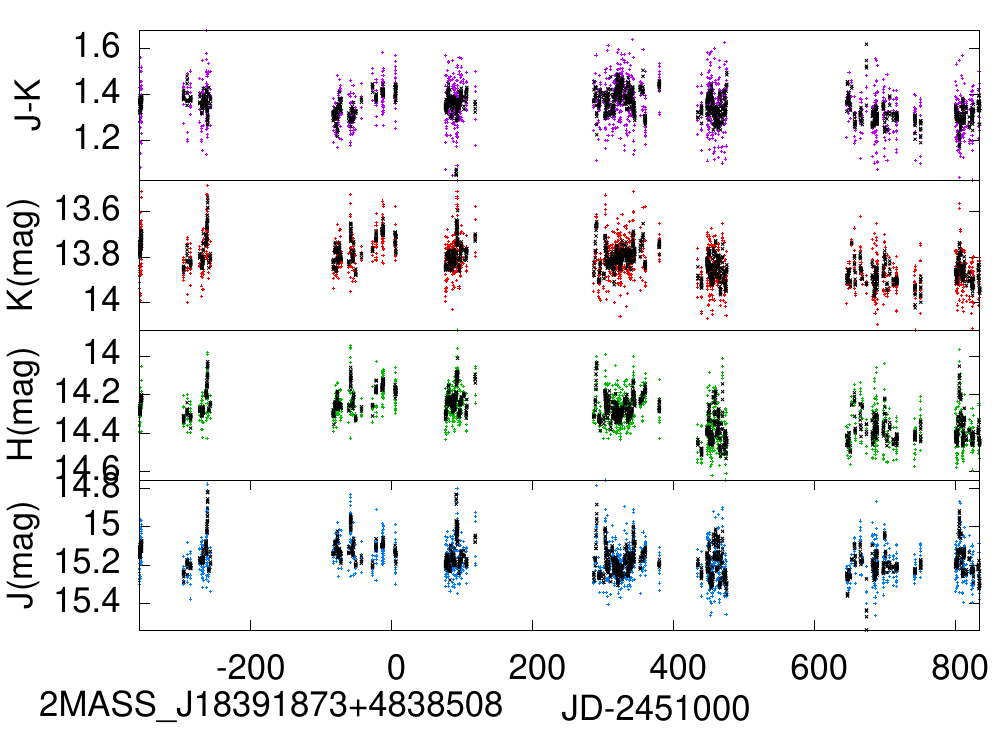}
\includegraphics[width=2.1in, trim= 0 0 0 0 ]{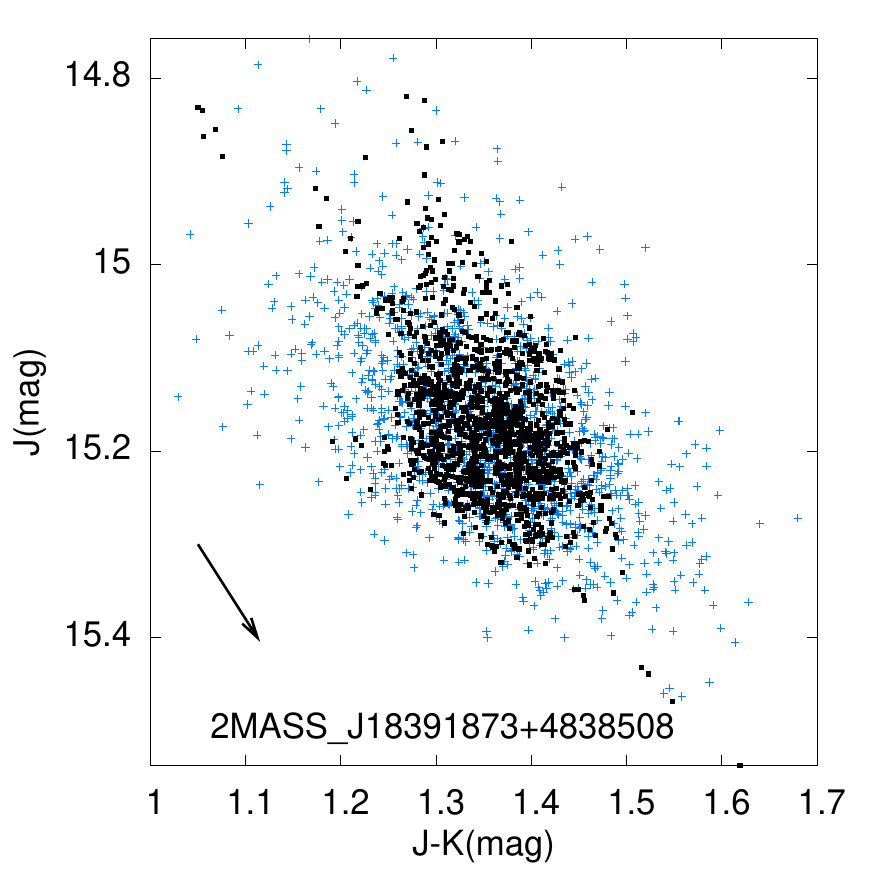}
\includegraphics[width=2.8in, trim= 0 0 0 0 ]{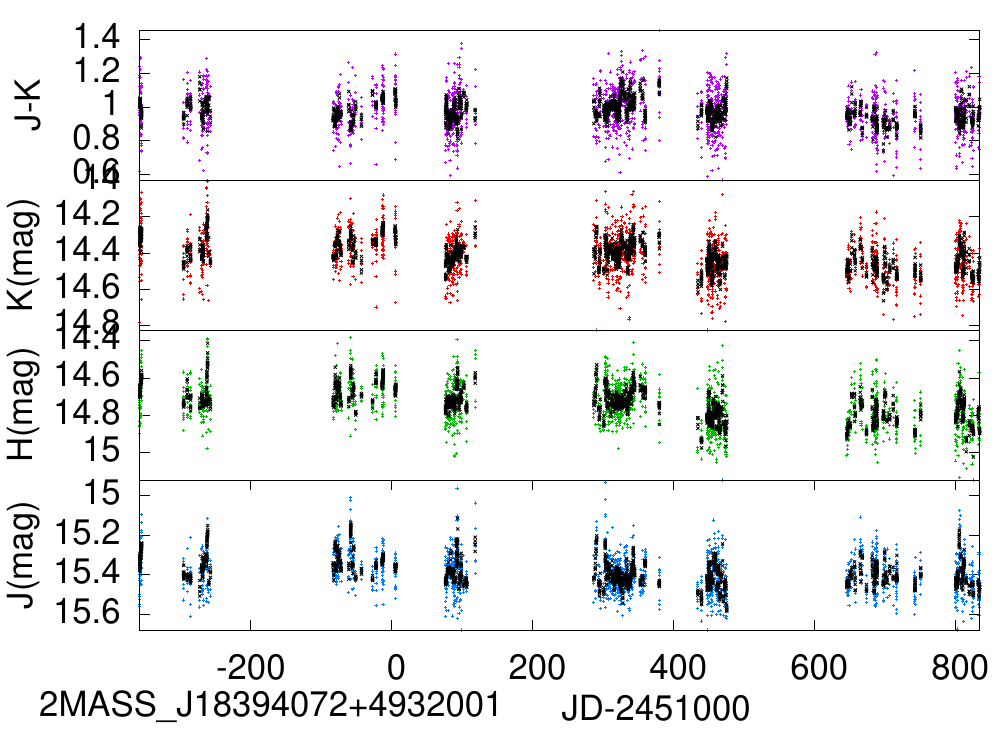} 
\includegraphics[width=2.1in, trim= 0 0 0 0 ]{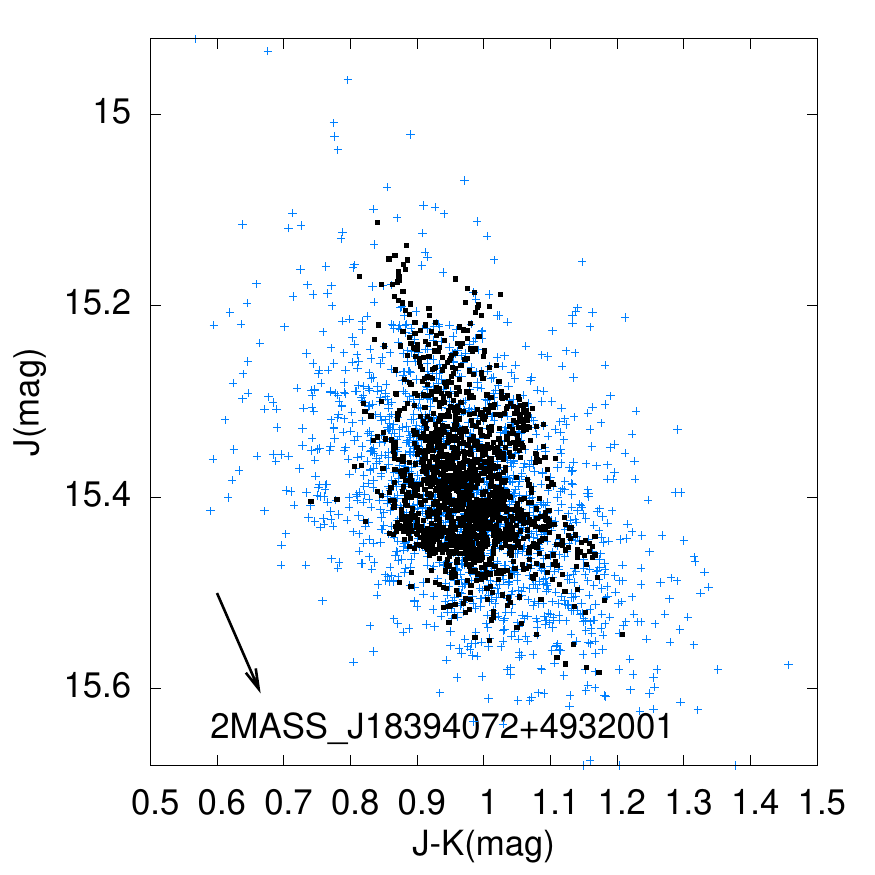} 
\includegraphics[width=2.8in, trim= 0 0 0 0 ]{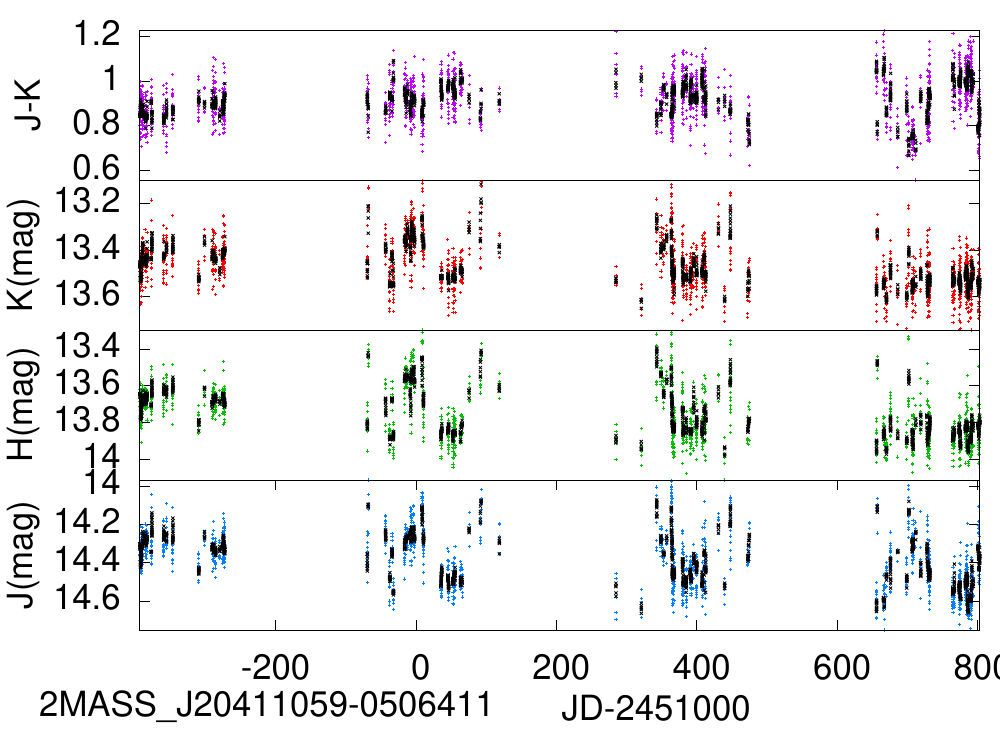}
\includegraphics[width=2.1in, trim= 0 0 0 0 ]{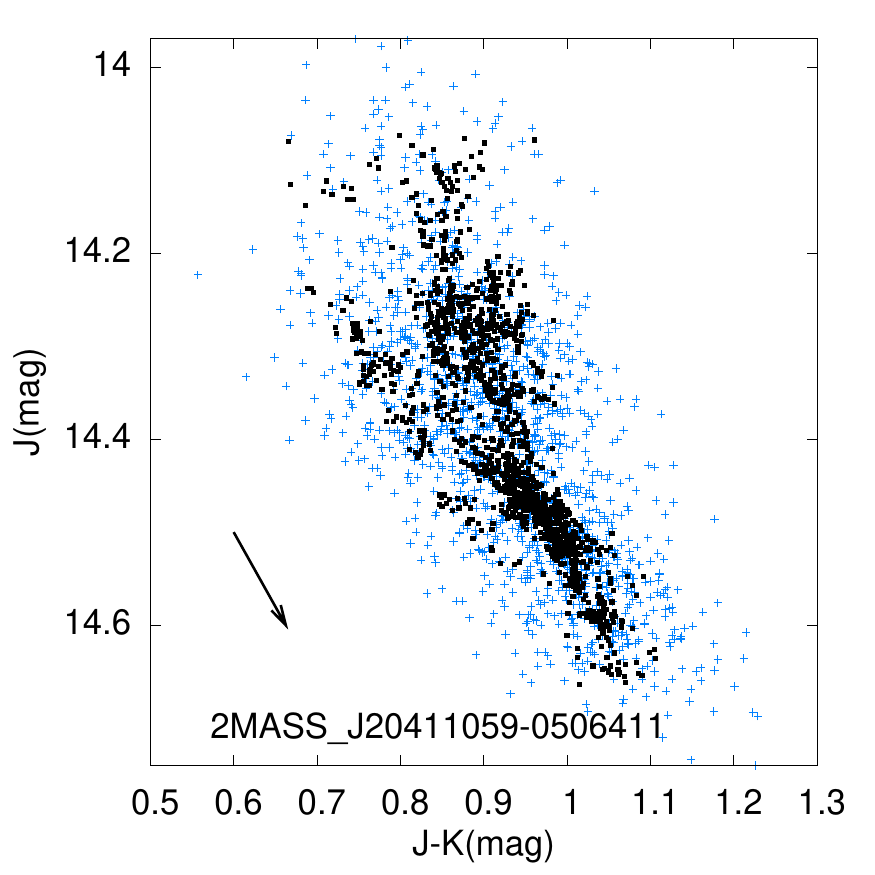}
\includegraphics[width=2.8in, trim= 0 0 0 0 ]{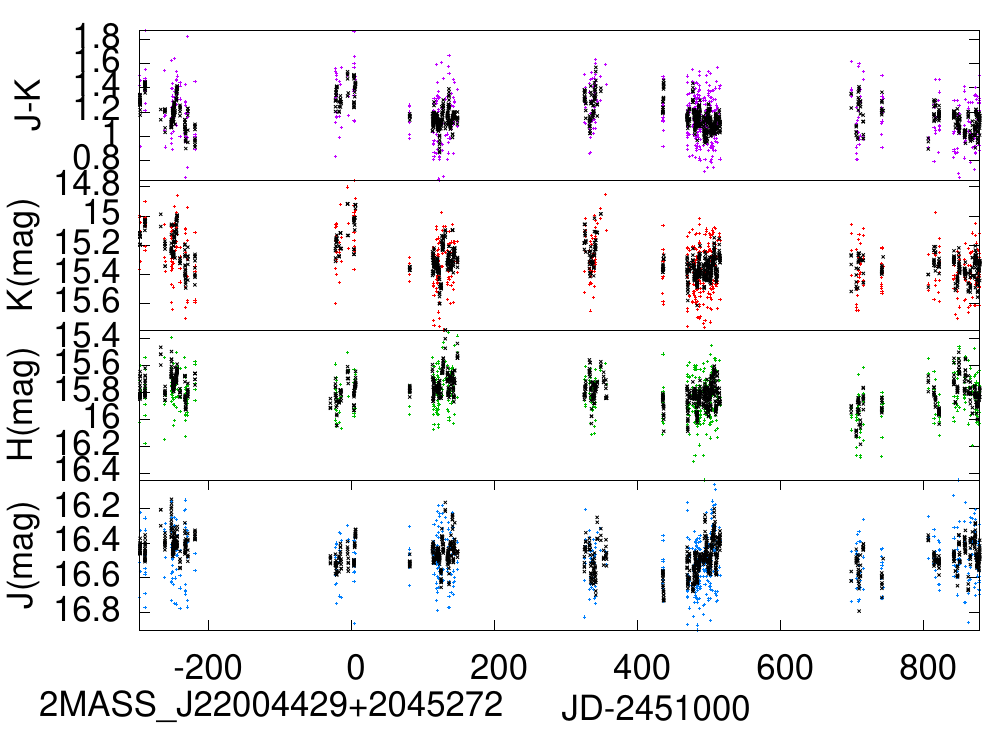}
\includegraphics[width=2.1in, trim= 0 0 0 0 ]{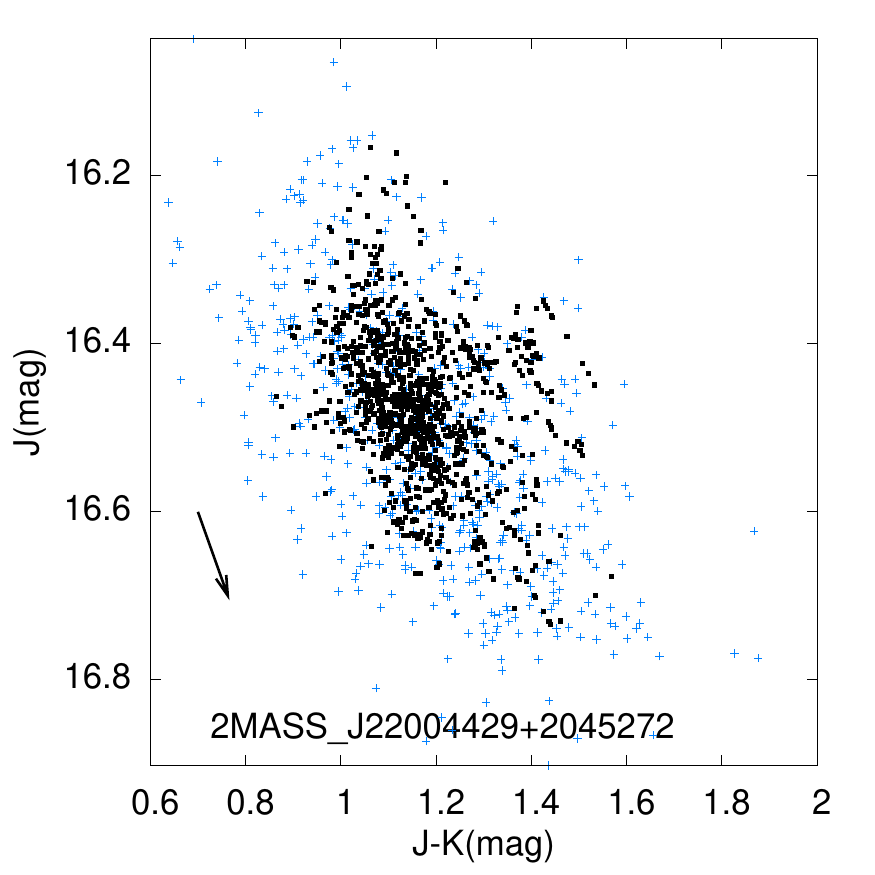}
\caption{Figure continued.
}
\end{figure*}

\setcounter{figure}{3}
\begin{figure*}
\includegraphics[width=2.8in, trim= 0 0 0 0 ]{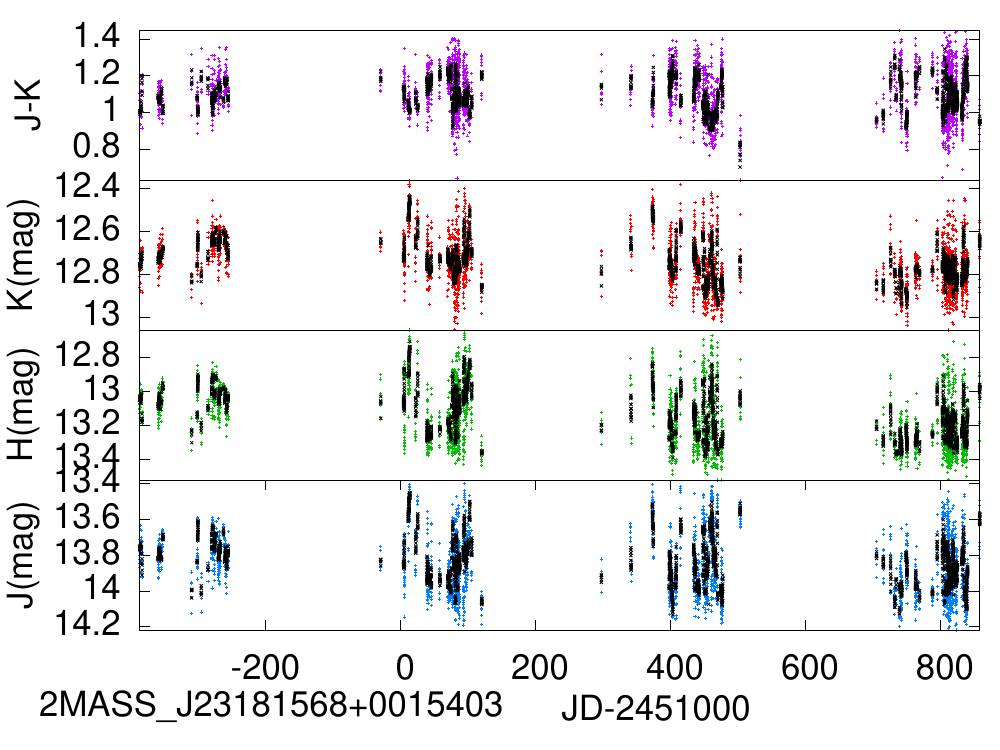}
\includegraphics[width=2.1in, trim= 0 0 0 0 ]{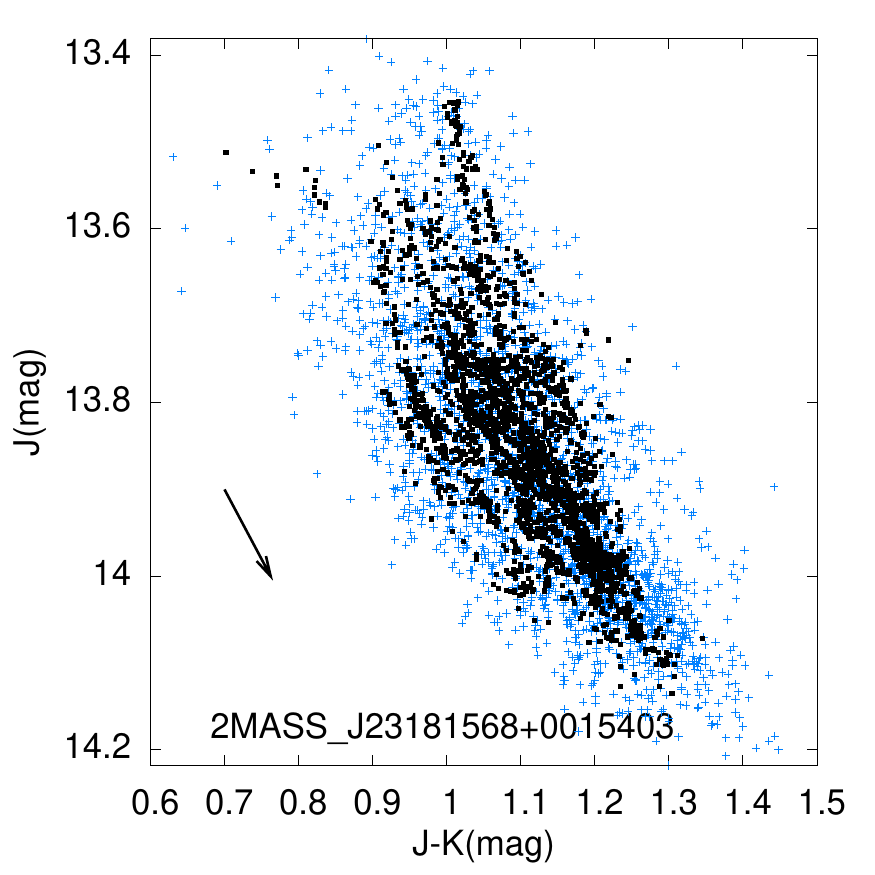}
\caption{Figure continued.
}
\end{figure*}

\begin{figure*}
\includegraphics[width=2.8in, trim= 0 0 0 0 ]{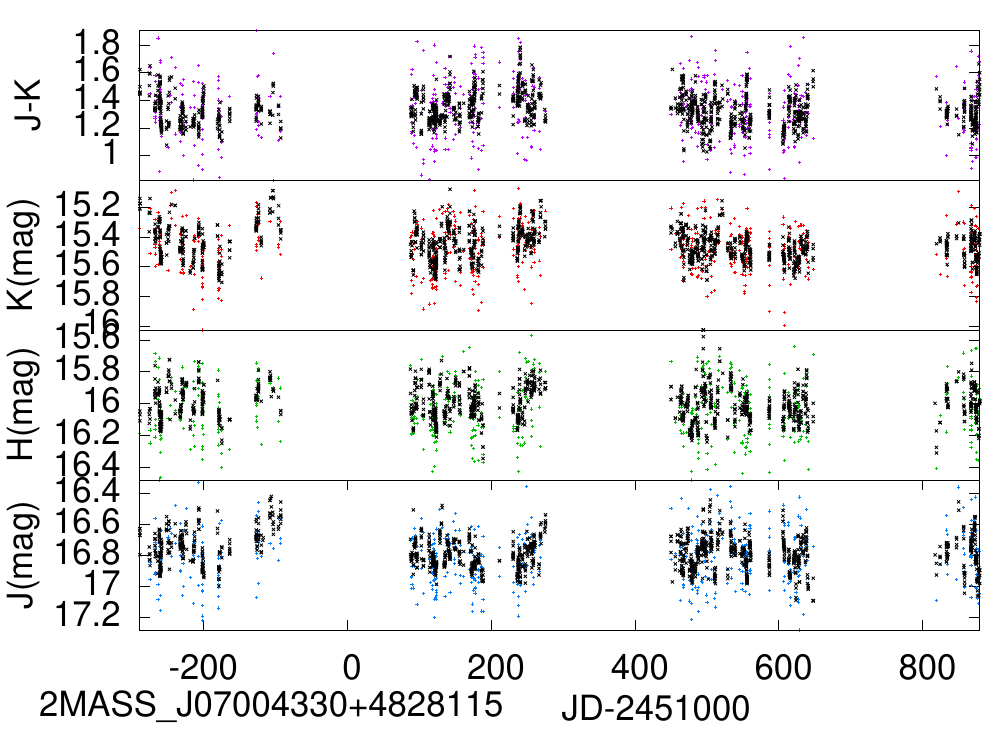} 
\includegraphics[width=2.1in, trim= 0 0 0 0 ]{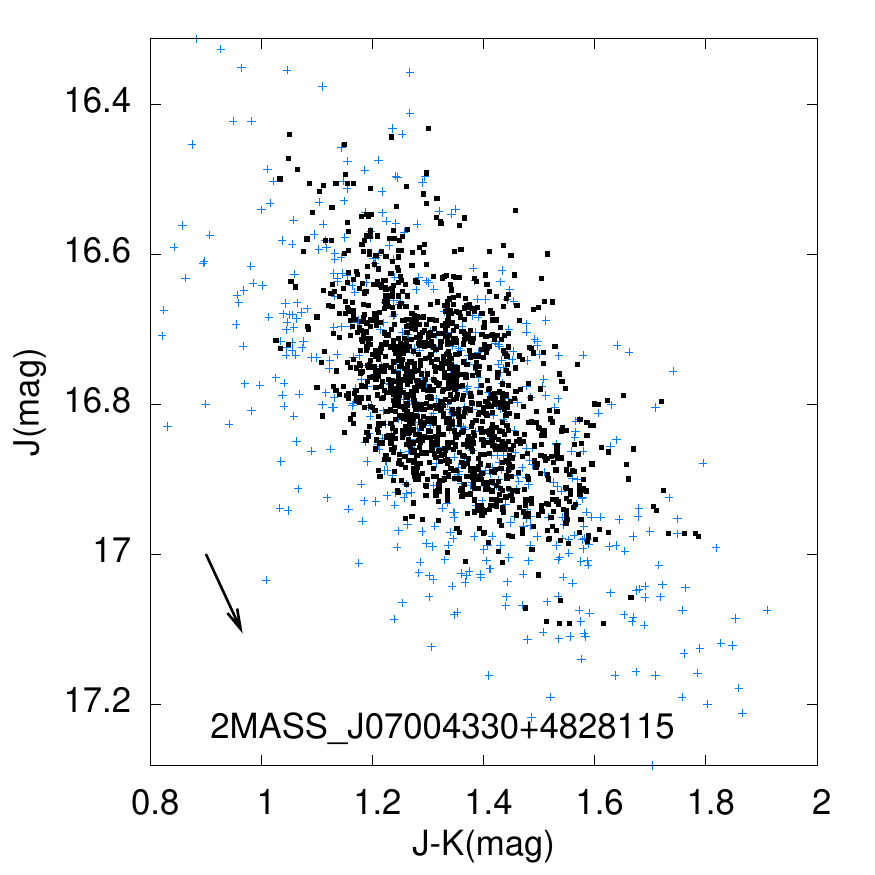} 
\includegraphics[width=2.8in, trim= 0 0 0 0 ]{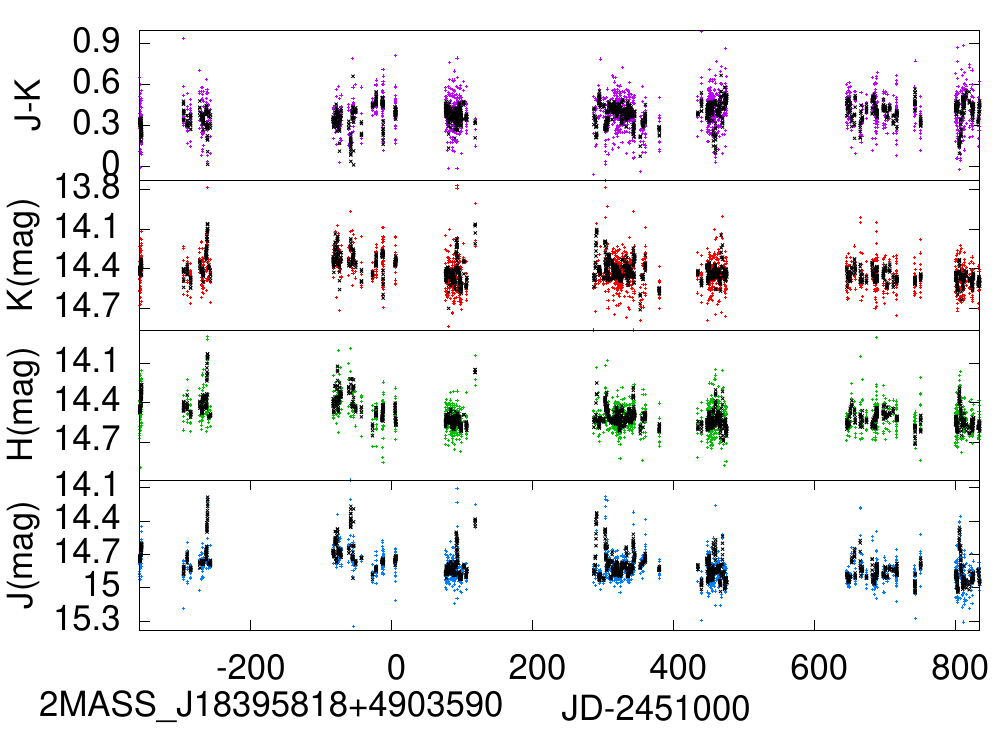}
\includegraphics[width=2.1in, trim= 0 0 0 0 ]{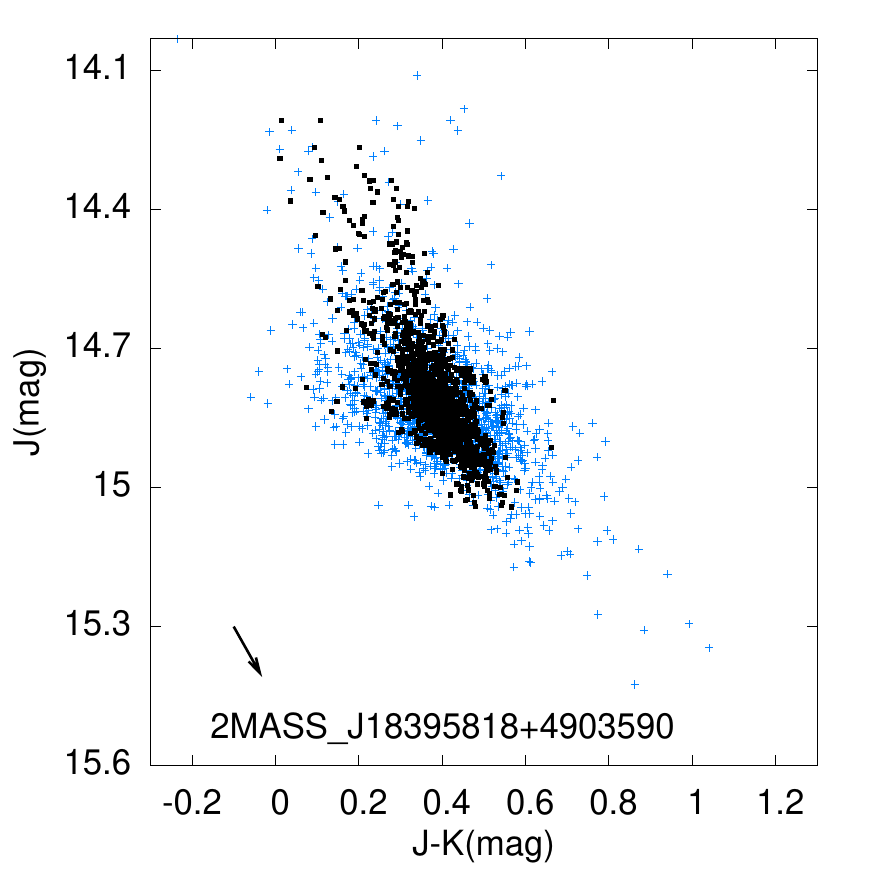}
\caption{Light curves and colors (similar to figure \ref{fig:aperiodic}) but of 
additional objects found in the red dimming search.  
These objects were not identified as extragalactic sources or young stellar
objects and
are listed in Table \ref{tab:additional}. 
}
\label{fig:additional}
\end{figure*}

\begin{table*}
 \begin{minipage}{190mm}
 \caption[]{Active galaxy candidates found in a search for red dimming events \label{tab:AGN}}
 \begin{tabular}{l l  l l l l l l  r l l l l}
 \hline
Object                             & $J$ & $\sigma_J$ & $H$ & $\sigma_H$ & $K$ & $\sigma_K$ & $J-K$ & [3.4]-[4.6] & [4.6]-[12]  &  $R$ & $b$ & Ref \\
      (1)                             &  (2) & (3)              & (4)    & (5)               & (6)   &  (7)              & (8)      & (9)          & (10)     & (11)  & (12) & (13) \\
\hline
2MASS J00241907$-$0138176&14.728 & 0.187  & 14.066 & 0.207 & 13.669 & 0.186 & 1.06 & 0.10  & 2.40  & 0.40 & 0.500$\pm$0.026& L11\\
2MASS J01545423+0048402&13.578 & 0.223 & 12.918 & 0.209 & 12.532 & 0.143 &  1.05 & 0.26  & 4.41  & 0.74 & 1.078$\pm$0.019& H12\\
2MASS J03320336+3658100&14.225 & 0.142 & 13.397  & 0.166 & 12.941 & 0.120 & 1.28 & 0.39 & 3.77  & 0.51 & 0.828$\pm$0.032& 2MES\\
2MASS J03410750+0646008&13.939  & 0.162 & 13.152 & 0.170 & 12.818 & 0.120 & 1.12  & -0.08 & 2.63  & 0.64& 1.104$\pm$0.031& H12\\
2MASS J08255140$-$3907590&15.586 & 0.209  & 14.386 & 0.166 & 13.386 & 0.110 & 2.2  & 1.05$^a$  & 2.99   &  0.90 & 1.402$\pm$0.012& - \\  
2MASS J14563958$-$4438030&13.563 & 0.244 & 12.852 & 0.178 & 12.485 & 0.121 & 1.08 & -0.07  & 0.65  & 0.48 & 0.827$\pm$0.038& P05\\
2MASS J14583517+3707387&16.645 & 0.163 & 15.940  & 0.178 & 15.393 & 0.187 &  1.25 & 0.29 & 3.50  & 0.54 & 0.397$\pm$0.018& SDSS\\
2MASS J15002879$-$0043174&16.702 & 0.183 & 15.914  & 0.189 & 15.190 &  0.174 & 1.51 & 0.39 & 2.87 & 0.58 & 0.510$\pm$0.026& SDSS\\
2MASS J16265218+0551529&14.324  & 0.199 & 13.652 & 0.187 & 13.352 & 0.132 & 0.07 & -0.08 & 2.42  &0.67 & 1.167$\pm$0.025 & P03\\
2MASS J18391873+4838508&15.162 & 0.124 & 14.288  & 0.113 & 13.815  & 0.095 & 1.35 & 0.11 & 2.16  & 0.50 & 0.524$\pm$0.023& P03\\ 
2MASS J18394072+4932001&15.382 &0.100  & 14.729 & 0.115 & 14.410 & 0.120 & 0.97 & 0.08  & 2.68   & 0.44 & 0.349$\pm$0.017& P03\\
2MASS J20411059$-$0506411&14.364 & 0.143 & 13.725 & 0.149 &  13.460 & 0.108 &  0.90 & -0.04 & 2.06 &0.66 & 0.975$\pm$0.028& P03 \\
2MASS J22004429+2045272&16.465 & 0.156 & 15.788 & 0.168 & 15.311  & 0.175 & 1.15 & 0.33  & 3.54  &0.56 & 0.417$\pm$0.022& SDSS\\
2MASS J23181568+0015403&13.819 & 0.172 & 13.115 & 0.178 & 12.737 & 0.117 & 1.08 & 0.18  & 2.92 &0.72 &  1.022$\pm$0.019 & S10 \\
 \hline
\end{tabular}
 \\
Here we list extragalactic objects found in a search for red dimming events.
Columns: 
(1): 2MASS point source catalog identifier.
(2)-(7):  Mean and standard deviations, in magnitudes, computed from the light curves in $J, H,$ and $K$ bands, respectively.
(8): $J-K$ Color in magnitudes was computed from the mean magnitudes of the light curve.
(9,10):  Colors in Vega magnitudes from the WISE survey \citep{wise}.
The four bands of the WISE survey are $3.4,4.6, 12$ and 22~$\mu$m.
(11): The cross correlation coefficient, $R$, (equation \ref{eqn:R}) between color and magnitude.
(12): The slope, $b$, of the best fitting line on the color/magnitude plot (equation \ref{eqn:b}).
(13): Reference where the object was listed as extragalactic.
L11 - \citep{lavaux11}; 
H12 - \citep{huchra12};
2MES - in the 2MASS Extended Sources catalog;
P05 - \citep{paturel05}; 
P03 - \citep{paturel03}; 
SDSS - classified as a galaxy with SDSS imaging;
S10 - \citet{schawinski10}.
$^a$2MASS J08255140-3907590 is identified here as extragalactic based on its red  [3.4]-[4.6] color.
2MASS J03320336+3658100 is a variable radio source \citep{ofek11}.
\end{minipage}
\end{table*}

\begin{table*}
\begin{minipage}{160mm}
 \caption[]{Additional objects found in a search for red dimming events \label{tab:additional}}
 \begin{tabular}{l l  l l l l l l  l l l l}
 \hline
Object  & $J$ & $\sigma_J$ & $H$ & $\sigma_H$ & $K$ & $\sigma_K$ & $J-K$ &  [3.4]-[4.6] & [4.6]-[12] & $R$ & $b$ \\
      (1)  &  (2) & (3)              & (4)    & (5)               & (6)   &  (7)              & (8)      & (9)          & (10)            & (11) & (12) \\
      \hline
2MASS J07004330+4828115 &16.773 & 0.177 & 16.003  & 0.175  & 15.453 & 0.170 &  1.32 & 0.34 & 3.03 &  0.60 & 0.499$\pm$0.021 \\
2MASS J18395818+4903590 & 14.790 & 0.169 & 14.500 & 0.126 & 14.411 & 0.137 & 0.38 & -       & -        & 0.55 & 0.588$\pm$0.022\\
\hline
\end{tabular}
 \\
Here we list  objects found in a search for red dimming events that were not identified as extragalactic or young stellar objects.
Columns are the same as in Table \ref{tab:AGN}.
2MASS J18395818+4903590 was not in the WISE all sky data release catalog.
\end{minipage}
\end{table*}

\section{Young Stellar objects from the $\rho$ Ophiucus region}

\begin{figure*}
\includegraphics[width=2.0in, trim= 0 0 0 0 ]{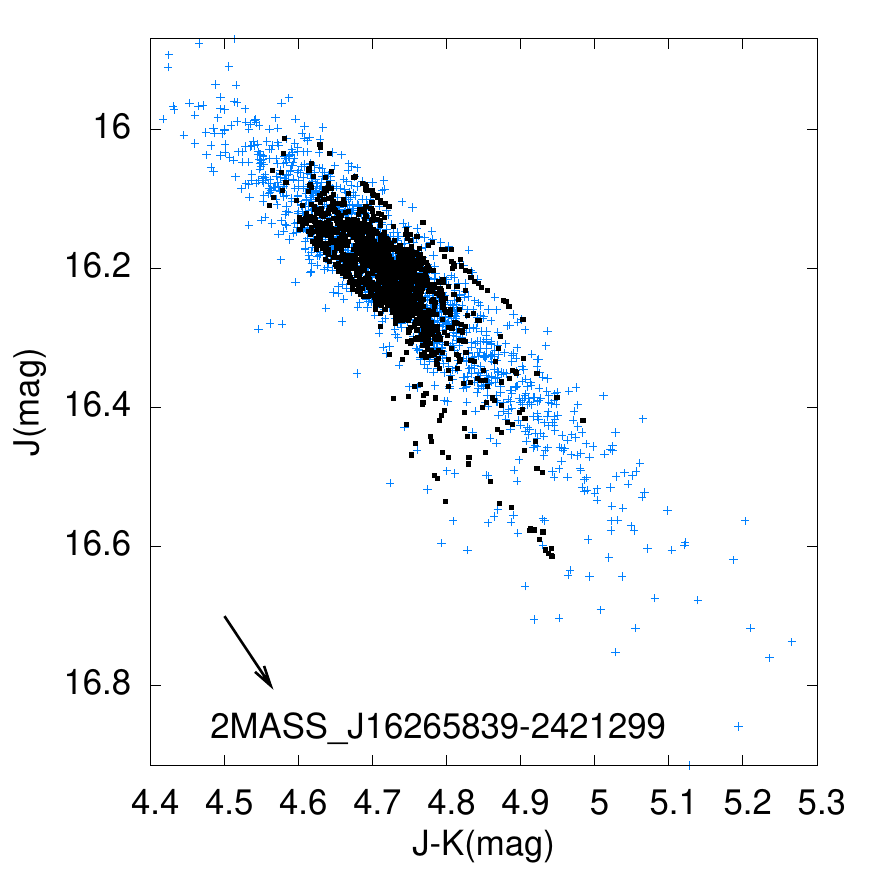} 
\includegraphics[width=2.0in, trim= 0 0 0 0 ]{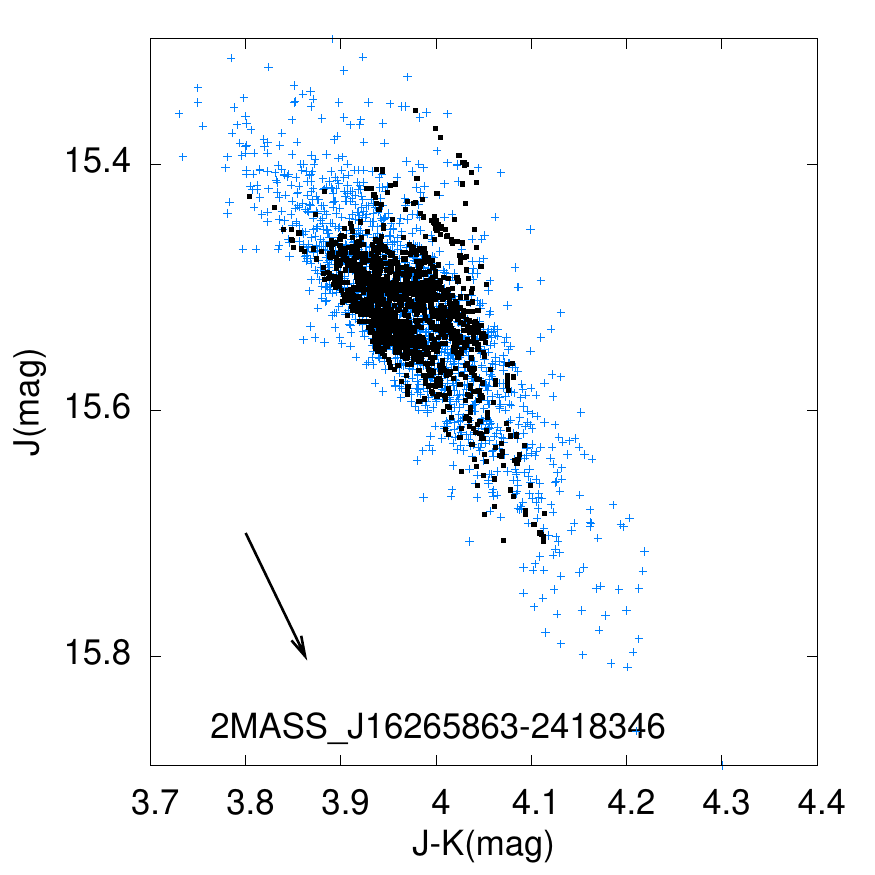}
\includegraphics[width=2.0in, trim= 0 0 0 0 ]{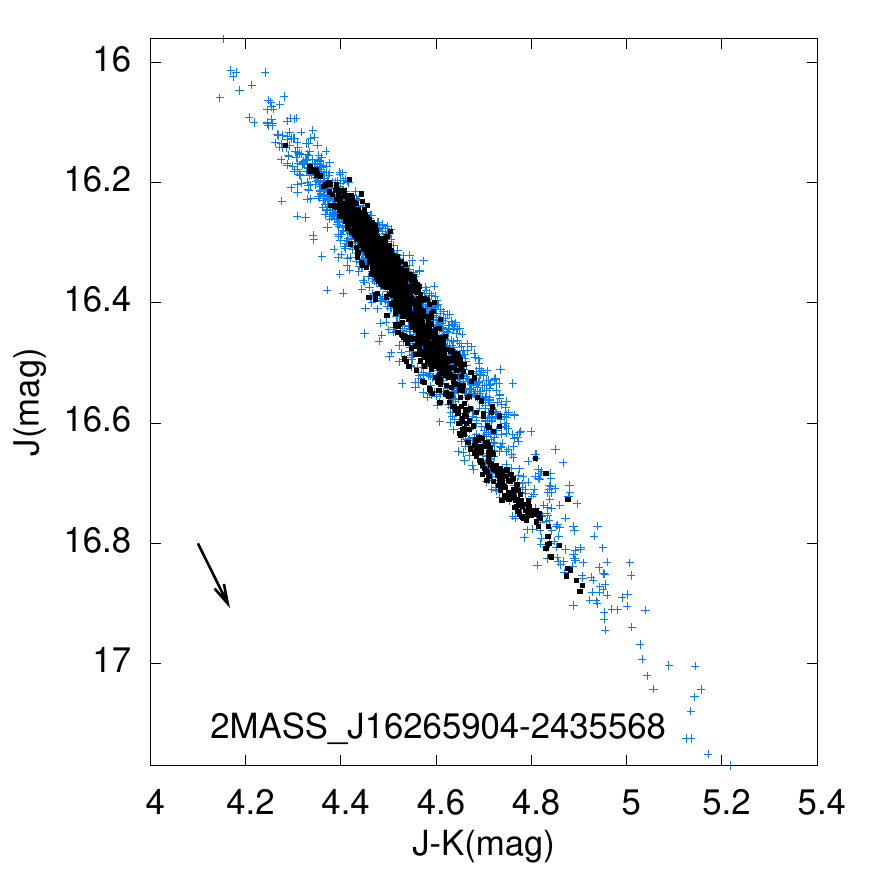}
\includegraphics[width=2.0in, trim= 0 0 0 0 ]{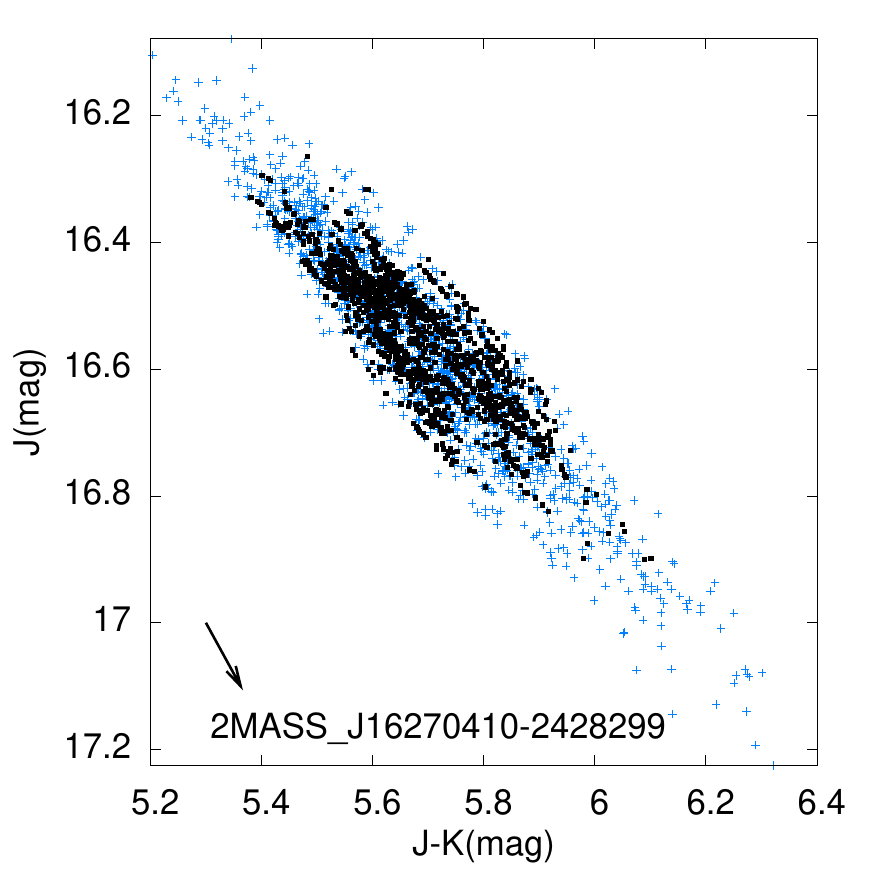} 
\includegraphics[width=2.0in, trim= 0 0 0 0 ]{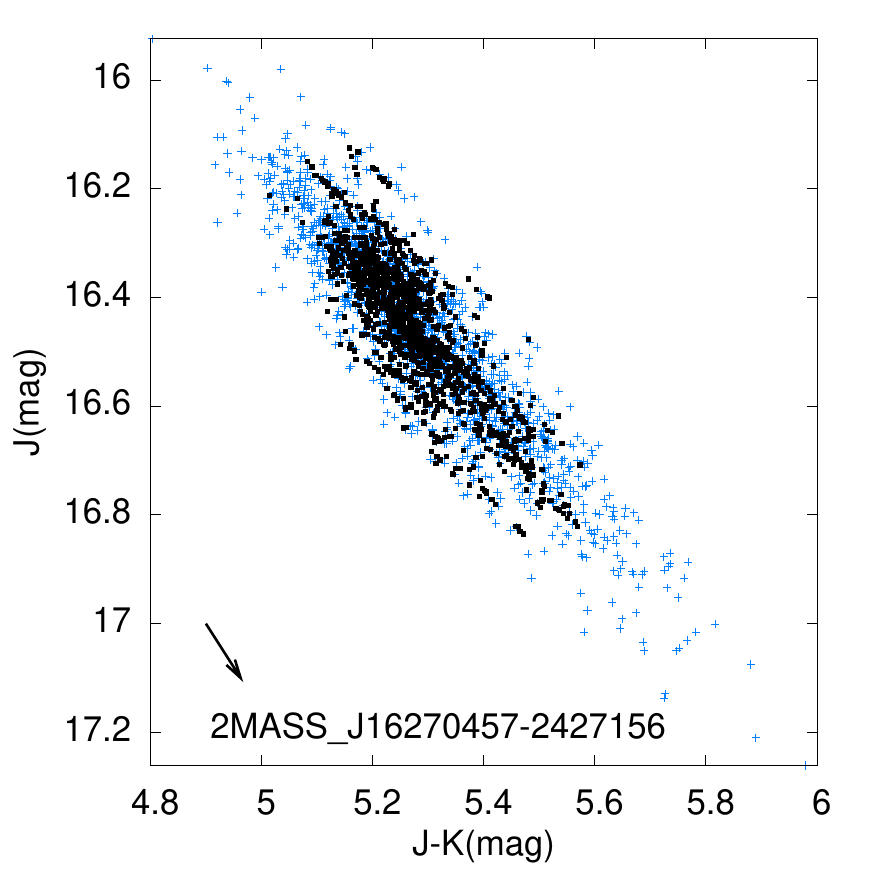}
\includegraphics[width=2.0in, trim= 0 0 0 0 ]{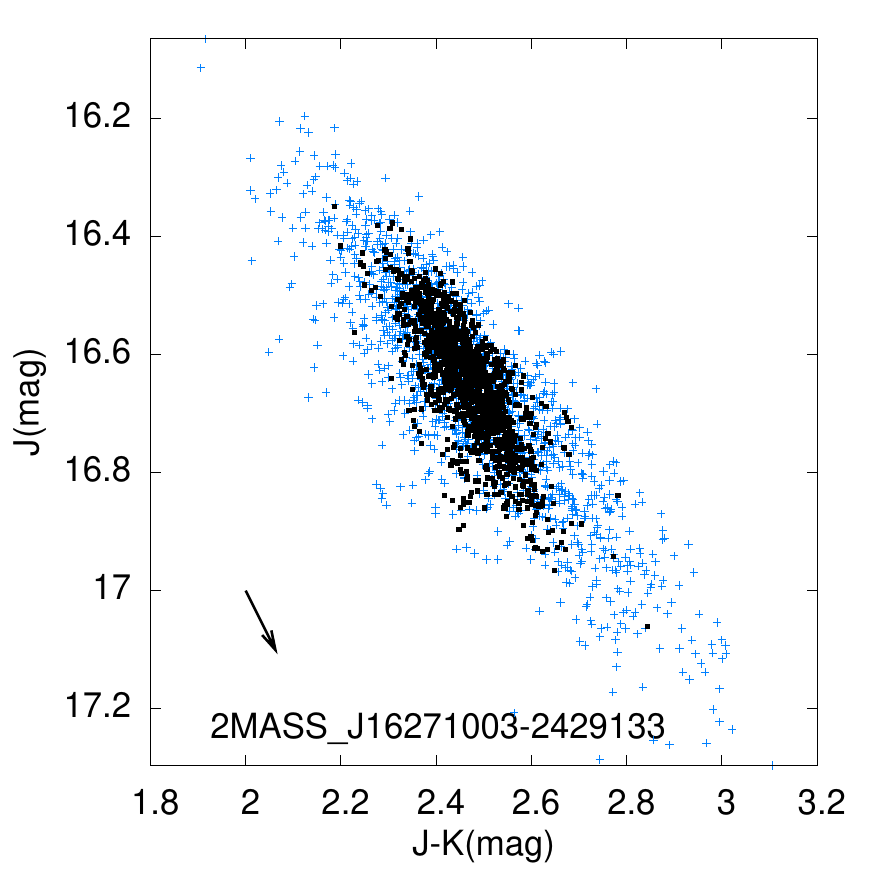} 
\includegraphics[width=2.0in, trim= 0 0 0 0 ]{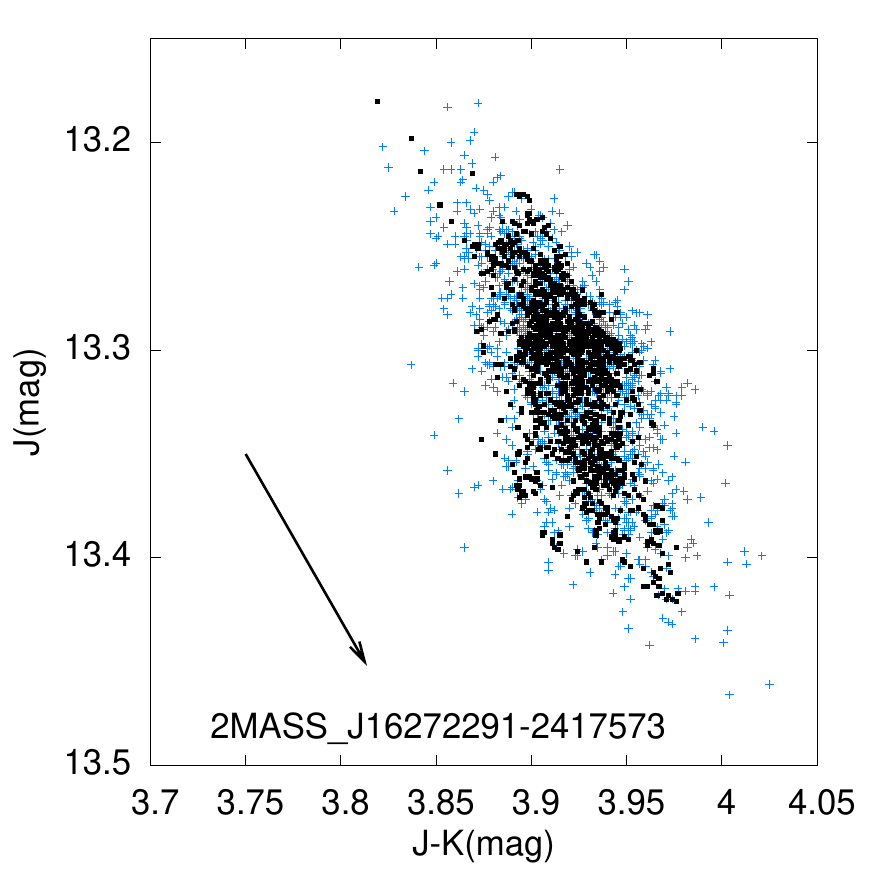} 
\includegraphics[width=2.0in, trim= 0 0 0 0 ]{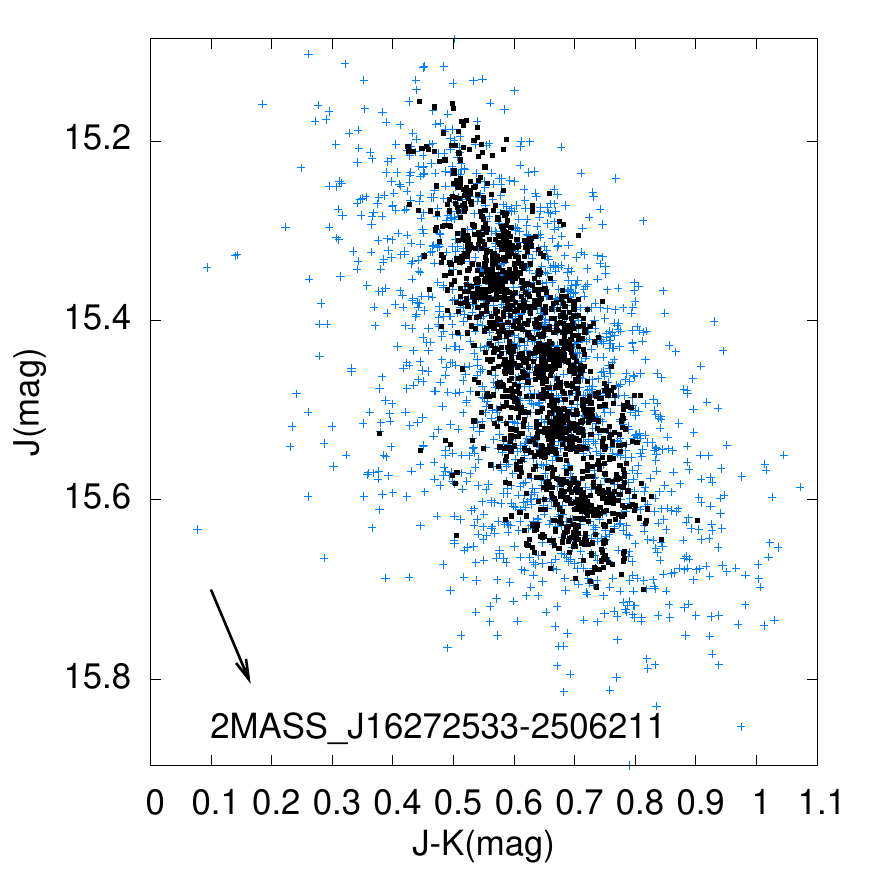} 
\includegraphics[width=2.0in, trim= 0 0 0 0 ]{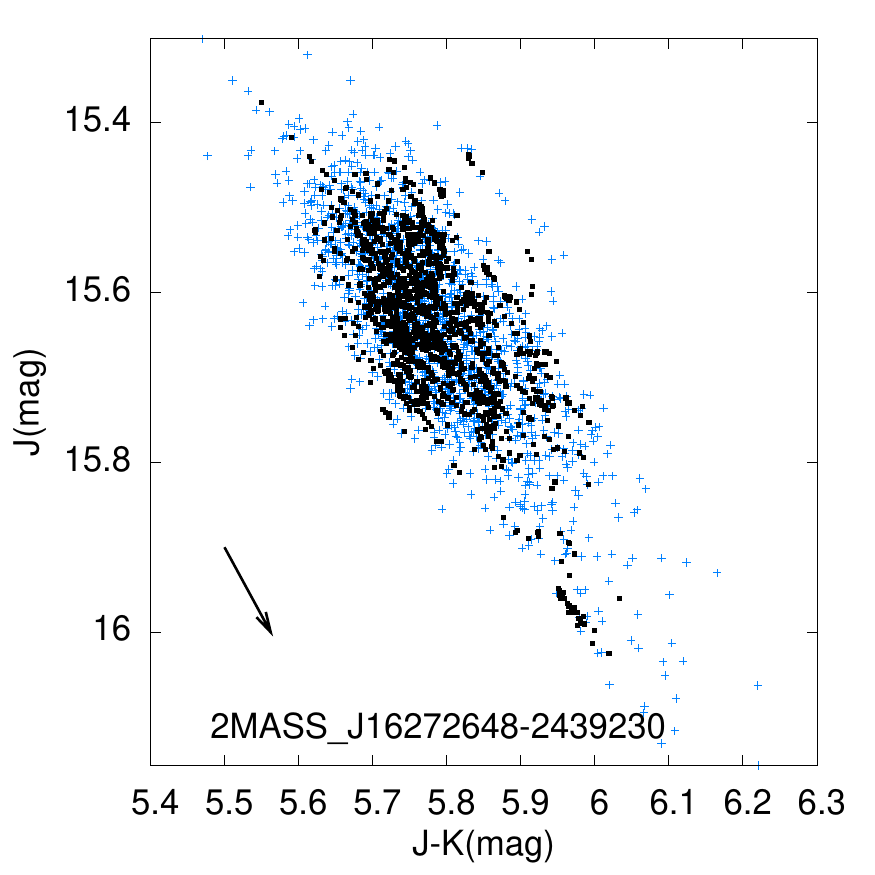} 
\includegraphics[width=2.0in, trim= 0 0 0 0 ]{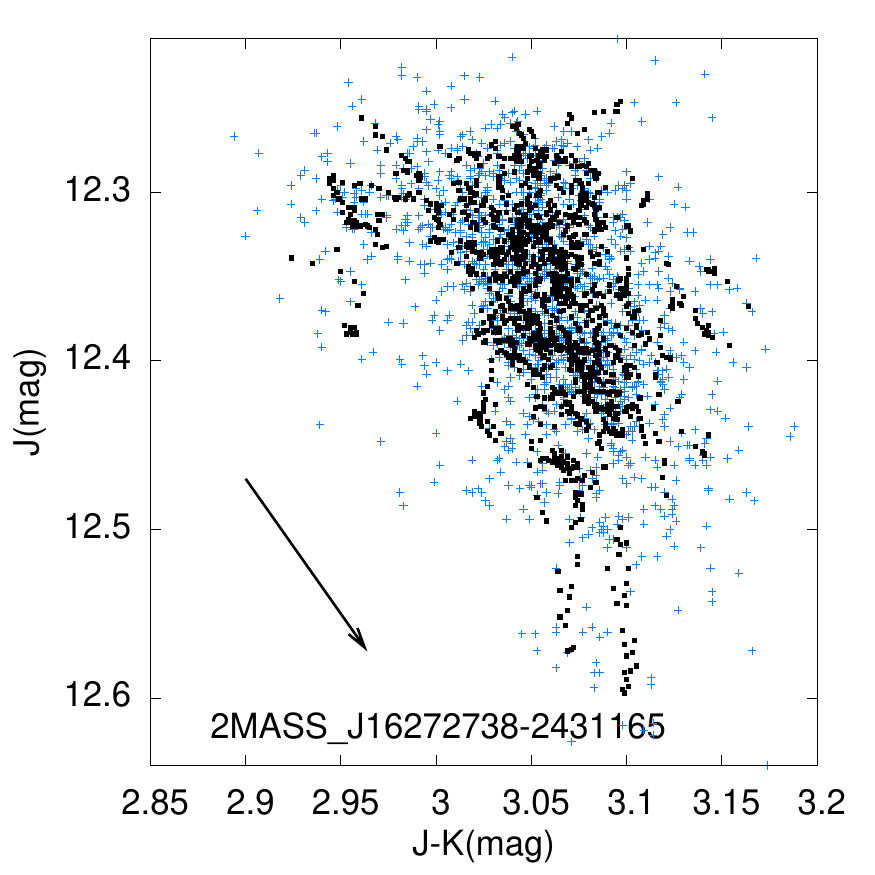} 
\includegraphics[width=2.0in, trim= 0 0 0 0 ]{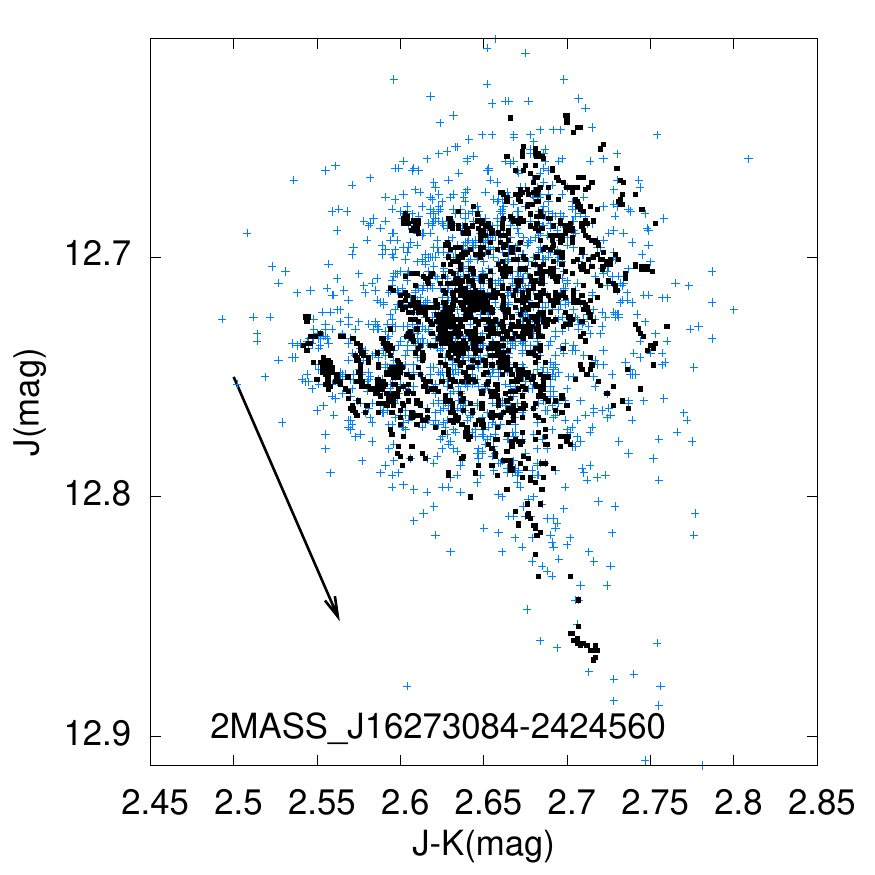} 
\caption{Color/magnitude plots (similar to those in Figure \ref{fig:aperiodic}) 
but of young stellar objects in the $\rho$ Ophiucus star
formation region found only in the color-blind dimming event search.  
}
\label{fig:YSOCB}
\end{figure*}

\begin{figure*}
\includegraphics[width=2.0in, trim= 0 0 0 0 ]{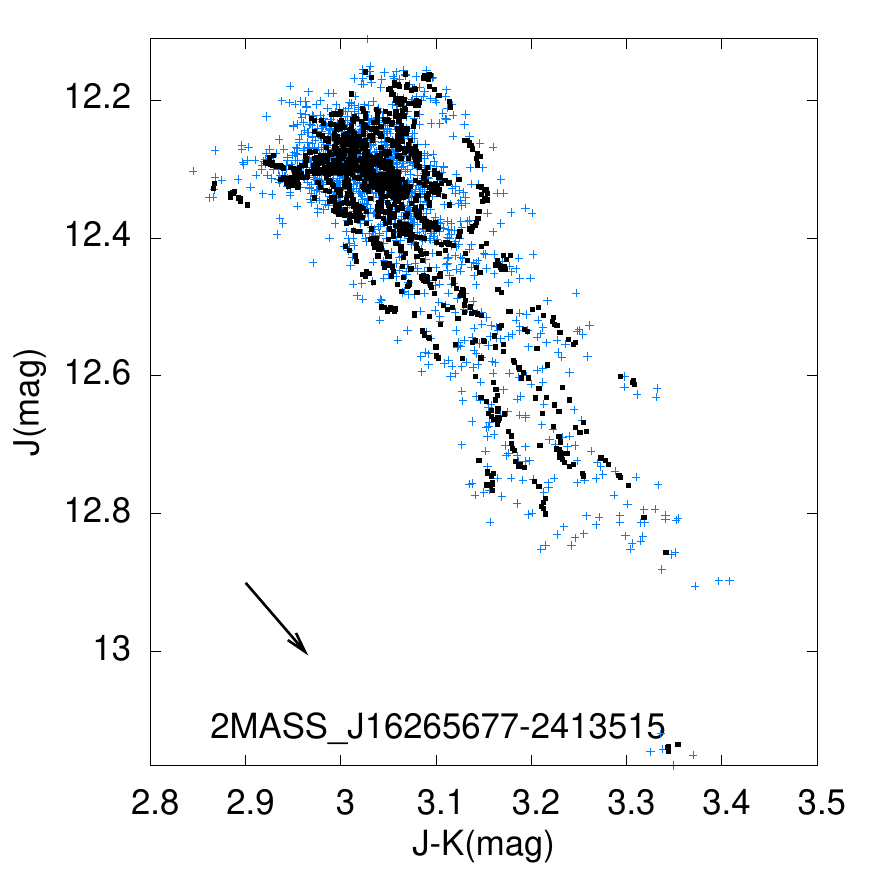}
\includegraphics[width=2.0in, trim= 0 0 0 0 ]{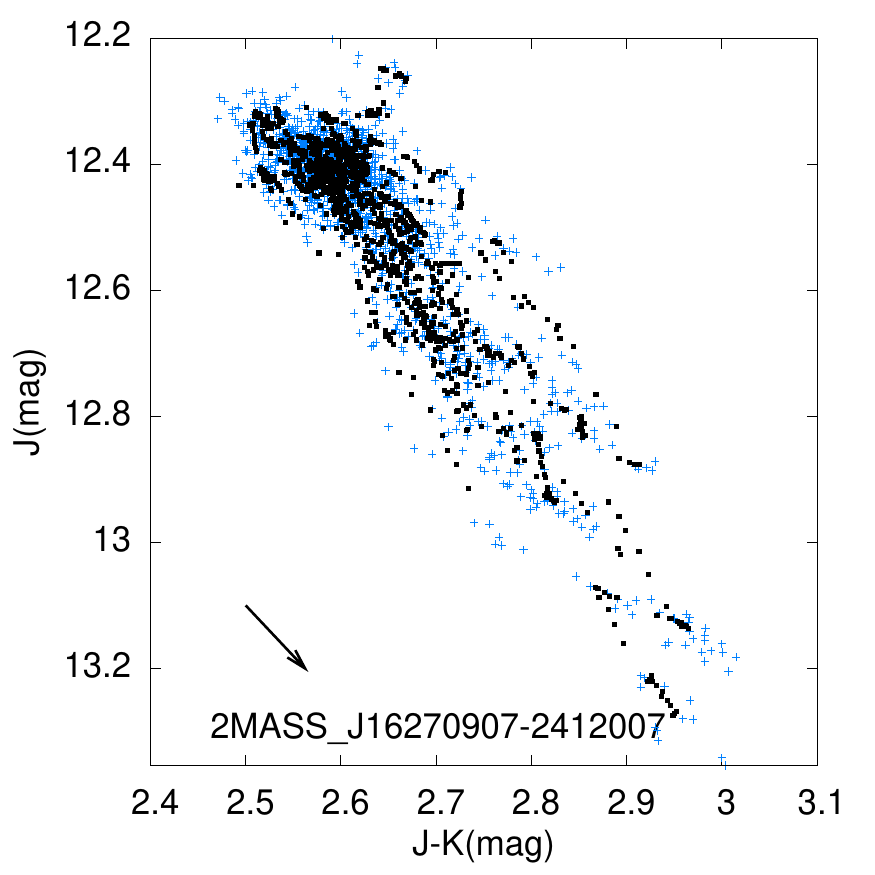}
\includegraphics[width=2.0in, trim= 0 0 0 0 ]{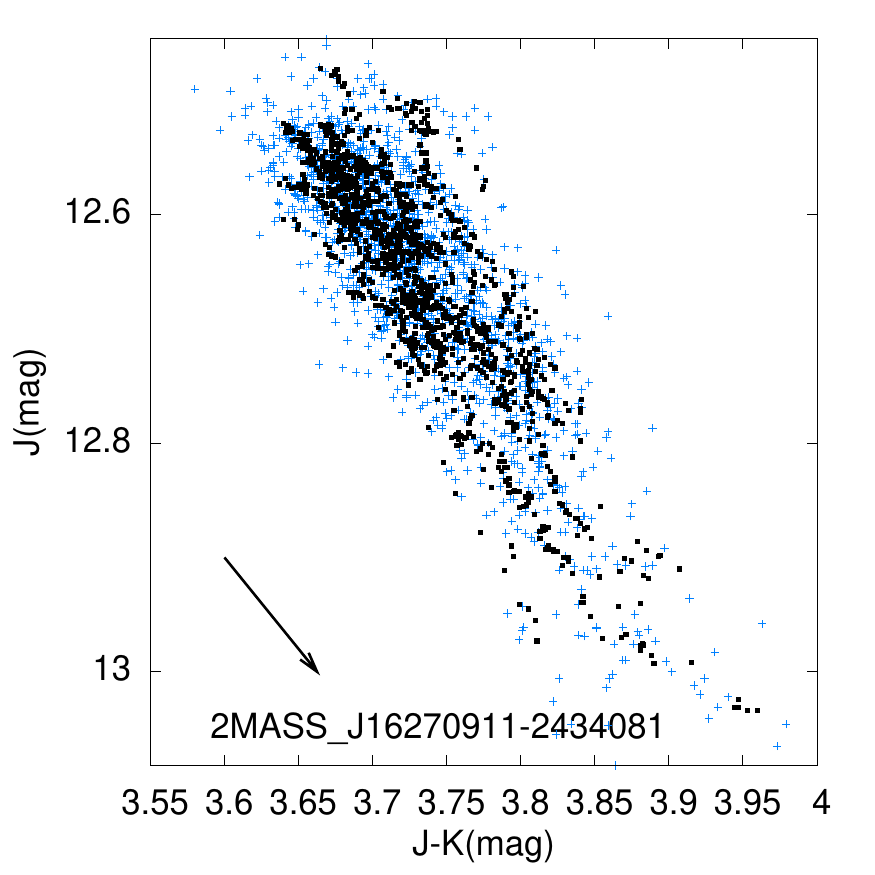}
\includegraphics[width=2.0in, trim= 0 0 0 0 ]{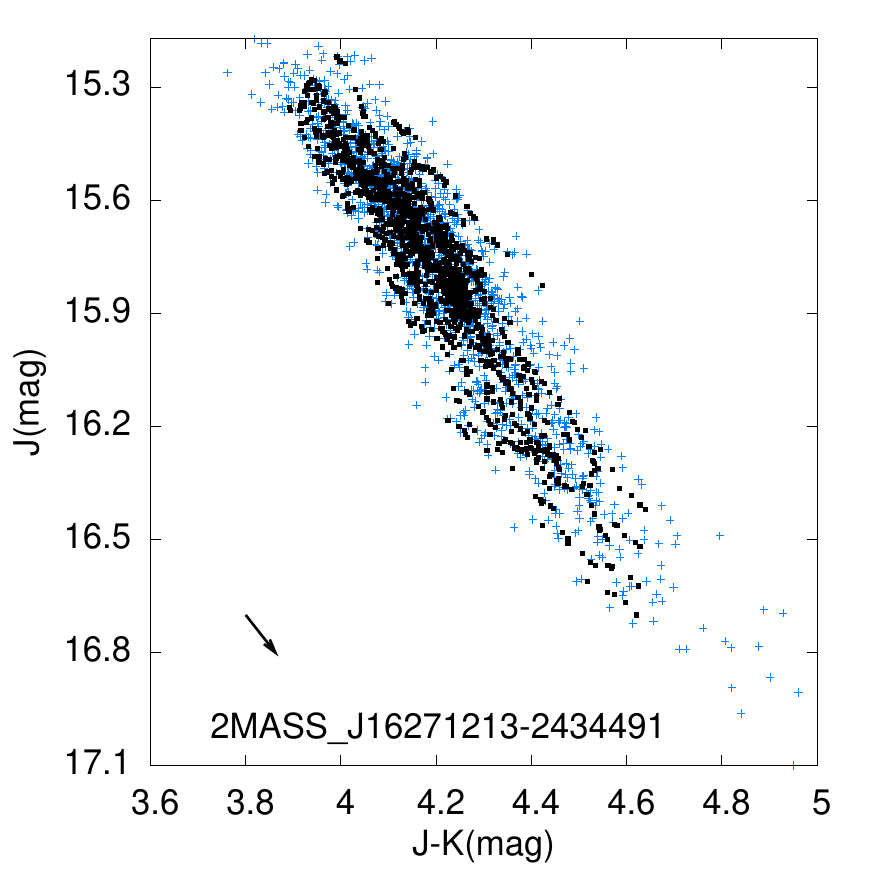}
\includegraphics[width=2.0in, trim= 0 0 0 0 ]{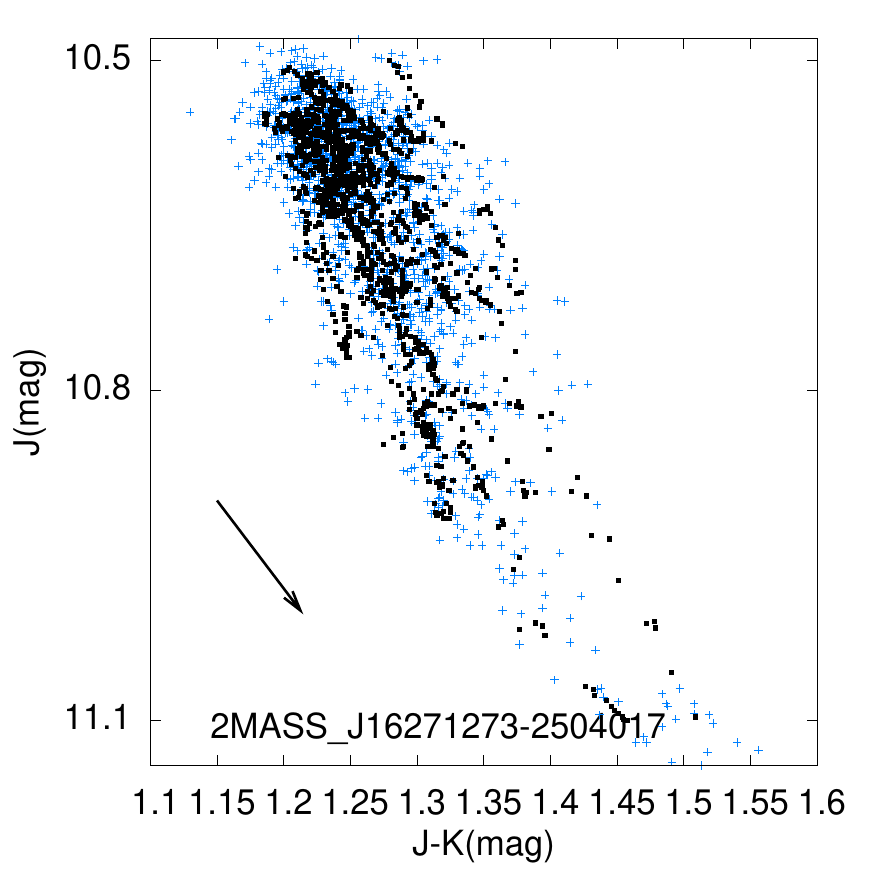}
\includegraphics[width=2.0in, trim= 0 0 0 0 ]{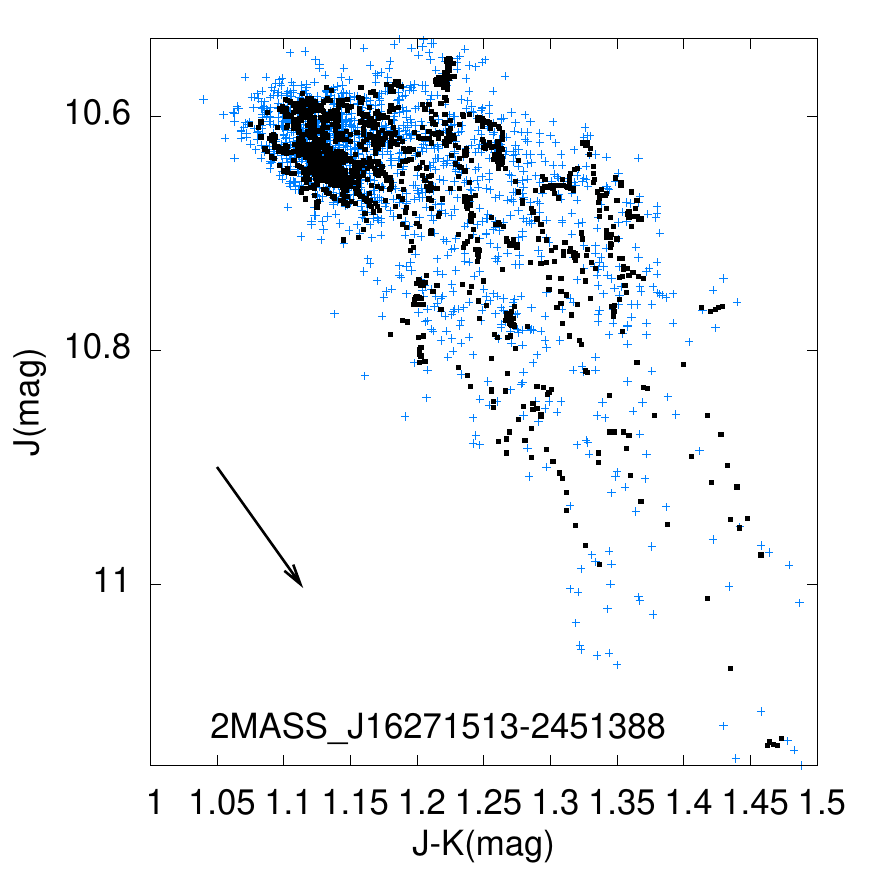}
\includegraphics[width=2.0in, trim= 0 0 0 0 ]{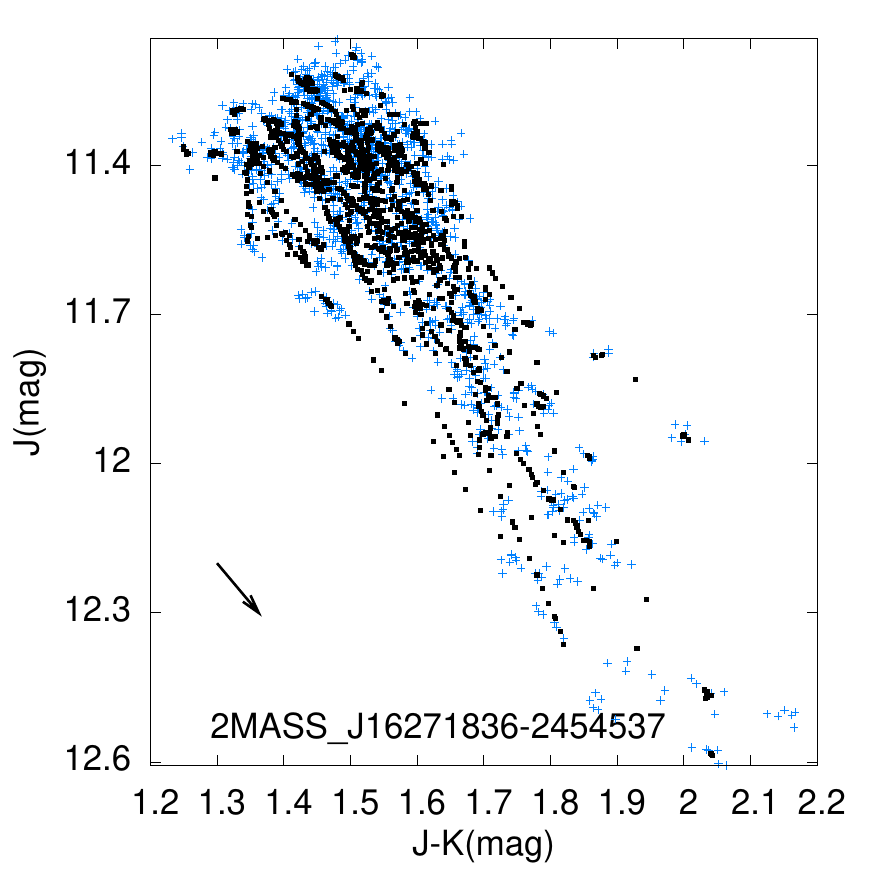}
\includegraphics[width=2.0in, trim= 0 0 0 0 ]{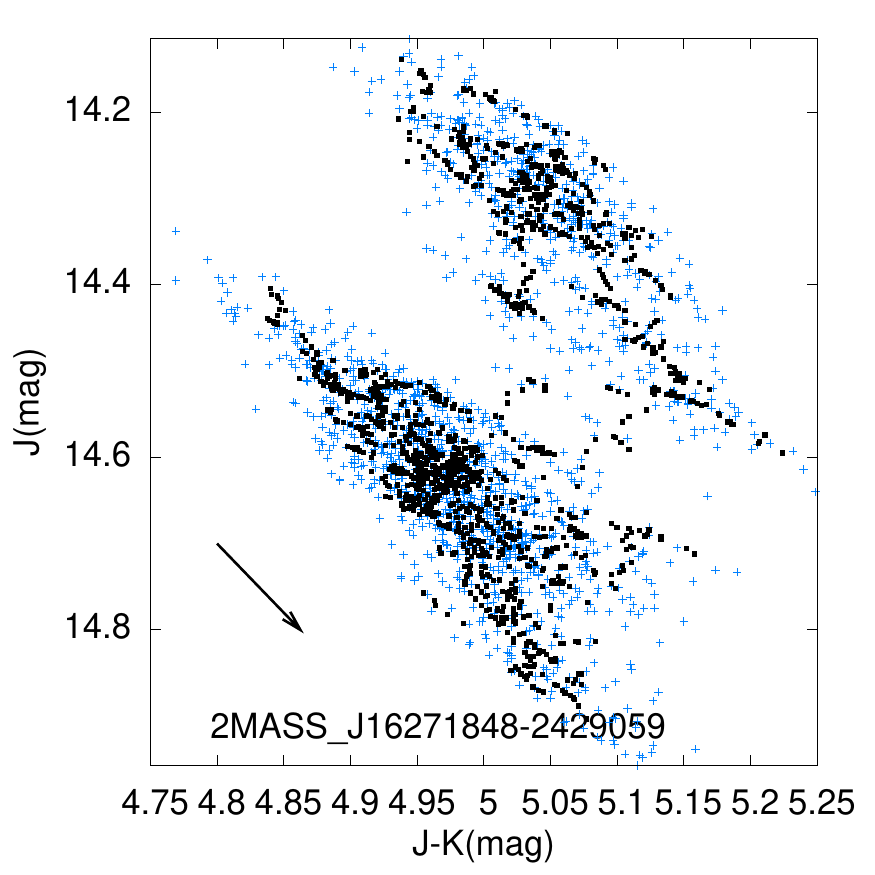}
\includegraphics[width=2.0in, trim= 0 0 0 0 ]{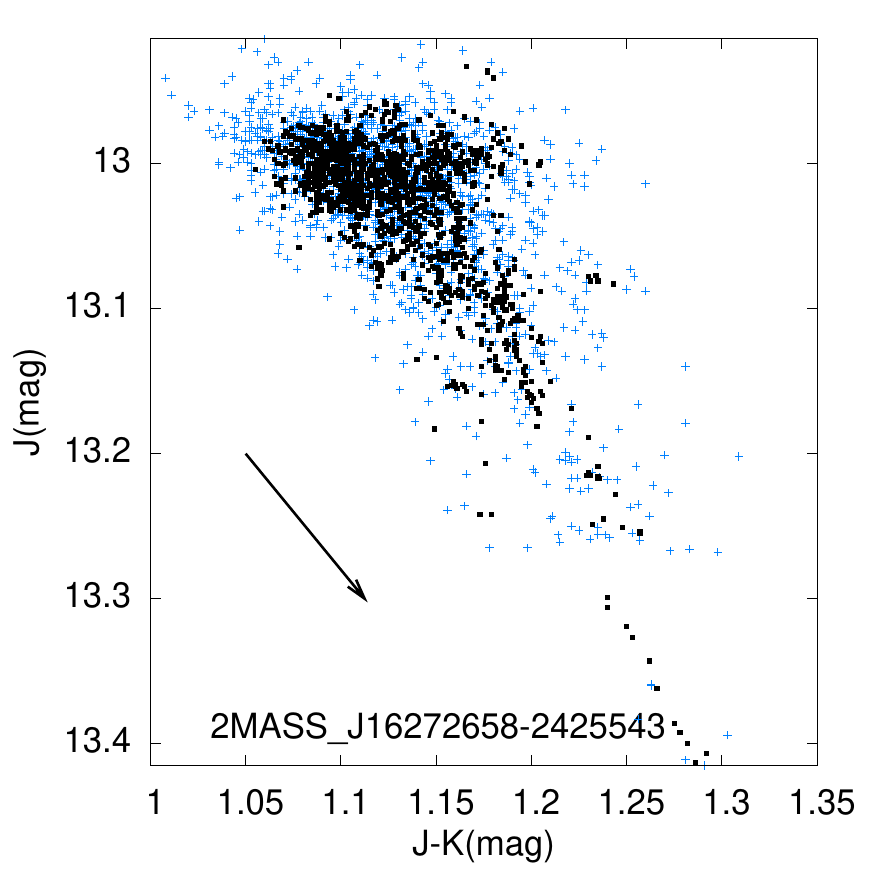}
\includegraphics[width=2.0in, trim= 0 0 0 0 ]{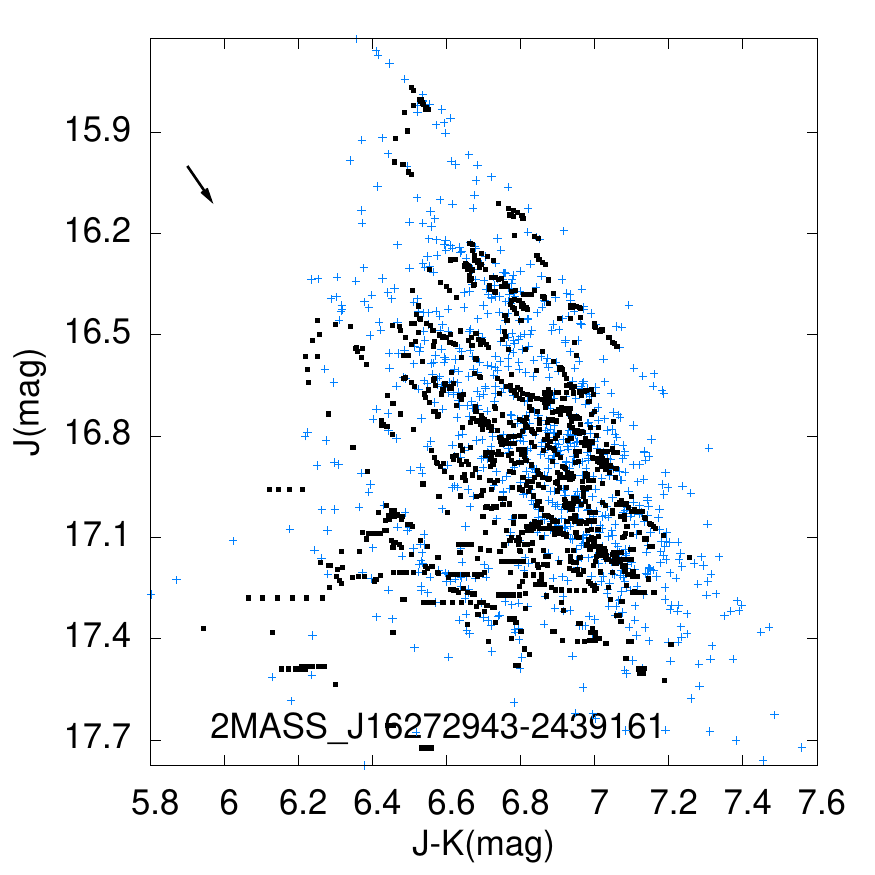}
\caption{Color/magnitude plots (similar to those in Figure \ref{fig:aperiodic}) 
but of young stellar objects in the $\rho$ Ophiucus star
formation region found in both dimming event searches.  These young stellar objects exhibit deep and red dimming events.
Note the bimodal distribution in the color/magnitude plot of  WL4 (2MASS J16271848-2429059).
}
\label{fig:YSO}
\end{figure*}

Our dimming searches revealed 21 objects in the $\rho$ Ophiucus molecular cloud region that were
previously studied by \citet{parks13}.
These young stellar objects are listed in Table \ref{tab:YSO} separated into two groups,
11 found only in the color insensitive search, followed by 10 found in both searches.
Their color magnitude distributions ares shown in Figures \ref{fig:YSOCB}
and \ref{fig:YSO}.
These young stellar objects are comparable to stars exhibiting brief, 
sharp drops in stellar flux,  denoted as AA Tau \citep{herbst94} or ÒdipperÓ variables, seen 
in the Orion Nebula and NGC 2264 stellar clusters \citep{moralescalderon11,cody14}
(also see light curves from the Cygnus OB7 association \citealt{wolk13}).

 Among the young stellar objects found in the red-dimming search,
all but 1 of the 10 young stellar objects found in the red-dimming search were found to be periodic by \citet{parks13}
and all but 1 of the 10  are class II objects, but since we used both $J$ and $K$ band for the search we necessarily
excluded extremely red young stellar objects.  
The exception is YLW 16b, classified as a long timescale variable and a class I young stellar object \citep{parks13}.
Here YSO classes were assigned based on those by \citet{bontemps01,gutermuth09}.
 Among the young stellar objects only found in the color insensitive dimming search, all are class II
objects except ISO-Oph 135 which is a class III object.
The objects found only in the color insensitive search included irregular and long timescale variables as well 
as periodic variables. 

\citet{parks13} classified the periodic behavior based on the color variations seen in the period-folded light curves.
They classified the periodic behavior as eclipse-like, inverse eclipse like, or sinusoidal with hot or cold star spots.
If dimming was associated with redder colors, then extinction was assumed, whereas if dimming
was not associated with redder colors, periodicity was associated with cool or hot star spots, depending upon 
the fraction of phase spent bright compared to dim.
The aperiodic variables were divided into long time scale variables (LTVs) and irregulars
and their brightness variations classified as due to accretion if they were colorless or  extinction if
dimming was associated with reddening.
These classifications are listed in our Table \ref{tab:YSO}.
We expected to only find eclipse-like or extinction-like classifications in our red dimming event selected sample, 
however some of them were classified as having 
inverse-eclipse-like period-folded light curves  and some with sinusoidal with colors variations expected from star spots.
 Our color magnitude diagrams show that {\it all} the young stellar objects predominantly vary along the
 reddening direction, even those only found in the color insensitive search.
 The objects with the largest variations in both color and brightness tend to be those that have the strongest
 correlations between reddening and brightness and these were found in the red dimming search.
 The color magnitude distributions are not bowed as were some of the color/magnitude distributions of the active galaxies.
Not all the young stellar objects displayed strong correlations in color vs magnitude,
for example Iso-Oph 151 (2MASS J16273084-2424560) has a broad color magnitude  distribution.
While features are seen in the distribution along the reddening direction the distribution is broad implying
that more than one emission process is affecting the brightness and color.

 One of the young stellar objects, 
2MASS J16271848-2429059,  WL4  has a bimodal color/magnitude plot. 
Its bimodal distribution is similar to that of RWA6 in the Cygnus OB7 association \citep{wolk13}.
On the color/magnitude plot in Figure \ref{fig:YSO} there are two groups
of points, both elongated along the extinction direction.  The brighter, but redder, group of points 
is centered at about $J$=14.3 mag and $J-K=5.05$ mag and fainter but bluer group of points is centered at
$J=14.6$ and $J-K=4.95$ mag.
The object was classified as inverse eclipse/circumbinary disk by \cite{parks13} and with a period
of 65.61 days, but the period identified as twice as long by \citet{plavchan08b} who remarked on 
the two-state behavior of the object.
\citet{plavchan08b,parks13} proposed the following model:
WL4 is a triple system comprised of three approximately equal mass 0.6 $L_\odot$ young stellar objects.
A close double system, denoted W4a, W4b has a 
130.87 or 65.61 day period and the two stellar components WL4a, WL4b, 
alternate being obscured by a circumbinary disk.  The circumbinary disk is tilted with respect to
the orbital plane of the binary due to the third star, W4c, that is resolved by direct imaging.
Star spots on the unobscured third object, W4c, account for shorter timescale variability.
We ask: is this model consistent with the distribution of points seen on the color/magnitude diagram?
The equal number of points in both regions in the color/magnitude diagram are consistent with the
fraction of the orbital period spent in each state being large.  The only known systems
with eclipses lasting a significant fraction of the orbital period are like KH 15D, with a misaligned circumbinary disk.
The only addition we would make to this model is that
the color magnitude distributions suggest that both components
experience variations in extinction.

Two sources with only moderate value of $J-K$ ($\sim 3$ mag), 2MASS J16273084-2424580 (Iso-Oph 151)
and 2MASS J16272738-2431165 (WL 13) display long (two day) colorless dimming events $\sim 0.2$ mag deep.
Iso-Oph 151 is a class III object, but WL 13 has a similar $J-K$ low color compared to the other more embedded
young stellar objects.
Of the young stellar objects found in the dimming search, these two light curves have relatively low values of dispersion  
in $K$ band (the dispersion in $J$ band in some cases is high because the objects are faint). 
An obscuring disk need not cause reddening during eclipse if it is opaque or if it is comprised of large dust grains and so has
a gray extinction law (e.g., EE-Cep \citealt{galan12}).  The long and relatively colorless dimming events 
seen in these object may be due to dense or gray material passing in front of the object.
 

\begin{table*}
\begin{minipage}{190mm}
 \caption[]{Young stellar objects in the $\rho$ Ophiucus region  \label{tab:YSO}}
 \begin{tabular}{l l  l l l l l l l l l l l l }
 \hline
 Object  & ID  & variability type & $J$ & $\sigma_J$ & $H$ & $\sigma_H$ & $K$ & $\sigma_K$& $R$ & $b$  \\
(1)        & (2)  & (3)                   & (4) &      (5)           & (6) & (7)                & (8)   & (9)               & (10) & (11)  \\
\hline
2MASS J16265839$-$2421299 & YLW 1C       &P/S/Hot SS & 16.213 &0.129 & 13.165& 0.068 & 11.489& 0.056 & 0.91 & 1.003$\pm$0.011\\
2MASS J16265863$-$2418346 & ISO-Oph 87 & Irr/flares    &15.519 &0.079 & 12.995 & 0.048 &11.548& 0.050  & 0.80 & 0.814$\pm$0.015\\
2MASS J16265904$-$2435568 & WL 14          & LTV/Ext  & 16.397 &0.175 & 13.463 &0.083&  11.867& 0.049 & 0.97 & 1.130$\pm$0.007\\
2MASS J16270410$-$2428299 & WL 1            & LTV          &16.539 &0.167 & 13.028& 0.061&  10.872 &0.064  & 0.94 & 0.900$\pm$0.008\\
2MASS J16270457$-$2427156 & ISO-Oph 98 & P/S/Cool SS &16.445& 0.181& 13.075& 0.108 &11.178 &0.080  & 0.91 & 1.085$\pm$0.012\\
2MASS J16271003$-$2429133 & -                   & Irr/unknown &16.638 &0.175 & 15.111& 0.116& 14.179& 0.097 & 0.85 & 0.878$\pm$0.014\\
2MASS J16272291$-$2417573 & ISO-Oph 135& P/S/Cool SS & 13.310 &0.046 & 10.692 &0.039&  9.390 &0.035& 0.65 & 0.953$\pm$0.028\\
2MASS J16272533$-$2506211 & -                   &P/S/unknown &15.453& 0.145 & 14.985& 0.119 &  14.819 & 0.143  & 0.48 & 0.470$\pm$0.022\\
2MASS J16272648$-$2439230 & YLW 16C     & P/S/cool SS & 15.628& 0.113 & 12.046 &0.078&  9.852& 0.075  & 0.78 & 0.914$\pm$0.019\\
2MASS J16272738$-$2431165 & WL 13          & P/S/cool SS & 12.365 &0.066&  10.368 &0.066& 9.308 &0.061  & 0.45 & 0.668$\pm$0.034 \\
2MASS J16273084$-$2424560 & ISO-Oph 151& LTV/Acc   &12.729 & 0.040& 10.998& 0.044&  10.079& 0.063  & 0.02 & 0.014$\pm$0.021 \\
\hline
2MASS J16265677$-$2413515 & ISO-Oph 83  & P/S/Ext & 12.360 &0.148 & 10.385 &0.122 & 9.302 &0.100 & 0.77 & 1.395$\pm$0.030\\
2MASS J16270907$-$2412007 & ISO-Oph 106 & P/Ecl/Ext  & 12.501&0.191 & 10.755 &0.150 & 9.862 &0.113 & 0.90 & 1.783$\pm$0.023\\
2MASS J16270911$-$2434081 & WL10            &  P/S/cool SS & 12.655& 0.106 & 10.266 &0.087 &  8.923& 0.068 &0.82 & 1.514$\pm$0.027\\
2MASS J16271213$-$2434491 & WL11            &  P/S/hot SS? & 15.723 &0.292 & 13.180 &0.237 & 11.573 & 0.172&0.93 & 1.732$\pm$0.018\\
2MASS J16271273$-$2504017 & -                    & P/S/hot starspot? & 10.643 &0.115 & 9.763 &0.097 & 9.376& 0.082 & 0.77 & 1.636$\pm$0.035\\
2MASS J16271513$-$2451388 & -                   & P/Ecl/Ext  & 10.659 &0.085 & 9.809 &0.070 &  9.469 & 0.070  & 0.65 & 0.669$\pm$0.020\\
2MASS J16271836$-$2454537 & -                    & P/S/hot starspot? & 11.510 &0.267 & 10.583 &0.215 &9.966& 0.172 & 0.82 &1.526$\pm$0.027\\
2MASS J16271848$-$2429059 & WL4             & P/IEcl/CB & 14.526 &0.181 &11.341 &0.193 & 9.528 & 0.193  &0.03 & 0.078$\pm$0.061\\
2MASS J16272658$-$2425543 & -                  & P/Ecl/Ext  &13.028 &0.063 & 12.368& 0.053 & 11.897& 0.049  &0.63 & 0.845$\pm$0.026 \\
2MASS J16272943$-$2439161 & YLW 16b     & LTV/Ext & 16.721 &0.395 & 13.069 &0.536& 10.249& 0.471& 0.49 & 0.818$\pm$0.049\\
\hline
\end{tabular}
\\
The top 11 YSOs were found only in the sigma-limited dimming search and the bottom
10  YSOs found in both searches.
Columns:
(1): The 2MASS identifier of the star.
(2): The identifier used by \citet{parks13} in their variability study.
(3): The type of variability found by \citet{parks13} in their Table 5 for periodic variables, 
  their Table 7 for the aperiodic long timescale variables or in their Table 8 for irregular variables.
Stars are either Periodic (P),  long-timescale variables (LTV) or irregular (Irr).  
If periodic, then variability is either sinusoidal (S),  eclipse-like (Ecl), or inverse eclipse like (IEcl).
Color variations are either similar to hot starspots (hot SS)
or cool starspots (cool SS) or due to extinction  (Ext).  If an LTV then color variations
are either colorless (and so associated with accretion; Acc) or red and so due to extinction (Ext).  
CB stands for circumbinary disk.
(4)-(9):  Mean and standard deviations, in magnitudes, computed from the light curves in $J, H,$ and $K$ bands, respectively.
(10): The cross correlation coefficient, $R$, (equation \ref{eqn:R}) between color and magnitude.
(11): The slope, $b$, of the best fitting line on the color/magnitude plot (equation \ref{eqn:b}).
\end{minipage}
\end{table*}

\section{Synthesis}

In this section we first discuss objects that were found by both searches and then we discuss how 
different types of objects might be found or eliminated in automated searches.
We then discuss correlation coefficients measured from the color/magnitude distributions
for the different classes of objects found in our searches.

\subsection{Objects found in both searches}

Because our color insensitive dimming search (described in section \ref{S:CB})
used the standard deviation of the light curve, variable stars with large standard
deviations were often discarded. 
However, our red dimming event search described below was sensitive to variable stars with large
standard deviations.   

\begin{figure*}
\includegraphics[width=3.0in, trim= 0 0 0 0 ]{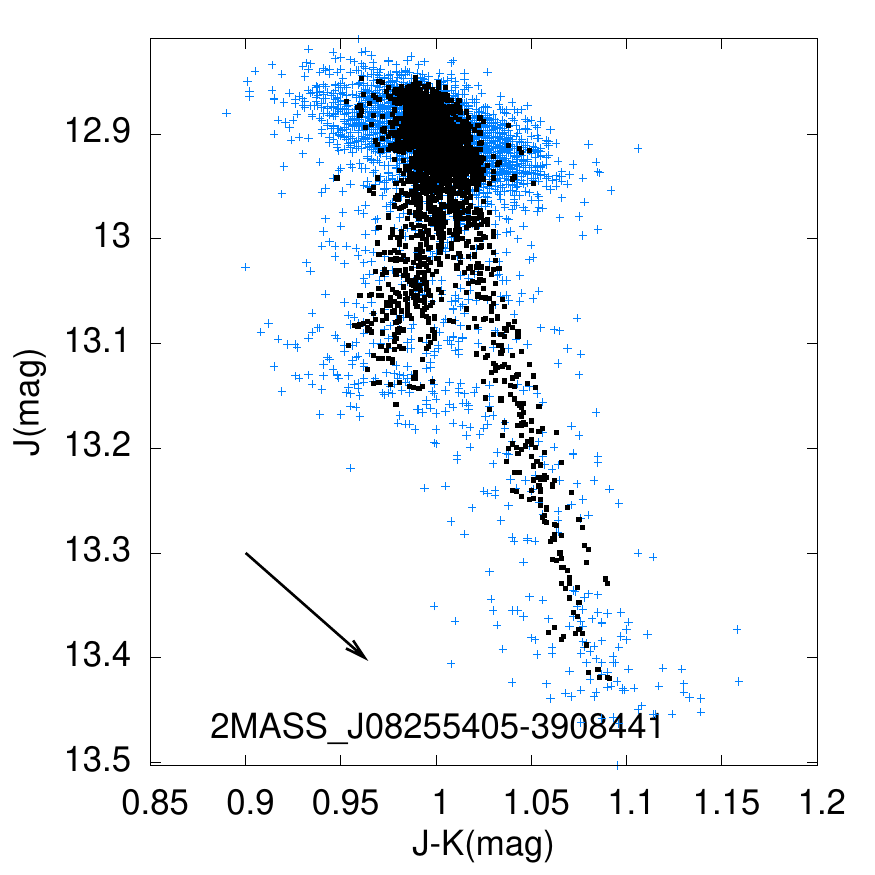}
\caption{Color/magnitude plot for 2MASS J08255405-3908441, the eclipsing binary with color variations in eclipse.
This object was found in both red-dimming and sigma-limited colorblind dimming search.
\label{fig:redEB}}
\end{figure*}

Which objects were found in both searches?  
Young stellar objects in the $\rho$-Ophiucus region that exhibited large and red brightness
variations were present in both searches.
The object 2MASS J18394250+4856177 and discussed in section \ref{S:CB}, was found in both searches.
This object exhibits rapid and correlated in color/magnitude variations, similar to the active galaxies,
but it was not identified as extragalactic or as having an infrared excess.
The previously known eclipsing binary
2MASS J08255405-3908441 was also found in both searches.
The eclipsing binary,  2MASS J08255405-3908441, has a period of 8.08986 days and was previously identified by \citet{plavchan08}.  
They noted the color variations ($\Delta (J-H) \sim 0.1$ mag) in primary eclipses (see their Figure 49) and attributed
 them to different effective temperatures in the primary and secondary stars.
 The color magnitude plot for 2MASS J08255405-3908441 is shown in Figure \ref{fig:redEB}
 and displays deviations in color in two directions, one for primary eclipse and the other for the secondary
 eclipse.  

 As only one eclipsing binary, was found in the search for red dimming events, the
color variability selection we used in the red dimming search can effectively eliminate most eclipsing binaries.
The sigma-limit used in the color-blind search eliminates many, but not all active galaxies and young stellar objects.
The red dimming event search did not find a population of objects, with a single or a few rare
 dimming events, rather the objects found tend to exhibit continuous variability. 
 By combining both searches with an absolute limit on the dispersion in the light curve we could eliminate
 all the active galaxies and young stellar objects.
 Such a search on a larger sample of stars might reveal an object such as 
 OGLE-LMC-ECL-11893, with a rare red eclipse.
 
\subsection{Cross correlation coefficients}

\begin{figure*}
\includegraphics[width=3.45in, trim= 0 0 0 0]{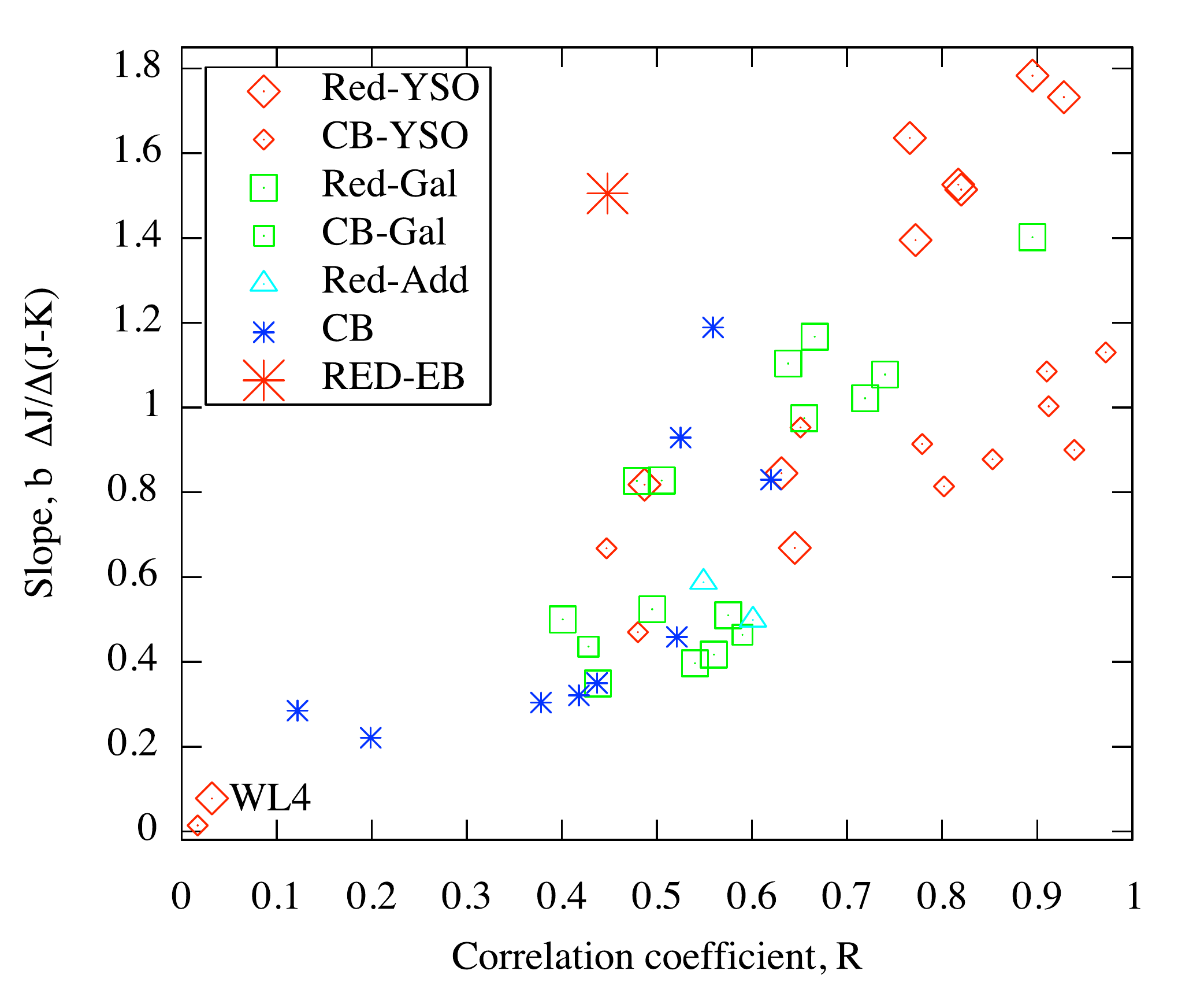} 
\includegraphics[width=3.45in, trim= 0 0 0 0]{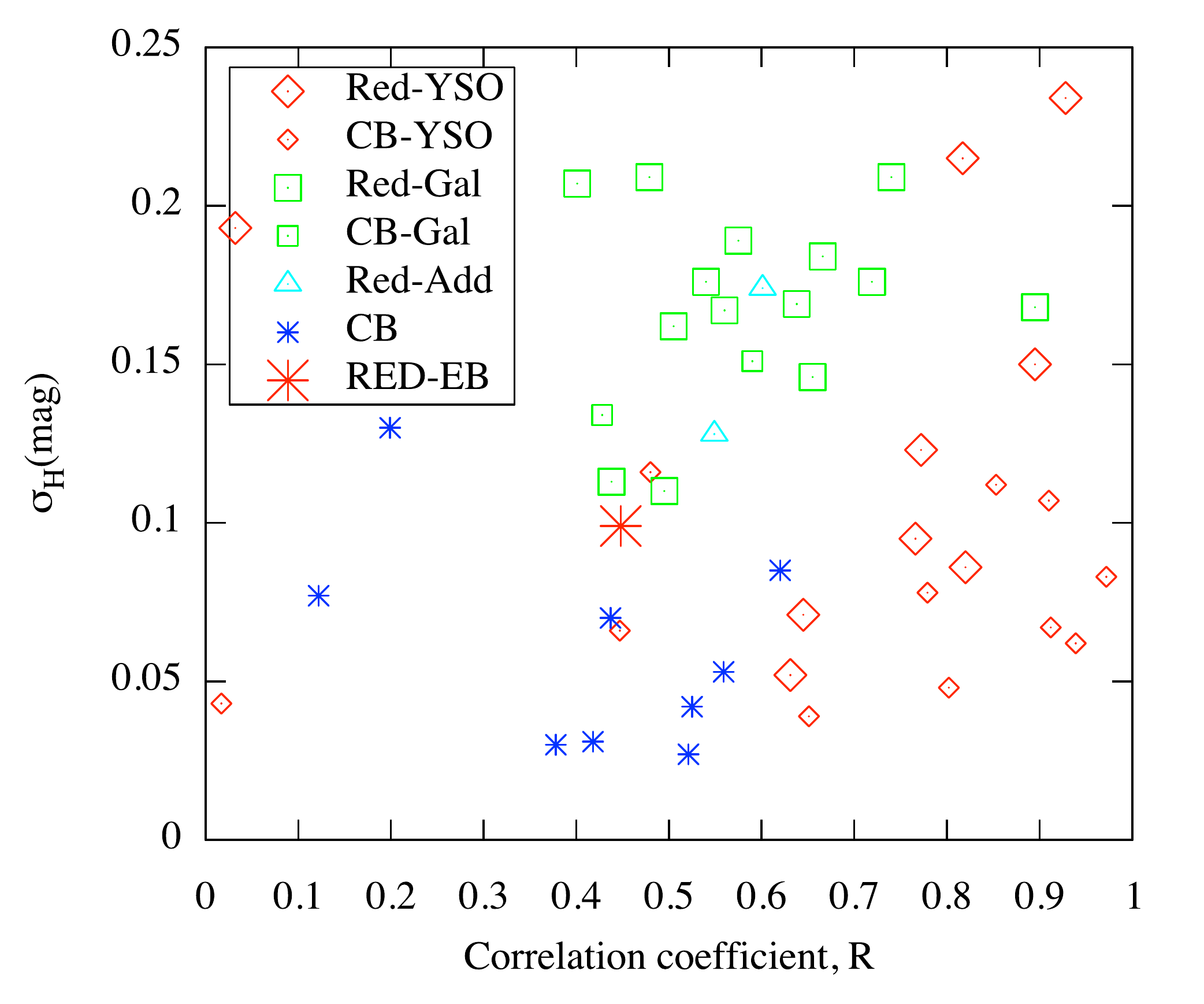} 
\caption{a) Slope on color/magnitude plot, $b$ vs cross correlation coefficient, $R$ for
objects
listed in Tables \ref{tab:aperiodic}, \ref{tab:YSO}, \ref{tab:AGN},  and \ref{tab:additional}).
For objects identified in the red-dimming search, 
large red diamonds (denoted Red-YSO in the key)  are the young stellar objects, 
large green squares are the extragalactic objects (Red-GAL in the key), 
turquoise triangles are additional objects not identified them as extragalactic (Red-Add in the key) 
For objects identified in the color insensitive, sigma-limited dimming search
small red diamond are the
young stellar objects only found in the color insensitive search (CB-YSO in the key),  
extragalactic objects are small green squares, (CB-Gal in  the key), 
blue stars are aperiodic objects (CB)
and the large red star is the single eclipsing binary,  2MASS J08255405-3908441,  (denoted RED-EB in the key) found in both searches.
A slope of 1.6 is consistent with variability caused by changes in extinction.
The young stellar objects have the largest slopes and correlation coefficients. 
The galaxies are intermediate and the objects found in the color independent search
have low correlation coefficients.
b) The same objects but plotting the standard deviation $\omega_H$ in $H$ band of the light
curve versus the cross correlation coefficient $R$.
\label{fig:corr}}
\end{figure*}

In Figure \ref{fig:corr}a we show the slope $b$ fit to the color/magnitude diagrams
versus correlation coefficients measured from the color/magnitude distributions.
Points plotted are listed in Tables \ref{tab:aperiodic}, \ref{tab:YSO}, \ref{tab:AGN},  and \ref{tab:additional}).
For objects identified in the red-dimming search, 
large red diamonds (denoted Red-YSO in the key)  are the young stellar objects, 
large green squares are the extragalactic objects (Red-GAL in the key), 
turquoise triangles are additional objects not identified as extragalactic (Red-Add in the key).
For objects identified in the color insensitive, sigma-limited dimming search
small red diamond are the
young stellar objects only found in the color insensitive search (CB-YSO in the key),  
extragalactic objects are small green squares, (CB-Gal in  the key), 
blue stars are aperiodic objects (CB)
and the large red star is the single eclipsing binary,  2MASS J08255405-3908441,  (denoted RED-EB in the key) found in both searches.
Using the same symbols and objects we plot the dispersion in the $H$ band light curve, $\sigma_H$, versus the 
cross correlation coefficient $R$ in Figure \ref{fig:corr}b.  We have used the $H$-band dispersion as some
of the young stellar objects were faint in $J$ band.

Points on the right of Figure \ref{fig:corr}a,b have high correlation coefficients and so when they dim, they become redder.
Points on the left have uncorrelated variations in color and brightness.
Points on the top of \ref{fig:corr}a have large changes in brightness for a small change in color.
A slope of 1.6 is consistent with variability caused by a variation in extinction.
The young stellar objects have the largest slopes and correlation coefficients. 
The galaxies are intermediate in both correlation coefficient and slope.
The intermediate value of slope is perhaps expected as the color variations are expected to be due to a contribution from 
a blue active source.  
Objects exhibiting low correlation coefficient and slope are most likely eclipsing binaries but we have failed to find their periods,
however, this group could contain more exotic objects.
The odd eclipsing binary with red eclipses stands out as having a high slope but a low correlation coefficient.

Two young stellar objects have a low correlation coefficients,  WL4 (2MASS J16271848-2429059)
that has a bimodal color/magnitude plot (see that in Figure \ref{fig:YSO}) and
Iso-Oph 151 (2MASS J16273084-2424560) displaying a significant, but rare dip in brightness.

The AGNs and young stellar objects tend to lie on different regions in Figure \ref{fig:corr}a, however,  
they are not completely separated.
However, the AGNs tend to have higher dispersions in their light curve, as seen from Figure \ref{fig:corr}b.
It is likely that most of the AGNs could be eliminated by placing a limit on the magnitude dispersion $\sigma_H$.
 
\section{Summary and Conclusion}

We have presented a search for dimming events in approximately 40,000 objects (mostly stars)  observed
over a period of about 4 years in the 2MASS Calibration fields.
We searched for day long dimming events that are significant compared to the standard deviation of the object's magnitude
distribution.
Our goal was to search for rare long dimming by an exotic object such as the occulting disk of OGLE-LMC-ECL-11893, 
consequently we
 did not restrict our search to objects with periodic variability.
In a color independent search for day-long dimming events we found 46 previously unidentified eclipsing binaries 
and 6 additional periodic variable stars.   We also found a few objects exhibiting dimming events but without strong color variations
during dimming.   These could be eclipsing binaries but we have not observed enough eclipses to identify their period, 
or they could be something more exotic.
We also found two active galaxies and a few objects displaying correlated color and magnitude variations suggesting
that they are also extragalactic. 
Young stellar objects in the $\rho$ Ophiucus field, previously studied by \citet{parks13} also displayed dimming events,
however no ordinarily non-variable or quiescent star in this region was detected with a dimming event.
Two less obscured objects (Iso-Oph 151, a class III source and WL 13 a class II source) variable young stellar objects in this field 
displayed 0.2 mag deep, colorless two-day long dimming events. 
We failed to find any ordinarily quiescent object with a dimming event lasting more than a day.
From the 4 year survey of approximately 40,000 objects (and excluding the young stellar objects) 
approximately 1 of 700 is an eclipsing binary, and approximately
1 of 4000 exhibits a dimming event but requires additional observations to classify.

We carried out a second search, looking for dimming events that are accompanied by reddening.
In this search we did not restrict the search by the standard deviation of the object's magnitude distribution.
This search primarily revealed two classes of objects, active galaxies that become bluer as they increase in brightness,
and class II young stellar objects in the $\rho$ Ophiucus region that exhibit large variations in brightness 
(up to a magnitude in $J$ band) that are also red ($\Delta (J-K) \sim 0.4$ mag).
The young stellar objects found in both searches display color variations consistent with interstellar extinction, though
in some cases there must be additional emission mechanisms influencing the brightness and color.
Of the 40000 objects searched we found 14 active galaxies,  corresponding to approximately one of 3000 searched, 
exhibiting dimming events and
correlations between brightness and color.

We did not find any dimming events greater than 0.2 magnitude deep and longer 
than a day outside the $\rho$ Ophiucus region.    We did not find an object similar to  
OGLE-LMC-ECL-11893, ordinarily without variability, but exhibiting long, deep and red dimming events, 
 in the 2MASS calibration fields (and including objects in the $\rho$ Ophiucus field).
 
 Because we failed to find any ordinarily quiescent object with a long (greater than a day) 
 dimming event, we consider
 how we would eliminate more common objects in a future larger scale search. 
The red dimming event search effectively eliminates all but one eclipsing binary comprised of two
stars with significantly different effective temperatures.
The only targets that were found in both searches were the young stellar objects 
in the $\rho$ Ophiucus star forming region, a single exceptional eclipsing binary and a rapidly
varying object that might be an active galaxy.

For objects found in our searches we computed correlation coefficients between color and magnitude
and fit a slope to points in the color magnitude distribution.  We find that young stellar objects, active galaxies
and objects displaying insignificant variations in color (and so are most likely eclipsing binaries) tend to be
found in different regions of this plot, however these two quantities are not sufficient to definitely separate the populations.
Most active galaxies can be eliminated with a limit on the dispersion in the magnitude distribution.
Perhaps combining additional color information or by using higher moments of the magnitude distributions,
or additional spectral information  the two populations could be better delineated and identified. 

There is a long history of searching for and following up objects that temporarily brighten, such 
as novae or gravitational micro-lenses,  but there have been few systematic searches for dimming events. 
 In this study we took care to discard multiple dimming events that occurred at the same time in the same calibration field.
However, differential photometric methods, where the brightness of nearby stars are contrasted in the data reduction pipeline,
 can be used to minimize the number of spurious dimming events (e.g., see discussion by \citealt{vaneyken11}).
 This would allow a more robust search for rare and single
 dimming events in a larger sample of stars and make it possible to develop strategies to follow up unidentified objects.

  New and ongoing surveys are increasing the number of stars photometrically monitored, improving upon 
the cadence of observations and on the numbers of photometric bands for photometric monitoring.    
An object similar to  OGLE-LMC-ECL-11893 could be found in larger and  multi-color survey that searched
for deep and red dimming events, a high cadence single band survey
 that examined the shape of the light curve during a single dimming event
or computed the magnitude distribution in an eclipse \citep{meng14}, or
an extended (many year) lower cadence single band survey
survey that examined the shape of the eclipse in the period-folded light curve  (\citealt{dong14}; and how it was found).
Rare events, such a transit of a planet hosting icy rings, like Saturn, or a disk with large dust grains (and a grayer extinction law) 
might not cause a red dimming event
but could induce an atypically shaped eclipse.  Differentiating this from an eclipsing binary would require high 
cadence and high photometric precision (needed to detect the shape of the eclipse).   

We note that our search criteria were not standard compared to those used in other surveys, so our search
for variable stars in the 2MASS calibration fields is not complete.  For example, we did not 
follow up bright stars identified with large dispersions computed from the magnitude distribution in the light curve.

We searched here for rare dimming events using a multicolor near-infrared survey. 
An optical survey could survey a larger number of fainter stars, however a near-infrared survey
would be particularly useful for studying star forming regions embedded in 
the Galactic plane.   Eclipsing disks associated with planet formation might better be found in 
a near-infrared survey that includes many young stars, whereas a search for ringed planets might be better 
done in the visible bands.

In our search of approximately 40000 light curves approximately 1 of 4000 objects (11 objects total) that have displayed
dimming events remain unidentified.   As more common objects such as eclipsing binaries and active
galaxies are classified in a continuously monitored field, it could become possible to 
discover more exotic objects displaying rare dimming events. 
Roughly scaling from our estimates, a search for dimming events in a billion good light curves 
would yield of order a million eclipsing binaries,
a few hundred thousand active galaxies and a few hundred thousand unidentified objects (that would include young stellar objects).   
Amongst these could be a bright counterpart to OGLE-LMC-ECL-11893 or an analog of Saturn's rings.

\subsection*{Acknowledgements}
We thank Judy Pipher for helpful discussions.
This work was in part supported by NASA grant NNX13AI27G.


\begin{thebibliography}{}

\bibitem[Bontemps et al.(2001)]{bontemps01}
Bontemps, S., et al. 2001, A\&A, 372, 173

   
 \bibitem[Bouvier et al.(2007)]{bouvier07} 	
Bouvier, J., Alencar, S. H. P., Boutelier, T., Dougados, C., Balog, Z., Grankin, K., Hodgkin, S. T., Ibrahimov, M. A., Kun, M., Magakian, T. Yu., Pinte, C.	 2007, A\&A, 463, 1017
	
\bibitem[Chadima et al.(2011)]{chadima11}
Chadima, P. et al.\ 2011, A\&A, 530A, 146

\bibitem[Ciocca(2013)]{ciocca13}
Ciocca, M.  2013, Journal of American Variable Star Observers, 41, 267	

\bibitem[Collier Cameron et al.(2006)]{colliercameron06}
Collier Cameron, A. et al. 2006, MNRAS, 373, 799

\bibitem[Cody et al.(2014)]{cody14}
Cody, A. M. et al. 2014, ApJ in press, \url{http://arxiv.org/abs/1401.6582}

\bibitem[Cutri et al.(2003)]{cutri03}
     Cutri R.M., et al.
     2003,
    University of Massachusetts and Infrared Processing and Analysis Center
     (IPAC/California Institute of Technology),
   VizieR Online Data Catalog:   2MASS All-Sky Catalog of Point Sources     (Cutri+ 2003)
   
\bibitem[Cutri et al.(2006)]{cutri06}
Cutri, R., et al., 2006, Explanatory Supplement to the All Sky Data Release and Extended Mission 
\url{http://www.ipac.caltech.edu/2mass/releases/allsky/doc/explsup.html}

\bibitem[Cutri et al.(2012)]{wise} 
Cutri, R. M. et al. 2012,	
	WISE All-Sky Data Release, 
VizieR On-line Data Catalog: II/311. Originally published in: 2012yCat.2311....0C

\bibitem[di Clemente et al.(1996)]{diclemente96}
di Clemente, A., Giallongo, E., Natali, G., Trevese, D., \& Vagnetti, F. 1996, ApJ, 463, 466 

\bibitem[Dimitrov \& Kjurkchieva(2010)]{dimitrov10}
Dimitrov, D. P., \&  Kjurkchieva, D. P. 2010, MNRAS, 406, 2559	

\bibitem[Dong et al.(2014)]{dong14}
Dong, S.,  Katz, B., Prieto, J. L., Udalski, A., Kozlowski, S.  Street, R. A.  2014, \url{http://arxiv.org/abs/1401.1195}


\bibitem[Egan et al.(2003)]{egan03}
     Egan M.P., Price S.D., Kraemer K.E., Mizuno D.R., Carey S.J., Wright C.O.,
     Engelke C.W., Cohen M., Gugliotti G. M. 2003,
     The Midcourse Space Experiment Point Source Catalog Version 2.3, 
   Air Force Research Laboratory Technical Report AFRL-VS-TR-2003-1589 (2003)


\bibitem[Galan et al.(2012)]{galan12}
Galan, C., Miko?ajewski, M., Tomov, T.,  et al. 2012, A\&A, 544, 53   


\bibitem[Graczyk et al.(2003)]{graczyk03}
Graczyk, D.,  Mikolajewski, M., Tomov, T., Kolev, D., \&  Iliev, I.\ 2003, A\&A, 403, 1089

\bibitem[Graczyk et al.(2011)]{graczyk11}
Graczyk, D., Soszy\`nski, I.,  Poleski, R., Pietrzy\`nski, G., Udalski, A. Szyma\`nski, M. K., Kubiak, M.,
Wyrzykowski, L., \& Ulaczyk, K. 2011,
Acta Astron., 61, 103

 \bibitem[Graczyk \& Eyer(2010)]{graczyk10}
Graczyk, D., \& Eyer, L. 2010, Acta Astron., 60, 109

\bibitem[Grinin et al(2008)]{grinin08} 
Grinin, V., Stempels, H. C., Gahm, G. F., Sergeev, S., Arkharov, A., Barsunova, O., Tambovtseva, L.	2008, A\&A,  489, 1233	

\bibitem[Guinan \& DeWarf(2002)]{guinan02}
Guinan, E. F., \& DeWarf, L. E. 2002, in Exotic Stars as Challenges to Evolution,
ed. C. A. Tout, \& W. van Hamme, ASP Conf. Ser., 279, 121

\bibitem[Gutermuth et al.(2009)]{gutermuth09}
Gutermuth, R. A., Megeath, S. T., Myers, P. C., Allen, L. E., Pipher, J. L., \& Fazio, G. G. 2009, ApJS, 184, 18
	

\bibitem[Herbst et al.(1994)]{herbst94}
Herbst, W., Herbst, D. K., Grossman, E. J., \& Weinstein, D. 1994, AJ, 108, 1906
 

\bibitem[Huchra et al.(2012)]{huchra12}
  Huchra, J.P., et al. 2012, ApJS, 199, 26
  
\bibitem[Jarrett et al.(2011)]{jarrett11}
Jarrett, T. H. et al. 2011, ApJ, 735, 112
	 

\bibitem[Kearns \& Herbst(1998)]{kearns98}	
Kearns, K. E., \& Herbst, W. 1998, AJ, 116, 261
	
\bibitem[Kloppenborg et al.(2010)]{kloppenborg10}
Kloppenborg, B., Stencel, R., Monnier, J. D., et al.\  2010, Nature, 464, 870

\bibitem[Lavaux \& Hudson(2011)]{lavaux11}
Lavaux,  G., \&  Hudson,  M.J. 2011, MNRAS, 416,  2840



\bibitem[Mamajek et al.(2012)]{mamajek12}
Mamajek, E. E., Quillen, A. C., Pecaut, M. J.,  Moolekamp, F., Scott, E. L., Kenworthy, M. A., 
Collier Cameron, A., \& Parley, N. R. 2012, AJ, 143, 72	



\bibitem[Mikolajewski et al.(2005)]{mikolajewski05}
Mikolajewski, M., et al. 
2005, Ap\&SS, 296, 445

\bibitem[Meng et al.(2014)]{meng14}
Meng, Z., Quillen, A. C., Bell, C. P. M., Mamajek, E. E., Scott, E. L. 2014, 
\url{http://arxiv.org/abs/1401.1248}

\bibitem[Mikolajewski \& Graczyk(1999)]{mikolajewski99}
Mikolajewski, M., \& Graczyk, D. 1999, MNRAS, 303, 521

\bibitem[Morales-Calderon et al.(2011)]{moralescalderon11} 
Morales-Calderon, M. et al. (2011), ApJ, 733, 50 

\bibitem[Nefs et al.(2012)]{nefs12}
Nefs, S. V., et al. 
2012, MNRAS, 425, 950	

\bibitem[Norton et al.(2011)]{norton11}
 Norton, A. J. et al. 2011 A\&A, 528, 90 

\bibitem[Ofek \& Frail(2011)]{ofek11}
Ofek, E.O., \& Frail, D.A. 2011, ApJ, 737, 45 

\bibitem[Palaversa et al.(2013)]{palaversa13}
Palaversa, L.  et al. 2013, AJ, 146, 101

\bibitem[Parks et al.(2013)]{parks13}
Parks, R. J., Plavchan, P.  White, R. J., \&  Gee, A. H. 2013, arxiv/1309.5300

\bibitem[Paturel et al.(2003)]{paturel03}
    Paturel G., Petit C., Prugniel P., Theureau G., Rousseau J., Brouty M.,
    Dubois P., Cambresy L. 2003, A\&A, 412, 45
    
\bibitem[Paturel et al.(2005)]{paturel05}
Paturel, G., Vaughlin, I., Petit, C., Borsenberger, J., Epchtein, N.; Fouque, P., Mamon, G.   2005A\&A, 430, 751

\bibitem[Plavchan et al.(2008)]{plavchan08}
Plavchan, P.,  Jura, M.,   Kirkpatrick, J. D., Cutri, R. M.,  Gallagher, S.C. 2008, ApJS, 175, 191

\bibitem[Plavchan et al.(2008b)]{plavchan08b}
Plavchan, P., Gee, A. H., Stapelfeldt, K., Becker, A.	 2008, ApJ, 684, L37	

\bibitem[Rodriguez et al.(2013)]{rodriguez13}
Rodriguez, J. E., Pepper, J., Stassun, K. G., Siverd, R. J., Cargile, P., Beatty, T. G., \& Gaudi, B. S.	
2013, AJ, 146, 112

\bibitem[Roeser et al.(2010)]{PPMXL}
Roeser, S., Demleitner, M., Schilbach, E
2010, AJ, 139, 2440

\bibitem[Rucinski(2007)]{rucinski07}
Rucinski, S. M. 2007, MNRAS, 382, 393


 \bibitem[Sakata et al.(2010)]{sakata10}
 Sakata, Y., et al. 2010, ApJ, 711, 461
 
\bibitem[Schawinski et al.(201)]{schawinski10} 
 Schawinski, K.,  et al.  2010, ApJ, 711, 284

 \bibitem[Skrutskie et al.(2006)]{skrutskie06}  
Skrutskie, M., et al., 2006, AJ, 131, 1163



\bibitem[Udalski et al.(2003)]{udalski03}
Udalski, A. 2003, Acta Astron., 53, 291

\bibitem[van Eyken et al.(2011)]{vaneyken11}
van Eyken, J. C., et al.  
2011, AJ, 142, 60	

\bibitem[Webb \& Malkan(2000)]{webb00}
Webb, W. \& Malkan, M. 2000, ApJ,  540, 652

\bibitem[Winkler(1997)]{winkler97}
Winkler, H. 1997, MNRAS, 292, 273

\bibitem[Winn et al.(2004)]{winn04}
 Winn, J. N., Holman, M. J., Johnson, J. A., Stanek, K. Z., \& Garnavich, P. M. 2004, ApJ, 603, L45
 
\bibitem[Wolk et al.(2013)]{wolk13}	
Wolk, S. J., Rice, T. S., Aspin, C. 2013, ApJ, 773, 145

\bibitem[Wright et al.(2012)]{wright12}  
Wright, E. L., Eisenhardt, P. R. M., Mainzer, A. K., et al. 2012, AJ, 140,
1868

\bibitem[Yan et al.(2013)]{yan13}  
Yan, L., Donoso, E., Tsai, C.-W., et al. 2013, AJ, 145, 55
 
\newpage

\end{thebibliography}
\end{document}